\documentclass[a4paper,fleqn,usenatbib]{mnras}

\usepackage{amsmath,amssymb}    

\usepackage{mathptmx}

\usepackage{etoolbox}
\makeatletter
\patchcmd\@combinedblfloats{\box\@outputbox}{\unvbox\@outputbox}{}{%
   \errmessage{\noexpand\@combinedblfloats could not be patched}%
}%
\makeatother

\usepackage[T1]{fontenc}
\usepackage{ae,aecompl}

\usepackage{graphicx}   
\usepackage{amssymb}    

\usepackage{natbib, aas_macros}
\usepackage{graphicx, color, url}
\usepackage{grffile} 

\newcommand{\kms}{\mathrm{km\ s^{-1}}\,}

\title[The dynamics of stellar disks in live halos]{The dynamics of stellar disks in live dark-matter halos}
\author[M. S. Fujii et al.]{M. S. Fujii$^{1}$\thanks{E-mail:
fujii@astron.s.u-tokyo.ac.jp (MSF)}, J. B\'edorf$^{2}$, J. Baba$^{3}$, and S. Portegies Zwart$^{2}$\\
$^{1}$Department of Astronomy, Graduate School of Science, The University of Tokyo, 7-3-1 Hongo, Bunkyo-ku, Tokyo, 113-0033, Japan\\
$^{2}$Leiden Observatory, Leiden University, NL-2300RA Leiden, The Netherlands\\
$^{3}$National Astronomical Observatory of Japan, Mitaka-shi, Tokyo 181-8588, Japan}

\date{Accepted . Received ; in original form }

\begin{document}
\label{firstpage}
\pagerange{\pageref{firstpage}--\pageref{lastpage}} \pubyear{2002}
\maketitle

\begin{abstract}

Recent developments in computer hardware and software enable
researchers to simulate the self-gravitating evolution of galaxies at
a resolution comparable to the actual number of stars.  Here we
present the results of a series of such simulations.  We performed
$N$-body simulations of disk galaxies with between 100 and 500 million
particles over a wide range of initial conditions.  Our calculations
include a live bulge, disk, and dark matter halo, each of which is
represented by self-gravitating particles in the $N$-body code.  The
simulations are performed using the gravitational $N$-body tree-code
{\tt Bonsai} running on the Piz Daint supercomputer.
We find that the time scale over which the bar forms increases
exponentially with decreasing disk-mass fraction and that the bar formation
epoch exceeds a Hubble time when the disk-mass fraction is $\sim0.35$.
These results can be explained with the swing-amplification theory.
The condition
for the formation of $m=2$ spirals is consistent with that for the
formation of the bar, which is also an $m=2$ phenomenon.  We further argue
that the non-barred grand-design spiral galaxies
are transitional, and that they evolve to barred
galaxies on a dynamical timescale.
We also confirm that the disk-mass fraction and shear rate
are important parameters for the morphology of disk galaxies.
The former affects the number of spiral arms and the bar formation
epoch, and the latter determines the pitch angle of the spiral arms.

\end{abstract}


\section{Introduction}

Simulations serve as a powerful tool to study the dynamical evolution
of galaxies. Galaxies are extremely complicated, in particular due to
the coupling between a varying environment and their internal
evolution.  Even the relatively simple self-gravity of an isolated
galaxy poses enormous challenges, in particular because of the
non-linear processes that govern the formation of spiral arms and
bar-like structures.  Many of these processes have been attributed to
perturbations from outside, and it is not a priori clear to what
extent internal dynamical processes play a role in the formation of
axis-asymmetric structures in disk galaxies.  Self-gravitating disks
are prone to form spiral arms and/or bars, but the precise conditions
under which these form are not well understood.

In earlier simulations~\citet{1971ApJ...168..343H} demonstrated, using
$\sim 7\times 10^4$ cells with near-neighbour interactions, that a 
stellar disk
without a (dark matter) halo leads to the formation of a bar within a
few galactic rotations.  In a subsequent
study~\citet{1973ApJ...186..467O} concluded that a dark-matter halo is
required to keep the disk stable.  For spiral galaxies,
\citet{1984ApJ...282...61S} performed simulations of two-dimensional 
stellar disks with $\sim 7000$ cells and $2\times 10^4$ particles
that developed multiple spiral arms. 
They suggested that spiral arms
tend to kinematically heat-up the disk, and that in the absence of an
effective coolant, such as ambient gas or star formation, this heating
would cause the spiral structures to disappear within a few galactic
rotations. \citet{1985ApJ...298..486C} also performed a series of 
simulations and found that the number of spiral arms decrease as
the disk-to-halo mass ratio decreases. 
In contradiction to the Lin-Shu quasi-stationary density-wave theory
\citep{1964ApJ...140..646L}, these simulations suggested that 
spiral arms are transient and develop from small perturbations
amplified by the self-gravity in a differentially rotating disk
\citep{1965MNRAS.130..125G, 1966ApJ...146..810J}.
Today this mechanism is known as ``swing amplification''
\citep{1981seng.proc..111T,2016ApJ...823..121M}.

The number of particles in simulations has gradually
increased with time as computers have become more powerful.  
After \citet{1984ApJ...282...61S} and
\citet{1985ApJ...298..486C}, the formation and evolution of bar
structures were often studied using three dimensional $N$-body
simulations. \citet{1990A&A...233...82C} showed that bars induce a
peanut-shaped (boxy) bulge using a three-dimensional Particle-Mesh
method with at most $\sim 8\times 10^6$ cells.  $N$-body simulations
with up to $2\time 10^8$ particles were performed by
\citet{2014ApJ...785..137S}, but they adopted a rigid potential for
the dark matter halo.  Performing such simulations with a `live'
particle halo has been hindered by the sheer computer power needed to
resolve baryonic and non-baryonic material simultaneously.

The importance of resolving the halo in such simulations
using particles was emphasized by \citet{2002ApJ...569L..83A}. They
found that once a bar formed its angular momentum is transferred to
the halo. This angular momentum transport can only be resolved in the
simulations if the halo is represented by particles that are
integrated together with the rest of the galaxy. The dynamics and
back-coupling of such a live halo also affects the evolution of the
bar.  In this study, we adopt a tree method
\citep{1986Natur.324..446B} for solving the equations of motion of all
particles in the simulations.  Several other simulations of barred
galaxies, in which halos were resolved using live particles, also
adopted a tree-code \citep{2005ApJ...631..838W,2008ApJ...679.1239W}.
\citet{2009ApJ...697..293D} performed a series of such $N$-body
simulations with up to $10^8$ particles. They confirmed that such a
large number of particles is required but also sufficient to obtain a
reliable solution for the morphology of the bar.  However, they
also found that bar formation in simulations with a larger number of
particles was systematically delayed.

Another problem of relatively low-resolution is the artificial heating
of the particles by close encounters.  In simulations of spiral
galaxies with multiple arms, \citet{2011ApJ...730..109F} demonstrated that
this effect becomes sufficiently small when the disk is resolved with
at least one million particles.  If a galactic disk needs to be
resolved with at least a million particles, it is understandable that
simulating an entire galaxy, including the dark matter halo, would
require at least 10 times this number in order to properly resolve the
disk and halo in an $N$-body simulation.  To overcome the numerical
limitations researchers tend to adopt a rigid background potential for
the galaxy's dark matter halo.
\citep[e.g.,][]{1984ApJ...282...61S,2013ApJ...763...46B,
  2013A&A...553A..77G}, as was also done in
\citet{2011ApJ...730..109F}.  In these simulations energy cannot be
transported self-consistently between the halo and the
disk, and vice versa.  Taking this coupling into account is
particularly relevant when studying the formation and evolution of
non-axisymmetric structures such as spiral arms and a bar in the disk;
bars tend to slow down due to angular-moment transfer with the halo
and grow faster when compared to models with a rigid
halo~\citep{2002ApJ...569L..83A}.

Up to now it has not been possible to carry out extensive parameter
searches with a sufficient number of particles that include a live
halo, simply because the amount of computer time required for such
studies exceeds the hardware and software capacity
\citep[e.g.][]{2009ApJ...697..293D}.  The current generation of
GPU-based supercomputers and optimized $N$-body algorithms
\cite{7328651} allows us to perform simulations with more than a
hundred billion particles \citep{2014hpcn.conf...54B} over a Hubble
time. The same developments allow us to perform an extensive parameter
over a wide range of initial conditions with a more modest number of
particles.

We developed the gravitational tree-code {\tt Bonsai} to perform such
simulations.  {\tt Bonsai}, uses the GPUs to accelerate the
calculations and achieves excellent efficiency with up to $\sim 19000$
GPUs \citep{2012JCoPh.231.2825B,2014hpcn.conf...54B}.  The efficiency
of {\tt Bonsai} allows us to run simulations with a hundred million
particles for 10\,Gyr in a few hours using 128 GPUs in parallel.  This
development allows us to perform simulations of disk galaxies for the
entire lifetime of the disk, and therefore to study the formation of
structure using a realistic resolution and time scale. The code is
publicly available and part of the {\tt AMUSE} framework
\citep{AMUSE}.

Using {\tt Bonsai} running on the CSCS Piz Daint supercomputer we
performed a large number of disk galaxy simulations with live
dark-matter halos at a sufficiently high resolution to suppress
numerical heating on the growth of the physical instabilities due to
the self-gravity of the disk.  With these simulations, we study the
relation between the initial conditions and the final disk galaxies.

\section{$N$-body simulations}

We performed a series of $N$-body simulations of galactic stellar
disks embedded in dark matter halos. In this section, we describe our
choice of parameters and the $N$-body code used for these simulations.

\subsection{Model}
Our models are based on those described in \citet{2008ApJ...679.1239W} and 
\citet{2005ApJ...631..838W}. 
We generated the initial conditions using GalactICS \citep{2005ApJ...631..838W}.
The initial conditions for generating the dark mater halo are taken from the
NFW profile \citep{1997ApJ...490..493N}, which has a density profile following:
\begin{eqnarray}
\rho_{\rm NFW}(r) = \frac{\rho_{\rm h}}{(r/a_{\rm h})(1+r/a_{\rm h})^3},
\end{eqnarray}
and the potential is written as
\begin{eqnarray}
\Phi_{\rm NFW} = -\sigma_{\rm h}^2\frac{\log (1+r/a_{\rm h})}{r/a_{\rm h}}.
\end{eqnarray}
Here the gravitational constant, $G$, is unity, 
$a_{\rm h}$ is the scale radius, $\rho_{\rm h}\equiv\sigma^2/4\pi a_{\rm h}^2$
is the characteristic density, and $\sigma_{\rm h}$ is the characteristic 
velocity dispersion. We adopt $\sigma_{\rm h}=340$ (km\,s$^{-1}$), 
$a_{\rm h}=11.5$ (kpc).
Since the NFW profile is infinite in extent and mass, the 
distribution is truncated by a halo tidal radius using
an energy cutoff $E_{\rm h}\equiv\epsilon_{\rm h}\sigma_{\rm h}^2$,
where $\epsilon _{\rm h}$ is the truncation parameter with $0<\epsilon _{\rm h}<1$.
Setting $\epsilon _{\rm h}=0$ yields a full NFW profile
\citep[see][for details]{2005ApJ...631..838W}. 
We choose the parameters of the dark matter halo such that the resulting
rotation curves have a similar shape.
The choice of parameters is summarized in Table~\ref{tb:params}.

For some models we assume the halo to have net angular momentum. This
is realized by changing the sign of the angular momentum about the
symmetry axis ($J_{z}$). The rotation is parameterized using $\alpha
_{\rm h}$.  For $\alpha _{\rm h}=0.5$, the fraction of halo particles
which have positive or negative $J_{z}$ are the same, in which case the disk
has no net angular momentum. If $\alpha _{\rm h}>0.5$, the halo rotates in the same
direction as the disk.

For the disk component, we adopt an exponential disk for which the surface density
distribution is given by
\begin{eqnarray}
\Sigma (R) = \Sigma_{0} {\rm e}^{-R/R_{\rm d}}.
\end{eqnarray}
The vertical structure is given by ${\rm sech}^2(z/z_{\rm d})$,
where $z_{\rm d}$ is the disk scale height.
The radial velocity dispersion is assumed to follow 
$\sigma_{R}^2(R)=\sigma_{R0}^2\exp(-R/R_{\rm d})$, where 
$\sigma_{R0}$ is the radial velocity dispersion at the disk's center.
Toomre's stability parameter $Q$ 
\citep{1964ApJ...139.1217T,2008gady.book.....B} at
a reference radius (we adopt $2.2R_{\rm d}$), $Q_{0}$, is controlled
by the central velocity dispersion of the disk ($\sigma _{R0}$). 
We tune $\sigma_{R0}$ such that for our standard model (md1mb1) $Q_0=1.2$ .

For model md1mb1 we use as disk mass $M_{\rm d}$=$4.9\times 10^{10}$ $M_{\odot}$,
and the scale length $R_{\rm d}$=2.8 kpc. The disk's truncation radius is set to 30 kpc, the 
scale height $z_{\rm d}$=0.36 kpc and the radial velocity
dispersion at the center of the galaxy to $\sigma_{R0}$=$105$ km\,s$^{-1}$.
The disk is truncated at ($R_{\rm out}$) with a 
radial range for disk truncation ($\delta R$). We adopt 
$R_{\rm out}=30.0$ (kpc) and $\delta R=0.8$ (kpc).

For the bulge we use a Hernquist model~\citep{1990ApJ...356..359H},
but the distribution function is extended with an energy cutoff
parameter ($\epsilon_{\rm b}$) to truncate the profile much in the
same way as we did with the halo model.  The density distribution
and potential of the standard Hernquist model is
\begin{eqnarray}
\rho_{\rm H} = \frac{\rho_{\rm b}}{(r/a_{\rm b})(1+r/a_{\rm b})^3}
\end{eqnarray}
and
\begin{eqnarray}
\Phi_{\rm H} = \frac{\sigma_{\rm b}^2}{1+r/a_{\rm b}}.
\end{eqnarray}
Here $a_{\rm b}$, $\rho_{\rm b}=\sigma_{\rm b}^2/(2\pi a_{\rm b}^2)$, and
$\sigma _{\rm b}$ are the scale length, characteristic density, and
the characteristic velocity of the bulge, respectively.
We set $\sigma_{\rm b}$=300 km\,s$^{-1}$, bulge scale length $a_{\rm b}$=0.64 kpc,
and the truncation parameter ($\epsilon _{\rm b}$=0.0).
This results in a bulge mass of $4.6\times 10^9M_{\odot}$,
which is consistent with the Milky Way model proposed by
\citet{2010ApJ...720L..72S}, and reproduces
the bulge velocity distribution obtained by BRAVA observations~\citep{2012AJ....143...57K}.
We do not assume an initial rotational velocity for the bulge.

For the simulation models we vary the disk mass, bulge mass,
scale length, halo spin, and $Q_0$.  Since the adopted generator for
the galaxies is an irreversible process and due to the randomization
of the selection of particle positions and velocities we cannot
guarantee that the eventual velocity profile is identical to the input
profile, but we confirmed by inspection that they are
indistinguishable.  The initial conditions for each of the models are
summarized in Table \ref{tb:params}.  The mass and tidal radius for
the bulge, disk, and halo as created by the initial condition
generator are given in Table \ref{tb:mass_radius}.

In each of the models we fix  the number of particles used for 
the disk component to $8.3 \times 10^6$. For the bulge and halo
particles we adopt the same particle mass as for the disk particles.
As a consequence the mass ratios between the bulge, halo and disk are
set by having a different number of particles 
used per component (Table~\ref{tb:mass_radius}).

\begin{table*}
\begin{center}
  \caption{Model names and their parameters.  The columns represent,
    1: Model name, 2: Halo scale radius, 3: Halo characteristic
    velocity dispersion, 4: Halo truncation parameter, 5: Halo
    rotation parameter, 6: Disk mass, 7: Disk scale radius, 8: Disk
    scale height, 9: Disk radial velocity dispersion at the center of
    the disk, 10: Bulge scale length, 11: Bulge characteristic
    velocity, 12: Bulge truncation parameter.  In the model names,
    `md' and `mb' indicate disk and bulge masses. `Rd' and `rb'
    indicate the disk and halo scale radii. `s' indicates models with
    halo spin. `Q' indicates the initial $Q$ value. For these, the
    values of model md1mb1 are referred to as 1.  For all models,
    movies of the evolution are available in the online materials.
\label{tb:params}
}
\begin{tabular}{lccccccccccc}
  \hline
  (1)      &   (2)      &   (3)      &   (4)      &   (5)      &   (6)      &   (7)      &   (8)      &   (9)      &   (10)      &   (11)      &   (12)      \\ 
           &  \multicolumn{4}{l}{Halo} &  \multicolumn{4}{l}{Disk} &  \multicolumn{3}{l}{Bulge} \\
Parameters &  $a_{\rm h}$ & $\sigma_{\rm h}$ & $1-\epsilon_{\rm h}$ & $\alpha_{\rm h}$& $M_{\rm d}$ & $R_{\rm d}$ & $z_{\rm d}$ & $\sigma_{R0}$  & $a_{\rm b}$ & $\sigma_{\rm b}$ & $1-\epsilon_{\rm b}$ \\ 
Model   &  (kpc) & ($\kms$) &  &  & $(10^{10}M_{\odot})$ & (kpc) & (kpc) & ($\kms$)  & (kpc) & $(\kms)$\\
\hline \hline
md1mb1        &11.5 &  340 & 0.8 & 0.5 & $4.9$ & 2.8 & 0.36 & 105  & 0.64 & 300 & 1.0 \\
md1mb1s0.65    &11.5 &  340 & 0.8 & 0.65 & $4.9$ & 2.8 & 0.36 & 105  & 0.64 & 300 & 1.0 \\
md1mb1s0.8    &11.5 &  340 & 0.8 & 0.8 & $4.9$ & 2.8 & 0.36 & 105  & 0.64 & 300 & 1.0 \\
\\
md0.5mb1  & 8.2 &  310 & 0.88 & 0.5 & $2.5$ & 2.8 & 0.36 & 59.2  & 0.64 & 300 & 0.86 \\
md0.4mb1  & 7.6 &  300 & 0.91 & 0.5 & $2.0$ & 2.8 & 0.36 & 49.0  & 0.64 & 300 & 0.84 \\
md0.3mb1  & 7.0 &  287 & 0.92 & 0.5 & $1.5$ & 2.8 & 0.36 & 38.5  & 0.64 & 300 & 0.82 \\
md0.1mb1  & 6.0 &  285 & 0.97 & 0.5 &$0.49$ & 2.8 & 0.36 & 13.5  & 0.64 & 300 & 0.79 \\
\\
md0.5mb0  & 22.0 &  450 & 0.7 & 0.5 & $2.3$ & 2.8 & 0.36 & 62.6  & 0.64 & 500 & 0.86 \\
md0.5mb3  & 7.0 &  270 & 0.8 & 0.5 & $2.3$ & 2.8 & 0.36 & 59.0  & 0.64 & 500 & 0.79 \\
md0.5mb4  & 6.6 &  260 & 0.82 & 0.5 & $2.3$ & 2.8 & 0.36 & 58.3  & 0.64 & 545 & 0.80 \\
md0.5mb4rb3  & 13.5 &  360 & 0.8 & 0.5 & $2.3$ & 2.8 & 0.36 & 57.2  & 1.92 & 380 & 0.99 \\
\\
md1mb1Rd1.5      &  9.0 &  290 & 0.95 & 0.5 & $4.9$ & 4.2 & 0.36 & 74.2  & 0.64 & 300 & 0.85 \\
md0.5mb1Rd1.5     &  7.5 &  290 & 0.91 & 0.5 & $2.5$ & 4.2 & 0.36 & 39.8  & 0.64 & 300 & 0.8 \\
md0.5Rmb1d1.5s &  7.5 &  290 & 0.91 & 0.8 & $2.5$ & 4.2 & 0.36 & 39.8  & 0.64 & 300 & 0.8 \\
\\
md1.5mb5      & 13.0 & 280 & 0.9 & 0.5 & 7.3 &  2.8 & 0.36 & 138 & 1.0 & 550 & 0.8 \\
md1mb10       & 18.0 & 500 & 0.9 & 0.5 & 4.9 & 2.8 & 0.36  & 93.2 & 1.5 & 600 & 1.0 \\
\\
md0.5mb0Q0.5  & 22.0 &  450 & 0.7 & 0.5 & $2.3$ & 2.8 & 0.36 & 26.1  & 0.64 & 500 & 0.86 \\
md0.5mb0Q2.0  & 22.0 &  450 & 0.7 & 0.5 & $2.3$ & 2.8 & 0.36 & 105  & 0.64 & 500 & 0.86 \\
\hline
\end{tabular}
\end{center}
\end{table*}

\begin{table*}
\begin{center}
  \caption{Models: mass, radius, and number of particles per component.
    Column 1: Model name, 2: Disk mass, 3: Bulge mass, 4: Halo mass, 5: Disk outer radius, 6: Bulge outer radius, 7: Halo outer radius, 
8: Toomre's $Q$ value at the reference point ($2.2R_{\rm d}$), 
9: Bulge-to-disk mass ratio ($B/D$), 10: Number of particles for the disk, 11:  Number of particles for the bulge, 12: Number of particles for the halo.
\label{tb:mass_radius}}
\begin{tabular}{lccccccccccc}
\hline
  (1)      &   (2)      &   (3)      &   (4)      &   (5)      &   (6)      &   (7)      &   (8)      &   (9)      &   (10)      &   (11)      &   (12)      \\ 
Model    & $M_{\rm d}$ & $M_{\rm b}$ & $M_{\rm h}$ & $R_{\rm d, t}$ & $r_{\rm b, t}$ & $r_{\rm h, t}$ & $Q_0$   &  $M_{\rm b}/M_{\rm d}$ & $N_{\rm d}$ & $N_{\rm b}$ & $N_{\rm h}$\\ 
   & ($10^{10}M_{\odot}$) & ($10^{10}M_{\odot}$) & ($10^{10}M_{\odot}$) & (kpc) & (kpc) & (kpc) &  &   &  &  & \\ 
\hline  \hline
md1mb1       & 4.97 &  0.462 & 59.7 & 31.6 & 3.17 & 229 & 1.2  & 0.0930 & 8.3M & 0.77M & 100M \\
md1mb1s0.65       & 4.97 &  0.462 & 59.7 & 31.6 & 3.17 & 229 & 1.2   & 0.0930 & 8.3 M & 0.77M & 100M \\
md1mb1s0.8       & 4.97 &  0.462 & 59.7 & 31.6 & 3.17 & 229 & 1.2  & 0.0930 & 8.3M & 0.77M & 100M \\
\\
md0.5mb1      & 2.55 &  0.465 & 43.8 & 31.6 & 2.56 & 232 & 1.2  & 0.182  & 8.3M & 1.5M & 140M \\
md0.4mb1      & 2.05 &  0.463 & 41.4 & 31.6 & 2.52 & 261 & 1.2  & 0.226 & 8.3M & 1.9M & 170M \\
md0.3mb1      & 1.56 &  0.462 & 36.2 & 31.6 & 2.49 & 247 & 1.2 & 0.296 & 8.3M & 2.5M & 190M \\
md0.1mb1      & 0.546 & 0.466 & 33.3 & 31.6 & 2.44 & 340 & 1.2  & 0.853 & 8.3M & 7.1M & 510M \\
\\
md0.5mb0      & 2.53 &  0.0 & 100.0 & 31.6 & - & 295 & 1.2  & 0.00 & 8.3M & - & 330M \\
md0.5mb3      & 2.61 &  1.37 & 39.7 & 31.6 & 2.81 & 120 & 1.2 & 0.525 & 8.3M & 4.4M & 130M \\
md0.5mb4      & 2.62 &  1.69 & 41.4 & 31.6 & 2.96 & 125 & 1.2 & 0.645 & 8.3M & 5.4M & 130M \\
md0.5mb4rb3   & 2.60 &  1.76 & 86.7 & 31.6 & 8.55 & 229 & 1.2 & 0.676 & 8.3M & 5.4M & 130M \\
\\ 
md1mb1Rd1.5      & 5.06 &  0.464 & 47.1 & 46.6 & 2.61 & 620 & 1.2 &  0.0916 & 8.3M & 0.77M & 78M \\
md0.5mb1Rd1.5    & 2.59 &  0.457 & 35.2 & 46.6 & 2.47 & 249 & 1.2 & 0.176 & 8.3M & 1.5M & 110M \\
md0.5mb1Rd1.5s & 2.59 &  0.457 & 35.2 & 46.6 & 2.47 & 249 & 1.2 & 0.176 & 8.3M & 1.5M & 110M \\
\\
md1.5mb5      & 7.52  &  2.09  & 104.6 & 31.6 & 3.53 & 269 & 1.2 & 0.279 & 8.3M & 2.3M & 120M \\
md1mb10       & 5.17  & 5.22  & 2050  & 31.6 & 11.6 & 264 & 1.2 &  1.0 & 8.3M & 8.4M & 400M \\
\\
md0.5mb1Q0.5   & 2.55 &  0.465 & 43.8 & 31.6 & 2.56 & 232 & 0.5 & 0.182 & 8.3M & 1.5M & 140M \\
md0.5mb1Q2.0   & 2.55 &  0.465 & 43.8 & 31.6 & 2.56 & 232 & 2.0  & 0.182 & 8.3M & 1.5M & 140M \\
\\
\hline
\end{tabular}
\end{center}
\end{table*}

\subsection{The {\tt Bonsai} optimized gravitational $N$-body tree-code}

We adopted the {\tt Bonsai} code for all calculations
\citep{2012JCoPh.231.2825B, 2014hpcn.conf...54B}.  {\tt Bonsai}
implements the classical Barnes \& Hut algorithm
\citep{1986Natur.324..446B} but then optimized for Graphics
Processing Units (GPU) and massively parallel operations. In {\tt Bonsai}
all the compute work, including the tree-construction, takes place on
the GPU which frees up the CPU for administrative tasks. By moving all
the compute work to the GPU there is no need for expensive data copies,
and we take full advantage of the large number of compute cores and
high memory bandwidth that is available on the GPU.  The use of GPUs
allows fast simulations, but we are limited by the relatively small
amount of memory on the GPU. To overcome this limitation we
implemented across-GPU and across-node parallelizations which enable
us to use multiple GPUs in parallel for a single simulation
\citep{2014hpcn.conf...54B}.  Combined with the GPU acceleration, this
parallelization method allows {\tt Bonsai} to scale efficiently from
single GPU systems all the way to large GPU clusters and
supercomputers~\citep{2014hpcn.conf...54B}.  We used the version of
{\tt Bonsai} that incorporates quadrupole expansion of the multipole
moments and the improved Barnes \& Hut opening angle criteria
\citep{2013MNRAS.436.1161I}.  We use a shared time-step of $\sim 0.6$
Myr, a gravitational softening length of 10\,pc and the opening angle
$\theta=0.4$.

Our simulations contain hundreds of millions of particles and
therefore it is critical that the post-processing is handled
efficiently.  We therefore implemented the post-processing methods directly in
{\tt Bonsai} and these are executed while the simulation is progressing. 
This eliminates the
need to reload snapshot data (which can be on the order of a few
terabytes) after the simulation.

The simulations in this work have been run on the Piz Daint supercomputer at
the Swiss National Supercomputing Centre. In this machine each compute node
contains an NVIDIA Tesla K20x GPU and an Intel Xeon E5-2670 CPU. Depending on the number
of particles in the simulation we used between 8 and 512 nodes per simulation.

\section{Results}

\subsection{The effects of disk and bulge masses}

\begin{figure*}
\includegraphics[width=1.8\columnwidth]{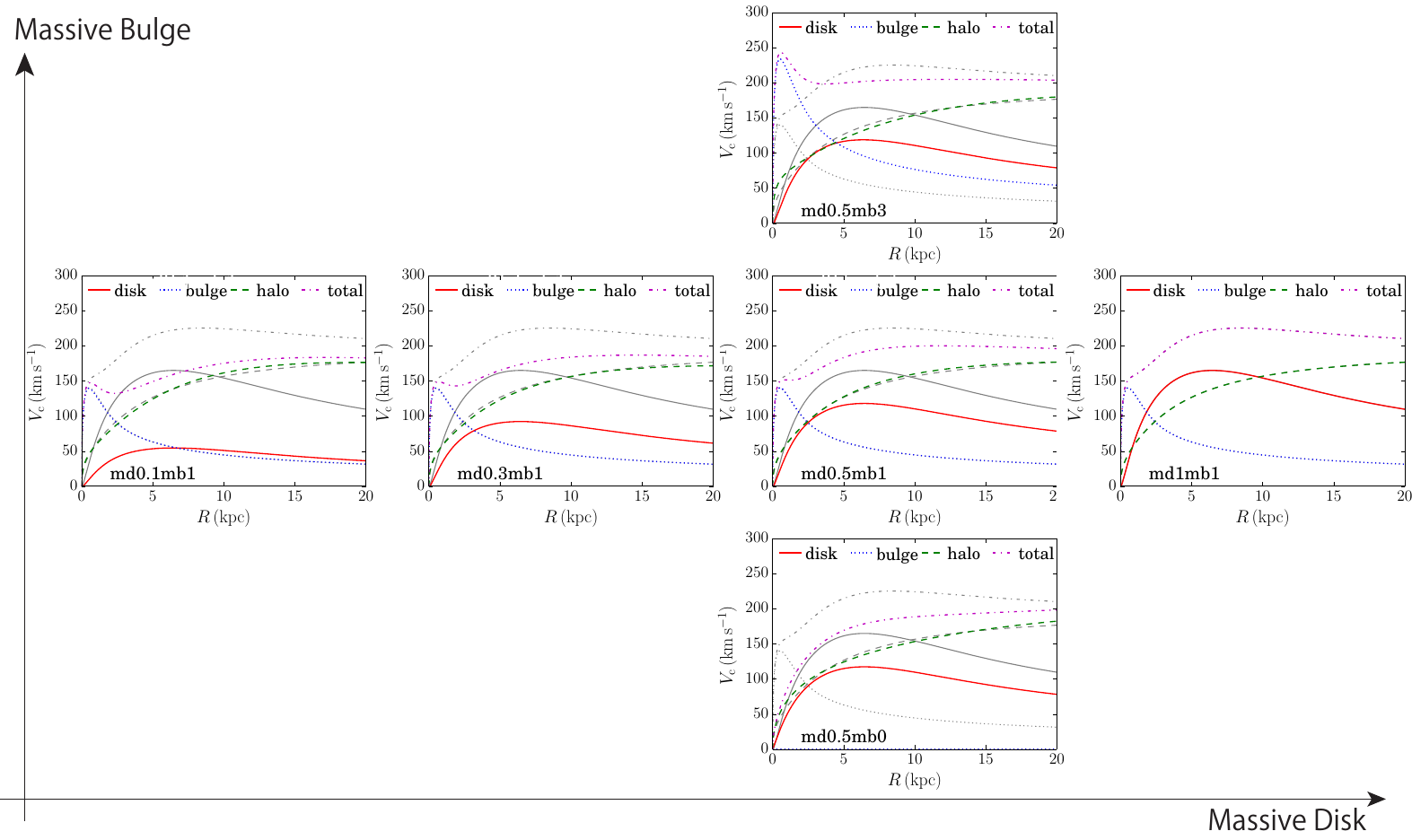}
    \caption{Rotation curves of the initial conditions for models 
      md1mb1, md0.5mb1, md0.3mb1, md0.1mb1, md0.5mb0, and md0.5mb3\label{fig:rotation_curves}.
      The gray solid, dotted, dashed, and dot-dashed curves indicate disk, bulge, halo, and total rotation curves of model md1mb1.}
\end{figure*}

\begin{figure*}
\includegraphics[width=1.8\columnwidth]{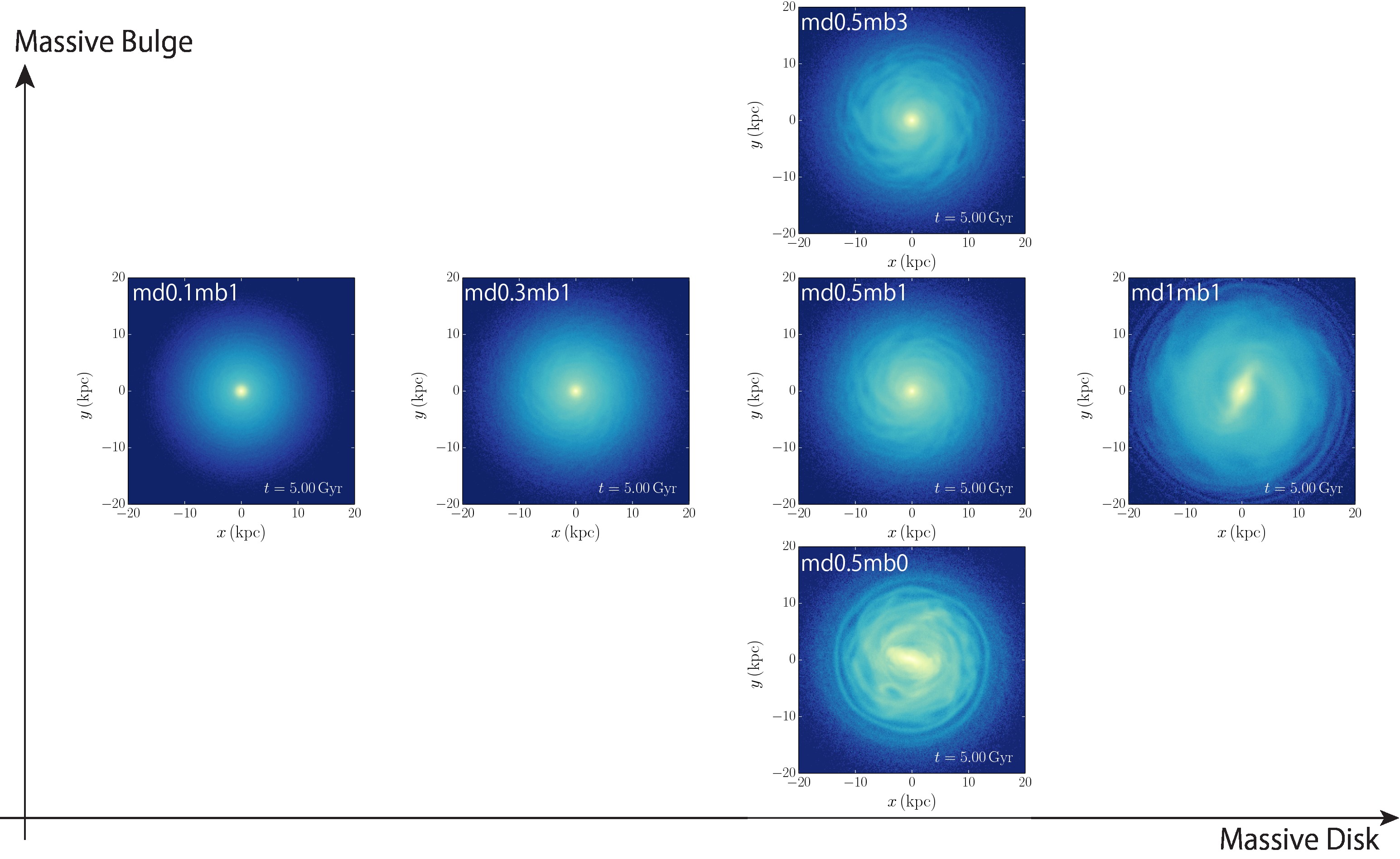}
    \caption{Snapshots (surface densities) at $t=5$ Gyr for models 
      md1mb1, md0.5mb1, md0.3mb1, md0.1mb1, md0.5mb0, and md0.5mb3.
      \label{fig:snapshots_5Gyr}}
\end{figure*}

\begin{figure*}
\includegraphics[width=1.8\columnwidth]{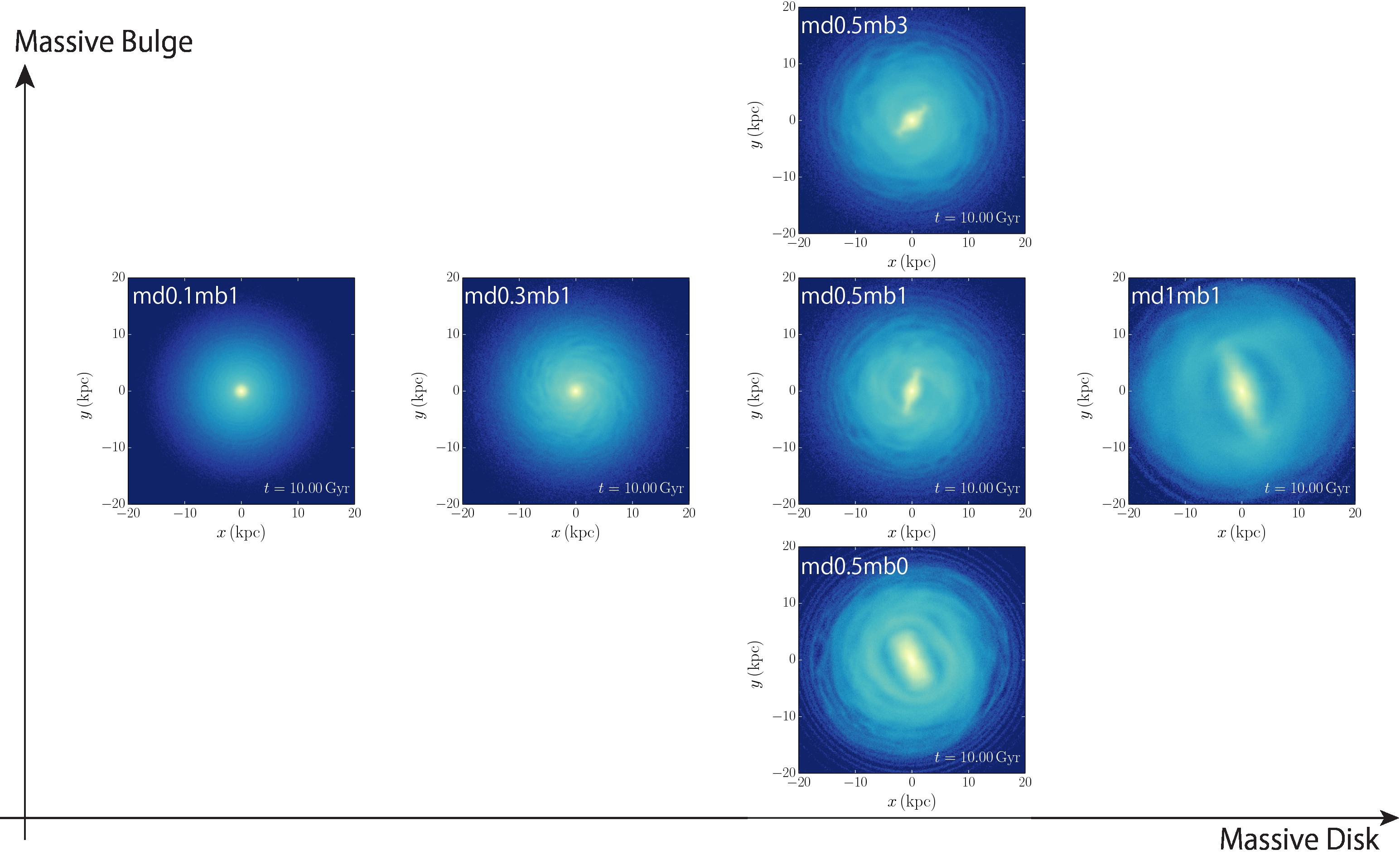}
    \caption{Snapshots (surface densities) at $t=10$ Gyr for models 
      md1mb1, md0.5mb1, md0.3mb1, md0.1mb1, md0.5mb0, and md0.5mb3.
      \label{fig:snapshots_10Gyr}}
\end{figure*}

We study the effect that the disk and bulge mass fractions have on the halo
and on the morphology of spiral arms and bars.  In
Fig.~\ref{fig:rotation_curves}, we summarize the initial 
rotation curves of
several models: models md1mb1, md0.5mb1, md0.3mb1, and md0.1mb1
(varying disk mass) and models md0.5mb0 and md0.5mb3 (varying bulge
mass).  We present the snapshots at $t=5$ and 10\,Gyr in
Figs.~\ref{fig:snapshots_5Gyr} and \ref{fig:snapshots_10Gyr}.  As
reported in previous studies, the number of spiral arms increase as
the disk mass decreases
\citep{1985ApJ...298..486C,2003MNRAS.344..358B,
2011ApJ...730..109F,2015ApJ...808L...8D}
and the formation of the bar is delayed when the bulge mass is increased
\citep{2013MNRAS.434.1287S}. This corresponds to the effect that
centrally concentrated potentials prevent the formation of bars
\citep{2001ApJ...546..176S}.

\subsection{Spiral Arms}
\subsubsection{Number of Spiral Arms}
We first focus on the number of spiral arms. As is shown in 
Fig.~\ref{fig:snapshots_10Gyr}, the number of spiral arms increases 
as the disk mass decreases. This relation can be understood by swing 
amplification theory \citep{1981seng.proc..111T}.
In a differentially rotating disk, the epicycle motions of particles 
are amplified and the amplification factor $X$ is written as 
\begin{eqnarray}
X \equiv \frac{k_{\rm crit}R}{m} = \frac{\kappa^2R}{2\pi G\Sigma m}.
\label{eq:X}
\end{eqnarray}
Here $R$ is the distance from the galactic center and $m$ is
the spiral multiplicity (the number of spiral arms). This $m$ is
  usually used for spirals, we adopt $m=2$ for bars, because bars are
  traced as a $m=2$ mode when using a Fourier decomposition. For
typical disk models the amplification is large for $1\lesssim
X\lesssim 2$ and rapidly drops for $2\lesssim X\lesssim
3$~\citep{1965MNRAS.130..125G,1966ApJ...146..810J,1981seng.proc..111T}.
The critical wave number $k_{\rm crit}$ (and also the critical wave
length $\lambda_{\rm crit}$) is obtained from the local stability in a
razor-thin disk using the tight-winding approximation
\citep{1964ApJ...139.1217T}:
\begin{eqnarray}
k_{\rm crit} &=& \frac{\kappa^2}{2\pi G\Sigma},\\
\lambda_{\rm crit} &=& \frac{2\pi}{k_{\rm crit}} = \frac{4\pi^2G\Sigma}{\kappa^2},
\end{eqnarray} 
where $\Sigma$ and $\kappa$ are the surface density and the epicyclic
frequency of the disk, respectively \citep[see also section 6.2.3 of
][]{2008gady.book.....B}.  

Equation (\ref{eq:X}) also predicts the number of spiral arms that
form in a disk.  By inverting equation (\ref{eq:X}) one obtains a
relation for $m$ as a function of the swing amplification factor
$X$:
\begin{eqnarray}
m =\frac{\kappa^2R}{2\pi G\Sigma X}.
\label{eq:mX}
\end{eqnarray}
Because the perturbations grow most efficiently for $X\sim1$--2, we
can relate $m$ as a function of $R$ (here, both $\kappa$ and $\Sigma$
are written as a function of $R$.). The
predicted number of spiral arms from swing amplification theory
has been validated using numerical simulations
\citep{1985ApJ...298..486C,2015ApJ...808L...8D}.

\begin{figure}
\includegraphics[width=\columnwidth]{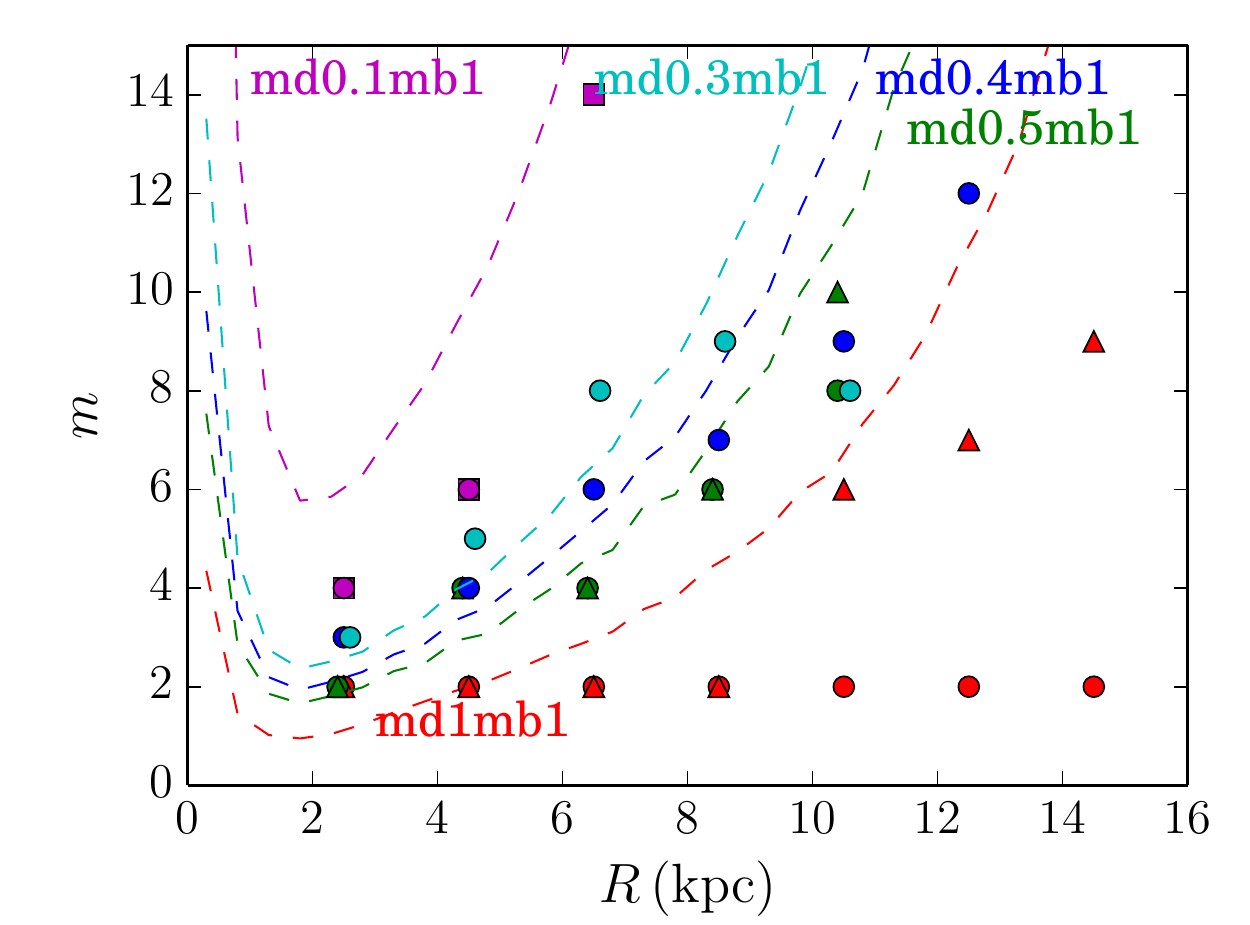}
    \caption{Theoretically predicted (using Eq.~\ref{eq:mX}, dashed 
    curves) and measured number of spiral arms (symbols) for models
    md0.1mb1 (magenta), md0.3mb1 (cyan), md0.4mb1 (blue), md0.5mb1 (green),
    and md1mb1 (red) from top to bottom. 
    Filled circles indicate the most frequently appearing number of arms (principal modes)
    over a 10\,Gyr period. 
    Triangle symbols indicates the principal mode before the formation of the bar.
    Square symbols indicate the principal mode for model md0.1mb1 excluding $m=2$.
    The symbols for md0.3mb1 and md0.5mb1 are shifted by 0.1\,kpc  
    to avoid overlapping points.\label{fig:m_measured}}
\end{figure}

For models with different disk masses, 
we estimate the number of spiral arms using equation (\ref{eq:mX})
and present the results in Fig.~\ref{fig:m_measured}. The dashed curves 
present the estimated number of spiral arms as a function of galactic radii 
where we adopt $X\sim2$ following 
\citet{1985ApJ...298..486C, 2014PASA...31...35D}.
Given the curves we expect fewer arms for the more massive models
and the number of arms increases for larger radii ($R$).

We also determine the number of spiral arms for each of the 
simulated galaxies and overplot the results in Fig.~\ref{fig:m_measured}.
We use a Fourier decomposition of the disks surface density:
\begin{eqnarray}
\frac{\Sigma(R,\phi)}{\Sigma_0(R)}=\sum_{m=0}^{\infty}A_m(R)e^{im[\phi-\phi_m(R)]},
\label{eq:Fourier}
\end{eqnarray}
where $A_m(R)$ and $\phi_m(R)$ are the Fourier amplitude and phase angle for
the $m$-th mode at $R$, respectively. 
We measure the amplitude at each radius up to 20 kpc using radial bins of 
$\Delta R=1$ kpc.
When a bar formed we obtained $m=2$ as strongest amplitude. 

Because the spiral arms are transient structures the dominant number 
of spiral arms,
those with the highest amplitude, changes over time~\citep{2011ApJ...730..109F}.
We therefore use the most frequently appearing 
number of spiral arms (hereafter, principal mode) 
as the number of arms ($m$) of the model.
The principal mode is measured between 2.5 and 14.5\,kpc at 2\,kpc intervals,
and for each the 1000 snapshots
between 0 and 10\,Gyr. The results are presented in Fig.~\ref{fig:m_measured}.
The $m=2$ mode will always become the dominant mode once a bar has formed (see 
red circles in the figure), but spiral arms might have formed before 
the bar formation.
We therefore also show the principal mode before the bar 
formation (triangular symbols). These results are roughly consistent 
with the number of spiral arms predicted by Eq.~\ref{eq:mX}:
the number of spiral arms increases as the galactic radius increases
and as the disk mass decreases.
For model md0.1mb1 we measure a principal mode of 2 at $R=6.5$\,kpc.
However, when we look at Fig.~\ref{fig:snapshots_10Gyr}, we 
see more than 2 faint spiral arms. We therefore also measured the strongest 
modes excluding $m=2$. These modes are indicated by the square symbols.
We perform the same analysis for all the other models and measure the
number of spiral arms (for the details of the individual evolution of these
models, see the following sections and Appendix~\ref{Sect:AppB}). 
The results are summarized in Table~\ref{tb:pitch_angle}.

Numerical results tend to deviate from theoretical predictions
as the number of spiral arms increases. The more spiral arms the 
fainter they become, which makes them harder to measure and identify
in the simulations. Our method has problems tracing these faint spiral arms
due to their lower amplitude. This is in particular the case for model md0.1mb1, for which
the amplitude becomes comparable to the particle noise 
(see Table~\ref{tb:pitch_angle}): it is therefore difficult to 
unambiguously detect spiral arms.

\begin{figure}
\includegraphics[width=\columnwidth]{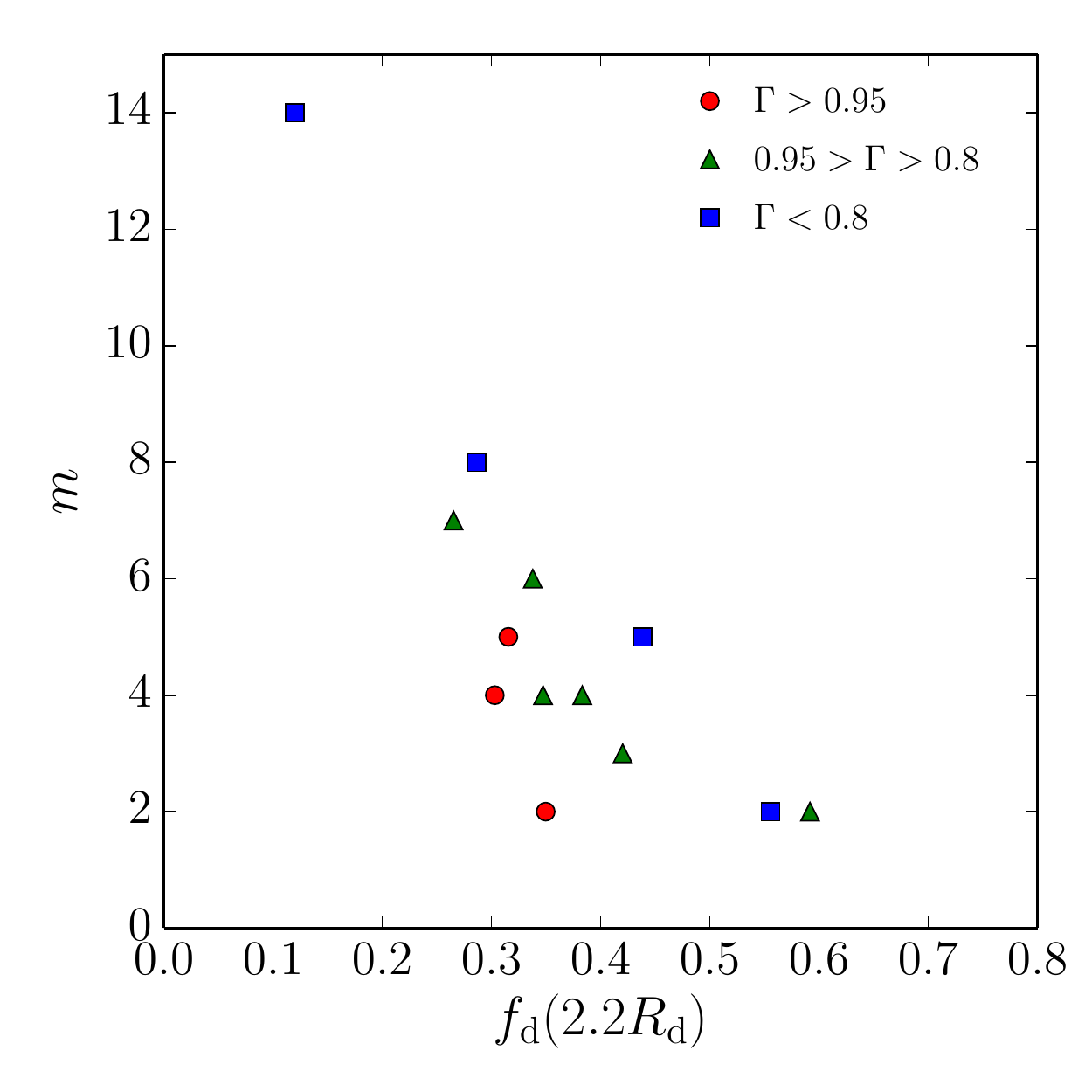}
\caption{The number of spiral arms ($m$) before the bar formation epoch.  
  The disk mass fraction ($f_{\rm d}={V_{\rm c, d}(R)^2/V_{\rm c, tot}(R)^2}_{R=2.2R_{\rm d}}$) is for
  the models with $Q_0=1.2$ and without halo spin. 
  We measure $m$ at 6.5\,kpc, which is close to $2.2R_{\rm d}$, for all models except models md1mb1Rd1.5 and md0.5mb1Rd1.5.
  For these models $R_{\rm d}$ is 1.5 times larger and we therefore measure $m$ at 9.5\,kpc.  
  The shear rate ($\Gamma$), measured in the initial conditions at $2.2R_{\rm d}$, is indicated by the different 
  symbols. 
  \label{fig:m_fdisk}  
  }
\end{figure}

In Fig.~\ref{fig:m_measured} we demonstrate how the number of spiral arms
changes with galactic radius. The number of spiral arms and 
the mass fraction of the disk are measured at $2.2R_{\rm d}$.
The relation between the measured number of spiral arms ($m$) at $2.2R_{\rm d}$
and the disk mass fraction ($f_{\rm d}$) is presented in Fig.~\ref{fig:m_fdisk}
where  ($f_{\rm d}$) is defined as:
\begin{eqnarray}
  f_{\rm d}\equiv \left( \frac{V_{\rm c, d}(R)}{V_{\rm c, tot}(R)} \right)^2_{R=2.2R_{\rm d}},
\label{eq:fd} 
\end{eqnarray}
where $V_{\rm c, d}$ and $V_{\rm c, tot}$ are the circular velocity of the disk
and of the whole galaxy, respectively. We find that $m$, before the bar 
formation, decreases as $f_{\rm d}$ increases. This matches the results of 
\citet{2015ApJ...808L...8D} (their figure 3).

The number of spiral arms is furthermore expected to depend on the shear rate:
\begin{eqnarray}
\Gamma = -\frac{d\ln \Omega}{d \ln R},
\label{Eq:ShearRate}
\end{eqnarray}
where $\Omega$ is the angular velocity. The value of $\Gamma$ indicates the 
shape of the rotation curve: $\Gamma=1$ indicates a flat rotation curve, $\Gamma>1$ indicates
a declining rotation curve, and $\Gamma<1$ indicates an increasing rotation curve.
\citet{1984PhR...114..319A} found that the swing amplification factor also depends on 
$\Gamma$. They computed the swing amplification factor as a function of $X$
for different $\Gamma$ values and found that the peak amplification factor
depends on $\Gamma$; $X\sim 1$ for $\Gamma = 0.5$ and $X\sim 2$ for $\Gamma=1.5$
(see their figure 26). 
Applying these results to Eq. (\ref{eq:mX}), a smaller value of $m$ is expected for 
a larger $\Gamma$.
In Fig.~\ref{fig:m_fdisk}, we show the relation between
$\Gamma$, disk mass fraction ($f_{\rm d}$) and the number of spiral arms ($m$). 
In this figure, we confirm that a larger shear rate 
results in a smaller number of spirals.

\begin{table*}
\begin{center}
\caption{Pitch angle and number of spiral arms\label{tb:pitch_angle}}
\begin{tabular}{lccccc}
\hline
Model    & Radius ($R$) & Shear rate ($\Gamma$)& Pitch angle ($i$) & Maximum amplitude & Number of arms ($m$)\\
      &      (kpc) &            &  (degree) &                     &  \\ 
\hline  \hline
md1mb1     & 10 & 0.991 & 18 & 0.198 & 2 \\
           & 12 & 1.04 & 18 & 0.164 & 2 \\
           & 14 & 1.07 & 19 & 0.161 & 2 \\
md1mb1s0.8 & 8  & 0.902 & 18 & 0.261 & 2 \\
           & 10 & 0.991 & 19 & 0.169 & 2 \\
           & 12 & 1.04 & 18 & 0.181 & 2 \\
           & 14 & 1.07 & 22 & 0.197 & 2 \\
md0.5mb1   & 6 & 0.804 & 25 & 0.101 & 4 \\
           & 8 & 0.875 & 25 & 0.121 & 6 \\
           & 10 & 0.944 & 27 & 0.0836  & 7 \\
           & 12 & 0.983 & 18 & 0.0561 & 2 \\
md0.4mb1   & 4 & 0.758 & 33 & 0.0372 & 5 \\
           & 6 & 0.796 & 33 & 0.0442 & 6 \\
           & 8 & 0.863 & 27 & 0.0378 & 7 \\
           & 10 & 0.926 & 26 & 0.0249 & 9 \\
md0.3mb1   & 4 & 0.774 & 32 & 0.0483 & 4 \\
           & 6 & 0.792 & 34 & 0.0453 & 7 \\
           & 8 & 0.847 & 26 & 0.0347 & 9 \\
           & 10 & 0.907 & 29 & 0.0189 & 10 \\
md0.1mb1   & 6 & 0.744 & 5 & 0.00965 & 1 \\
           & 8 & 0.777 & 3 & 0.0116 & 1 \\
           & 10 & 0.832 & 3 & 0.0111 & 2 \\
           & 12 & 0.885 & 3 & 0.0107 & 2 \\
md0.5mb0   & 8 & 0.833 & 15 & 0.185 & 2 \\
           & 10 & 0.888 & 11 & 0.169 & 2 \\
           & 12 & 0.905 & 12 & 0.199 & 2 \\
md0.5mb3   & 6 & 0.991 & 25 & 0.131 & 2 \\
           & 8 & 0.955 & 25 & 0.119 & 5 \\
           & 10 & 0.963  & 22 & 0.0959 & 3 \\
md0.5mb4   & 6 & 0.975 & 25 & 0.111 & 4 \\
           & 8 & 0.977 & 23 & 0.109 & 4 \\
           & 10 & 0.996 & 25 & 1.064 & 8 \\
md0.5mb4rb3 & 6 & 0.963 & 26 & 0.0924 & 5 \\
            & 8 & 0.961 & 25 & 0.0923 & 5 \\
            & 10 & 0.965 & 26 & 0.0725 & 5 \\
md1.5mb5   & 8 & 1.01 & 21 & 0.281 & 2 \\
           & 10 & 1.07 & 16 & 0.189 & 2 \\
           & 12 & 1.09 & 14 & 0.212 & 2 \\
           & 14 & 1.13 & 11 & 0.269 & 2 \\
md1mb10    & 6 & 1.10 & 27 & 0.287 & 2 \\
           & 8 & 1.07 & 22 & 0.269 & 2 \\
           & 10 & 1.05 & 18 & 0.211 & 2 \\
md1mb1Rd1.5   & 10 & 0.878 & 24 & 0.152 & 4 \\
           & 12 & 0.963 & 28 & 0.169 & 4 \\
           & 14 & 1.03 & 18 & 0.181 & 4 \\
md0.5mb1Rd1.5 & 10 & 0.850 & 29 & 0.0777 & 7 \\
           & 12 & 0.921 & 27 & 0.0671 & 8 \\
           & 14 & 0.977 & 24 & 0.0572 & 8 \\
md0.5mb1Rd1.5s & 10 & 0.850 & 26 & 0.0719 & 7 \\
           & 12 & 0.921 & 26 & 0.0804 & 7 \\
           & 14 & 0.977 & 24 & 0.0553 & 8 \\
\hline
\end{tabular}
\end{center}
\end{table*}

\subsubsection{Pitch angle}

The pitch angle is an important parameter in the discussion on the morphology 
of spiral galaxies.  We measure the pitch angle of our simulated 
galaxies using the Fourier transform method~\citep[see ][]{2013A&A...553A..77G, 2015MNRAS.454.2954B}.
Using the same Fourier decomposition (Eq.~\ref{eq:Fourier}) as for 
the bar amplitude we compute the phase angle,  $\phi_m(R)$.
Next, the pitch angle at $R$ for $m$ is obtained by using
\begin{eqnarray}
\cot i_m(R) = R\frac{d\phi (R)_m}{dR}.
\end{eqnarray}

In numerical simulations the pitch angle changes over
time~\citep{2013ApJ...763...46B, 2013A&A...553A..77G, 2015MNRAS.454.2954B}.
The pitch angle of spiral arms increases and
decreases repeatedly as the amplitude of transient spiral arms
increases and decreases \mbox{~\citep[see Figures 4 and 5
  in][]{2015MNRAS.454.2954B}}. Furthermore, the number of spiral arms also 
changes as a function of $R$ as we saw in previous sections. We 
therefore measure the most appearing pitch angle for the most 
appearing mode (principal mode) at each galactic radius.  
Following \citet{2015MNRAS.454.2954B},
we define the most frequently appearing pitch angle weighted by the
Fourier amplitude as the pitch angle. 
In Table \ref{tb:pitch_angle}, we report the measured
pitch angle and the number of spiral arms ($m$) for that angle.  
Note that we measure pitch angles at $\gtrsim 2.2R_{\rm d}$, 
except for barred galaxies where we, to avoid the bar's influence,
use an $R$ that is larger than the maximum bar length.

\begin{figure}
\includegraphics[width=\columnwidth]{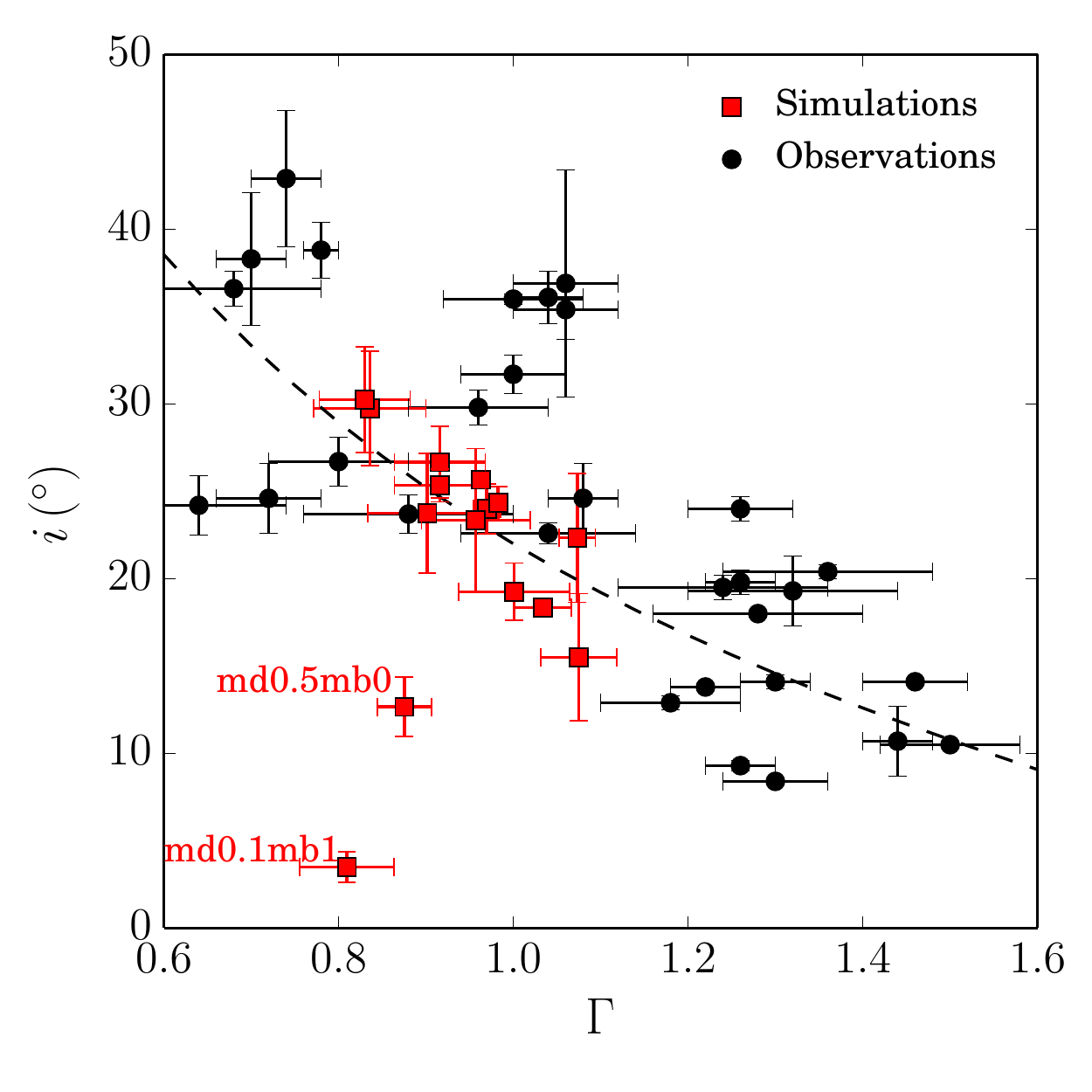}
\caption{The relation between shear rate ($\Gamma$) and pitch angle ($i$)
for the simulated galaxies shown in Table \ref{tb:pitch_angle} (red squares) 
and observed galaxies \citep{2006ApJ...645.1012S} (black circles). 
For simulated galaxies, the error bars on the $x$-axis indicate the range of 
shear rates depending on the radius at which we measured $\Gamma$ and $i$.
The error bars on the $y$-axis indicate the 
standard deviations of the measured pitch angles at each radius. 
The black dashed curve indicates the 
result of \citet{2014ApJ...787..174M} given by equation (\ref{eq:pitch_shear}). \label{fig:pitch_angle}}
\end{figure}

\citet{1966ApJ...146..810J} suggested that the pitch 
angle ($i$) is determined by the shear rate ($\Gamma$, see Eq.~\ref{Eq:ShearRate} and 
Table~\ref{tb:pitch_angle}) of the disk. The relation 
between the shear rate and pitch angle was recently investigated 
using both numerical simulations and analytic models 
\citep{2014ApJ...787..174M,2016ApJ...821...35M}.
The relation is also suggested by galaxy observations 
\citep{2005MNRAS.361L..20S,2006ApJ...645.1012S}.
In Fig.~\ref{fig:pitch_angle}, the relation between $\Gamma$
and $i$ is presented (averaged for each model). 
In order to compare our results with the theory, we also
present the relation between $\Gamma$ and $i$ as derived 
by \citep{2014ApJ...787..174M}:
\begin{eqnarray}
\tan i = \frac{7}{2}\frac{\sqrt{4-2\Gamma}}{\Gamma}.
\label{eq:pitch_shear}
\end{eqnarray}
In the figure this relation is presented with a dashed curve.
Except for models md0.5mb0 and md0.1mb1, the measured relation in our 
simulations is consistent with the theoretical curve.

We now briefly discuss these two outliers.
Model md0.5mb0 has, due to the lack of a bulge component,
a strong bar resulting in a ring structure around the 
bar end (see Fig.~\ref{fig:snapshots_10Gyr}).
Here the ring structure may affect the spiral structure in the 
outer disk region.
The pitch angle as measured before bar formation ($t<2.5$\,Gyr)
is $23^{\circ}$ and $m=$5--10 at $R=8$--12\,kpc which is 
consistent with the curve in Fig.~\ref{fig:pitch_angle}.

The other outlier, md0.1mb1, has a relatively low-mass disk which results
in low-contrast spiral structure (virtually invisible; see Fig.~\ref{fig:snapshots_10Gyr}).
The amplitude, $\sim 0.01$ (see Table~\ref{tb:pitch_angle}), is
comparable to the particle noise level (see Fig.~\ref{fig:A2_max_mdisk}).
This makes it difficult  
to accurately measure the number of spiral arms 
and their pitch angle using the scheme we adopted.

We also present the relation between the shear rate and 
the pitch angles of observed galaxies~\citep{2006ApJ...645.1012S} in 
Fig.~\ref{fig:pitch_angle} (black points). These points are also 
distributed around the theoretical curve with a scatter
larger than the simulated galaxies.

\subsection{Bar Formation}\label{Sect:bf}

\begin{figure*}
\includegraphics[width=\columnwidth]{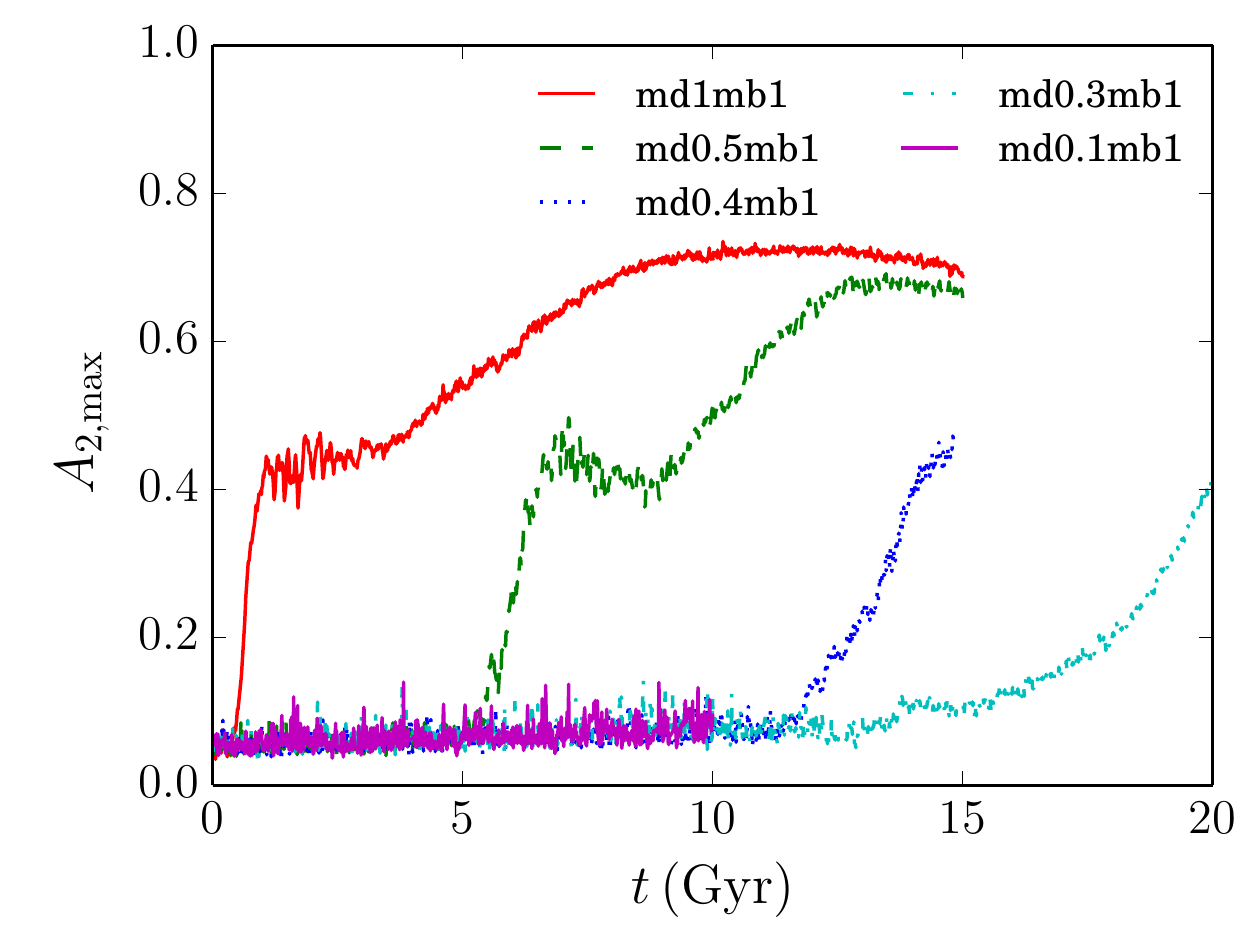}
\includegraphics[width=\columnwidth]{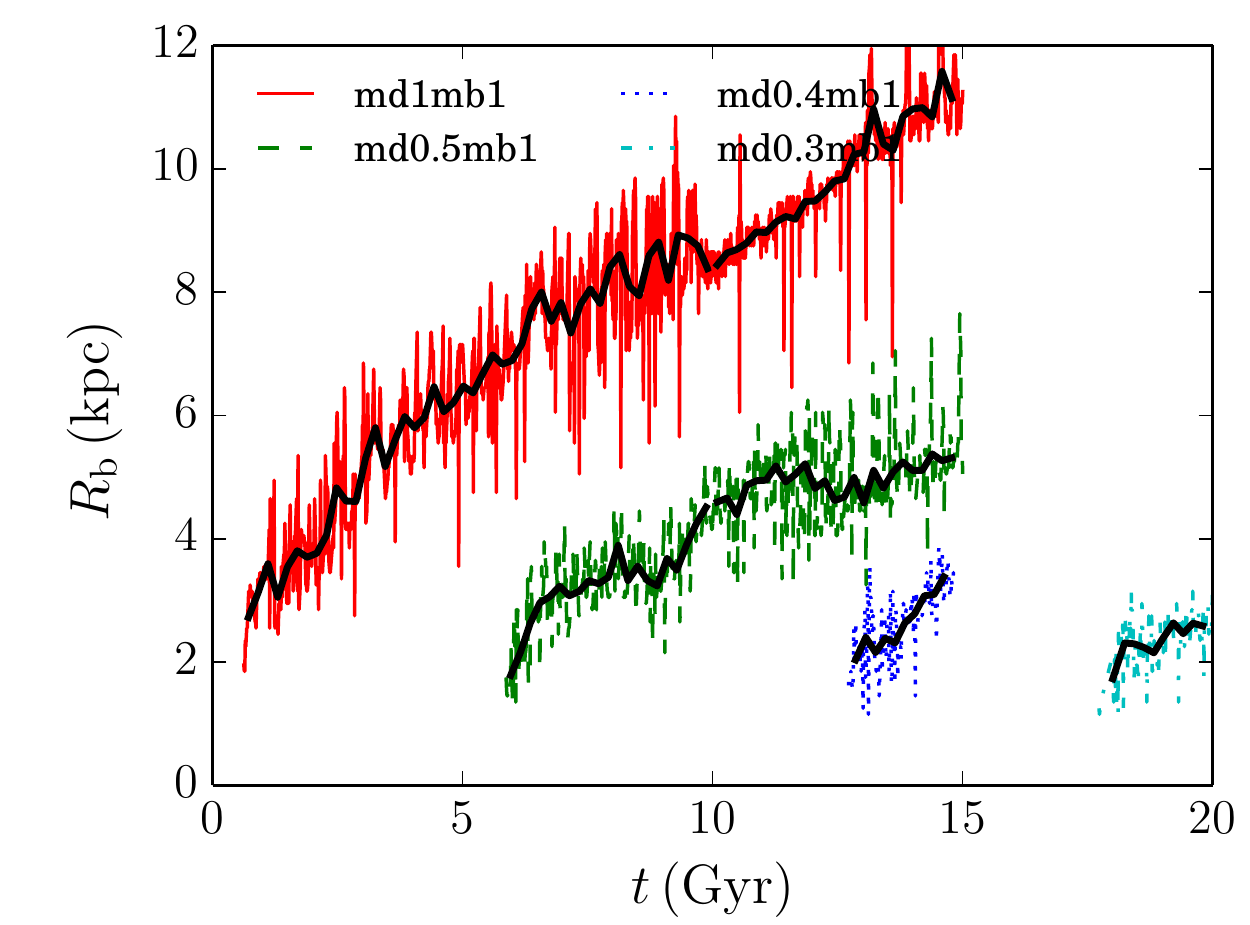}
\caption{
Time evolution of the maximum amplitude for $m=2$ (left)
and the bar length (right)
for models md0.1mb1, md0.3mb1, md0.4mb1, md0.5mb1, and md1mb1..
Black curves in the right panel indicate the bar length averaged over every
20 snapshots ($\sim 0.2$Gyr).
\label{fig:A2_max_mdisk}}
\end{figure*}

We will now investigate the formation of bars.  In order to define the
bar formation, we measure the time evolution of the length (radius) of the bar
and its amplitude which develops in our galaxy simulations. Because
this has to be done for one thousand snapshots for each galaxy
simulation we adopt a relatively simple method for measuring these
properties.  We measure the Fourier amplitude (Eq.\,\ref{eq:Fourier})
in radial bins of 1 kpc for the $m=2$ mode ($A_2(R)$), record the maximum
value and use this as the bar amplitude ($A_{\rm 2, max}$).  In the
left panel of Fig.~\ref{fig:A2_max_mdisk}, we present the time
evolution of $A_{\rm 2, max}$ for models md1mb1 to md0.1mb1.  Once a
bar begins to develop the amplitude increases exponentially and either
reaches a stable maximum (as is the case in model md1mb1) or decreases
slightly to increase again a few Gyr later (see model mb0.5mb1). Model
md0.1mb1 did not form a bar within 10\,Gyr.  

We also measure the bar
length using the method described in~\citet{2012MNRAS.425L..10S}
and \citet{2015PASJ...67...63O}. In this method we compute the phase angle
($\phi_{2}(R)$) and amplitude ($A_{2}(R)$) of the bar at each radius
using the Fourier analysis (Eq.~\ref{eq:Fourier}). 
As $R$ increases, $A_{2}(R)$ increases, reaches its maximum in the middle
of the bar, and then decreases. 
We define the radius at which $A_{2}(R)$ reaches its maximum value
as $R_{\rm max}$ and the phase at $R_{\rm max}$ as the phase 
angle of the bar ($\phi_{\rm 2, max}$). Starting at $R_{\rm max}$,
we compare $\phi_{2}(R)$ with $\phi_{\rm 2, max}$. 
When $\Delta \phi=|\phi_{2}(R)-\phi_{\rm 2, max}|>0.05\pi$, we consider 
that the bar has ended and define the radius as the bar's size, $R_{\rm b}$.
Hereafter, we refer to $R_{\rm b}$ as the length of the bar. The time evolution of
the bar length is presented in the right panel of
Fig.~\ref{fig:A2_max_mdisk}. The length of the bar grows continuously
until the end of simulation ($t=15$ Gyr).

We define the epoch of bar formation ($t_{\rm b}$) as the moment 
when $A_{\rm 2, max} > 0.2$ and $R_{\rm b} > $~1\,kpc. In our models the bar was always longer 
than 1 kpc when  $A_{\rm 2, max}>0.2$. 
In most cases the bar amplitude increases exponentially and therefore the critical 
amplitude has little effect on the moment the bar forms.
For models md0.4mb1 and md0.3mb1, which did not form a bar within 10 Gyr, we continued
the simulations until a bar formed after 13 and 18\,Gyr, respectively
(also see Fig.~\ref{fig:A2_max_mdisk} and Table~\ref{tb:bar_crit}).
We continued the simulations up to 15\,Gyr for md0.5mb4 and md0.5mb4rb3 
to confirm that they form a bar, which they do around  $\sim 10$Gyr.
 
\begin{figure*}
\includegraphics[width=\columnwidth]{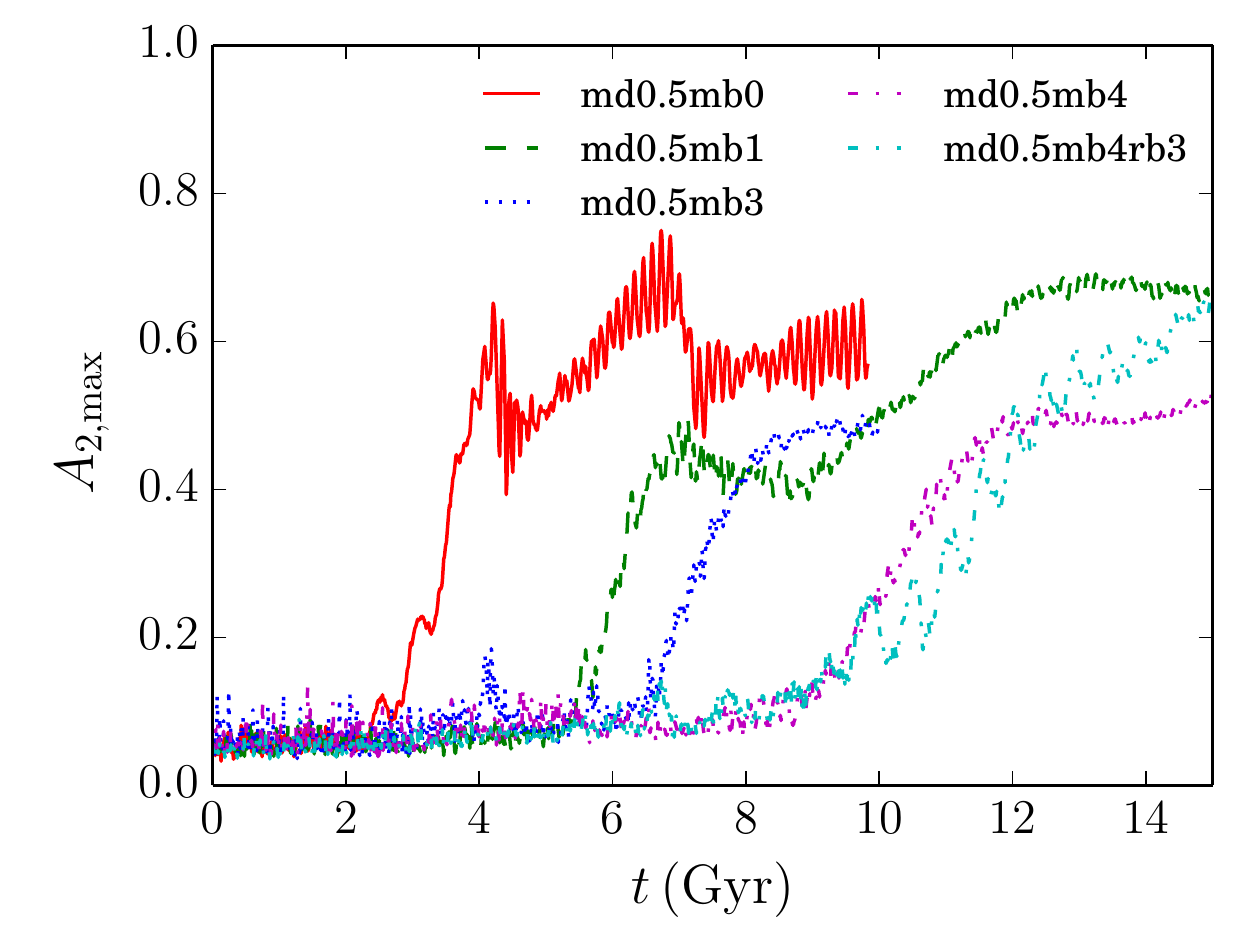}
\includegraphics[width=\columnwidth]{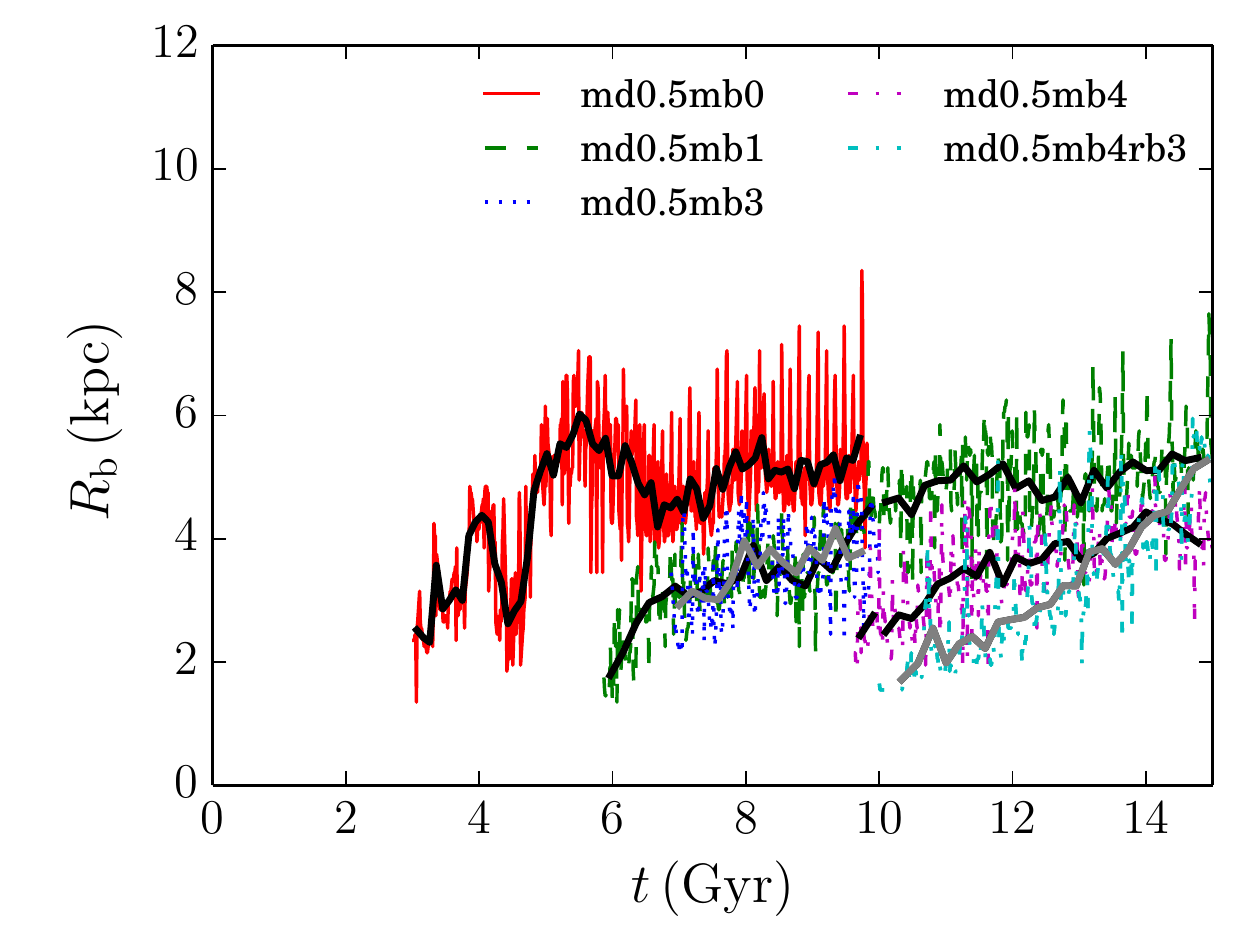}
\caption{Same as Fig.~\ref{fig:A2_max_mdisk} but for models md0.5mb0,
md0.5mb1, md0.5mb3, md0.5mb4, and md0.5mb4rb3.
\label{fig:A2_max_mbulge}}
\end{figure*}

We subsequently investigate the effect of the bulge mass on the bar formation. 
It was suggested that a massive central component, such as a bulge, have a
stabilizing effect 
on the disk and thereby prevents bar formation~\citep{2001ApJ...546..176S,2013MNRAS.434.1287S}.  
To test this we perform a set of simulations in which we vary the bulge mass.
We make the bulge 0 (md0.5mb0), 3 (md0.5mb3) and 4 (mb0.5mb4) times as massive 
as the bulge of model md0.5mb1. We further added model md0.5mb4rb3 with the same mass as md0.5mb4,
but with increased bulge scale length.
The amplitude evolution and bar length of these models is
presented in Fig.~\ref{fig:A2_max_mbulge}. 
When the bulge mass fraction increases the bar formation is delayed 
due to the the decreasing disk mass-fraction ($f_{\rm d}$)
(also see Table~\ref{tb:bar_crit}). 
These results are consistent with observations where the 
fraction of barred galaxies increase when the bulge to disk mass
ratio decreases and where the barred galaxies fraction even increases 
to $\sim 87$\% for the extreme case of bulge-less galaxies.
We further confirm that the bulge scale length does not effect the epoch of bar 
formation, but  the bar length at the end of the 
simulation (at 15\,Gyr). The final bar length for model md0.4mb4rb3 
is longer than that for md0.4mb4 (see Fig.~\ref{fig:A2_max_mbulge}). 
The bar formation epoch for all models is presented in Table~\ref{tb:bar_crit}.

Although increasing the bulge mass sequentially delays the formation of a bar,
the bulge mass fraction is not a critical parameter for the bar formation. We tested
some parameters and found that the disk-mass to the total mass fraction is a more
critical parameter for the bar formation epoch.

\begin{figure}
\includegraphics[width=\columnwidth]{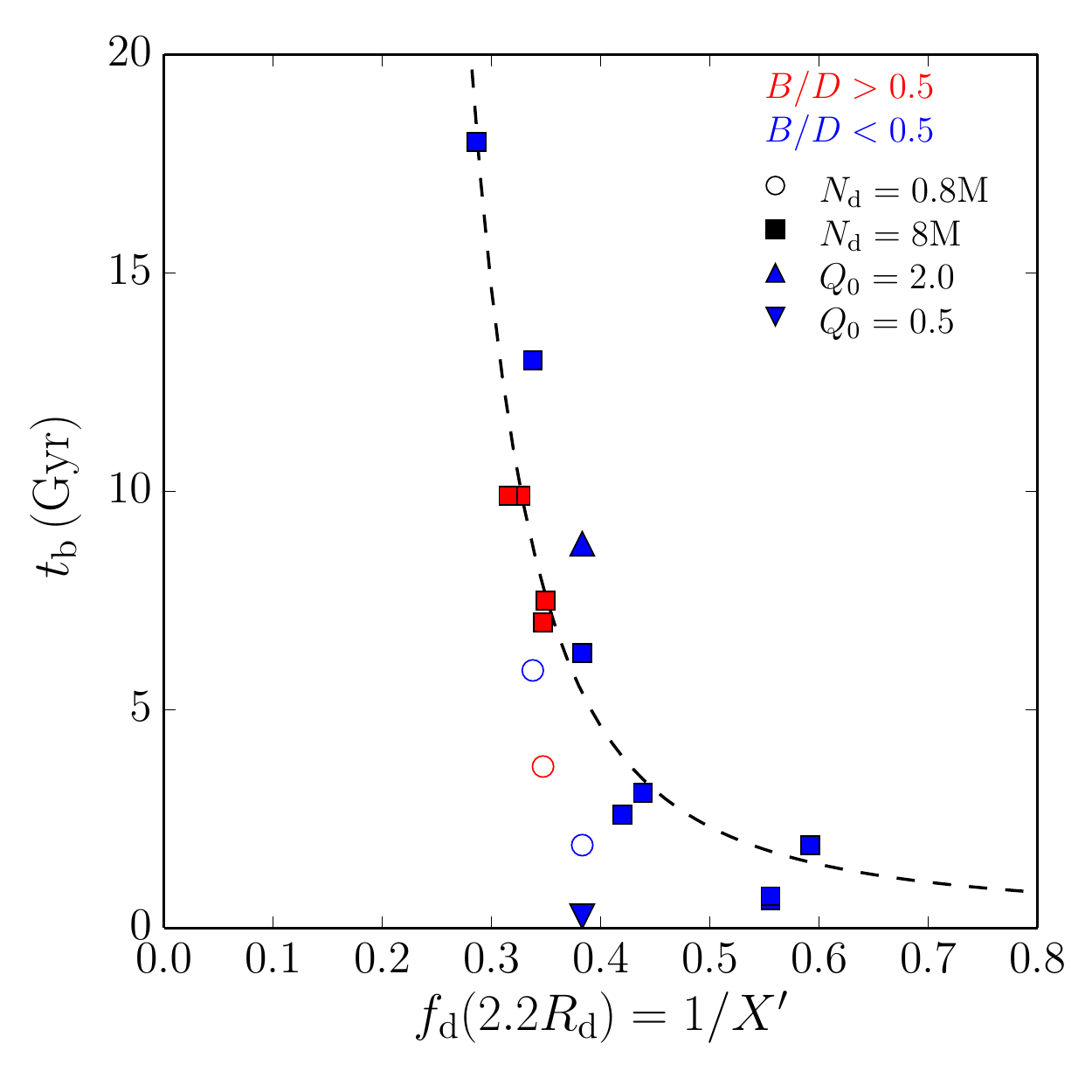}
\caption{Bar formation epoch ($t_{\rm b}$) and disk mass fraction 
($f_{\rm d}=1/X'$). 
Filled squares and open circles indicate models with $N_{\rm d}=8$M and 0.8M, respectively. 
Red (blue) indicates models with a bulge-to-disk mass ratio ($B/D$) of 
$>0.5$ ($<0.5$).
The dashed curve indicates a fit to the models with $N_{\rm d}=8$M and $Q_0=1.2$
(squares): $t_{\rm b}=0.146\pm 0.079 \exp [(1.38\pm 0.17)/f_{\rm d}]$.
\label{fig:bfe}}
\end{figure}

In Fig.~\ref{fig:bfe}, we present the relation
between the bar formation epoch and the disk mass fraction, $f_{\rm d}(=1/X'$),
where $X'$ is a parameter adopted by \citet{2008ApJ...679.1239W} as 
a bar formation criterion:
\begin{eqnarray}
 X'\equiv 1/f_{\rm d} = \left( \frac{V_{\rm c, tot}(R)}{V_{\rm c, d}(R)} \right)^2_{R=2.2R_{\rm d}}.
\label{eq:Xprime}
\end{eqnarray}
They argued that $X'\lesssim 3$ (for $f_{\rm d}\gtrsim 0.3$) is the
bar formation criterion in their simulation.  The epoch of bar
formation increases exponentially for decreasing disk mass-fraction,
although the scatter is large.  We fit an exponential function to our
results obtained with $N_{\rm d}=8$M and $Q_0=1.2$ and find that
$t_{\rm b}=0.146\pm 0.079 \exp [(1.38\pm 0.17)/f_{\rm d}]$.  The result
is indicated by the dashed black line in Fig.~\ref{fig:bfe}.

The resolution of the simulation in the number of particles is an important source for the scatter
 \citep{2009ApJ...697..293D}; 
a smaller number of particles for the same model results in faster bar formation.
We confirm this by performing simulations with an order of magnitude lower
resolution (0.8\,M disk particles, open circle symbols), and indeed find that the bar forms
earlier for these models in comparison with the high resolution models (Fig.~\ref{fig:bfe}, Table~\ref{tb:bar_crit}). 
Another parameter which is known to affect the epoch of bar formation is the value of $Q$.
In Fig.~\ref{fig:bfe} we also plot models md0.5mb1Q2.0 and md0.5mb1Q0.5, 
which are identical to model md0.5mb1, with the exception that $Q_0=2.0$ and 0.5, respectively.
As was shown in previous studies \citep[c.f.,][]{1986MNRAS.221..213A}, 
a larger value of $Q_0$ leads to a delay in the formation of the bar 
(see Appendix A2 for details).

\begin{figure}
\includegraphics[width=\columnwidth]{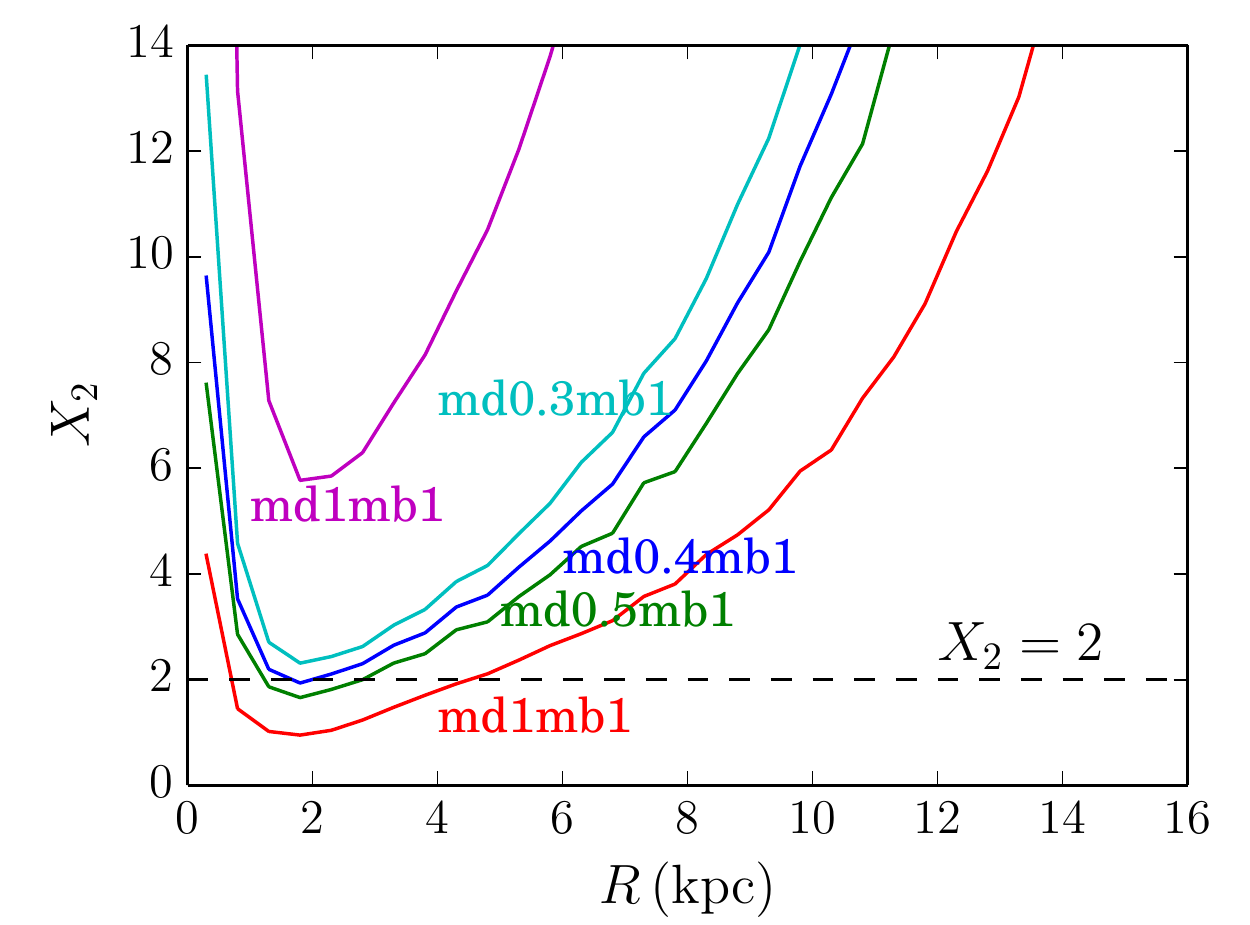}
\caption{$X$ values as a function of radius for $m=2$ mode. \label{fig:X_disk}}
\end{figure}

\begin{figure}
\includegraphics[width=\columnwidth]{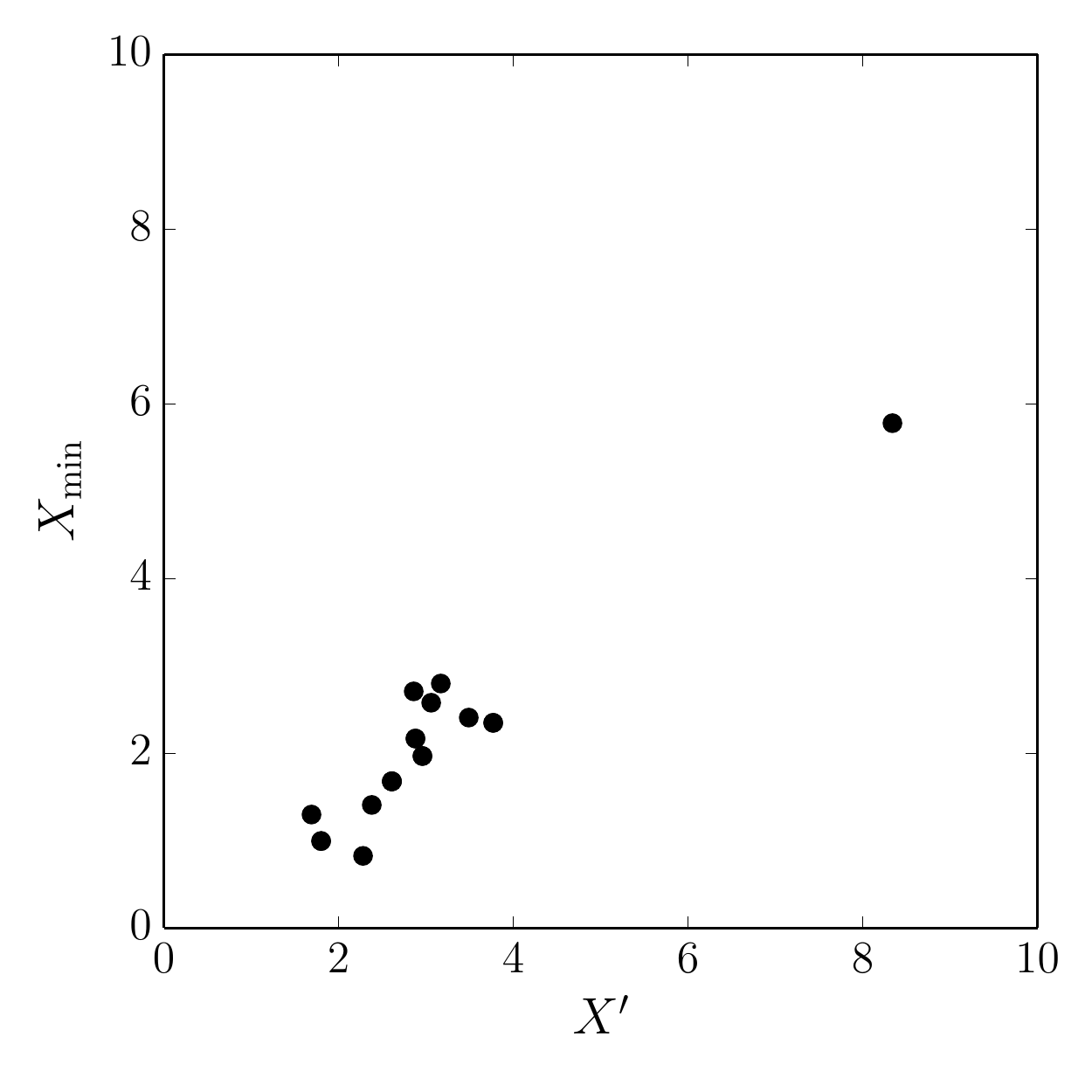}
\caption{Relation between $X'$ and the minimum value of $X$ ($X_{\rm min}$).\label{fig:X}}
\end{figure}

The relation between the moment of bar formation ($t_{\rm b}$) and 
the mass fraction of the disk ($f_{\rm d}$) can be
understood from Toomre's $X$ parameter (see Eq.~\ref{eq:X}). 
For a given value of $m$ we can calculate $X$ as a function of the disk radius $R$.
When we adopt $m=2$, i.e. the bar, we obtain $X$ for the bar mode ($X_2$) as a function of $R$.
This distribution is presented in Fig.~\ref{fig:X_disk}. Here, we see that $X_2$
reaches minimum values at $R\sim 2$ kpc. We find that the minimum value of $X_2$ ($X_{\rm min}$)
is roughly correlated with $X'(=1/f_{\rm d})$, and the relation between $X_{\rm min}$ and $X'$ is 
presented in Fig.~\ref{fig:X}. Thus, the disk fraction $f_{\rm d}$ is connected to 
Toomre's $X$.
As shown by \citet{1981seng.proc..111T}, the amplitude grows most efficiently for $1<X<2$
and decreases exponentially when $X$ increases from $\sim 2$ to $\sim 3$.
We find that models in which a bar forms have a minimum value of $X_2\lesssim 2$ 
(see Fig.~\ref{fig:X_disk}).
We conclude, based on these results, that there is no particular rigid criterion for bar
formation, but that the bar formation epoch starts to increase exponentially
when $f_{\rm d}\gtrsim0.3$, or equivalently, if $X'\lesssim0.3$.

We also test the bar formation criterion previously suggested by
\citet{1982MNRAS.199.1069E}, who
proposed that bar formation depends on the mass of the
disk ($M_{\rm d}$) within radius $R_{\rm d}$:
\begin{eqnarray}
\epsilon _{\rm m} \equiv \frac{V_{\rm c, max}}{(GM_{\rm d}/R_{\rm d})^{1/2}}<1.1.
\label{eq:epsilon}
\end{eqnarray}
Here $V_{\rm c, max}$ is the maximum circular velocity in the disk.
For this criterion, \citet{2008MNRAS.390L..69A} showed some exceptional
cases using $N$-body simulations of disk models with live halos.
In Table \ref{tb:bar_crit}, we present $\epsilon_{\rm m}$ (Eq.~\ref{eq:epsilon}),
and we confirm that in our simulations Efstathiou's criterion cannot always predict 
the bar formation.

\begin{table*}
\begin{center}
  \caption{Details on the formation of the bar.  The columns give the
    model name, a boolean indicating if a bar formed (Y) or not (N)
    within the simulation time period (0--20\,Gyr), the moment and
    criteria of the bar formation.
    \label{tb:bar_crit}}
\begin{tabular}{lccccc}
\hline
Model    & Bar formation & Bar formation epoch &  \multicolumn{3}{c}{Bar formation criteria}\\
      &                                  & $t_{\rm b}$ (Gyr)   &  $\epsilon_{\rm m}$ & $X_{\rm min}$ & $X'(\equiv 1/f_{\rm d})$\\ 
\hline  \hline
md1mb1     & Y & 0.64 & 0.824 & 0.997 & 1.80 \\
md1mb1s0.65 &  Y & 0.83 & 0.824 & 0.997 & 1.80 \\
md1mb1s0.8 &  Y & 0.73 & 0.824 & 0.997 & 1.80 \\
\\      
md0.5mb1   &  Y & 6.3 & 1.03 & 1.68 & 2.61 \\
md0.4mb1   &  Y & 13 & 1.00  & 1.97 & 2.96 \\
md0.3mb1   &  Y & 18 & 1.87 & 2.41 & 3.49 \\
md0.1mb1   &  N & - & 2.12 & 5.78 & 8.34 \\
\\
md0.5mb0   & Y & 3.1 & 1.08  & 0.827 & 2.28 \\
md0.5mb3   & Y & 7.0 & 1.26  & 2.17 & 2.88 \\
md0.5mb4   & Y  & 9.9 & 1.38  & 2.58 & 3.06 \\
md0.5mb4rb3 & Y  & 9.9 & 1.15  & 2.80 & 3.17 \\
\\
md1.5mb5   &  Y & 1.9 & 1.42 & 1.30 & 1.69 \\
md1mb10    & Y  & 7.5 & 1.60 &  2.71 & 2.86 \\
\\
md1mb1Rd1.5   & Y & 2.6 & 0.960  & 1.41 & 2.38 \\
md0.5mb1Rd1.5 & N & - & 1.25 & 2.35 & 3.77 \\
md0.5mb1Rd1.5s & N & - & 1.25 & 2.35 & 3.77 \\
\\
md0.5mb1Q0.5 &  Y & 0.27 & 1.03 & 1.68 & 2.61 \\
md0.5mb1Q2   &  Y & 8.8 & 1.03 & 1.68 & 2.61 \\
\hline
\end{tabular}
\end{center}
\end{table*}

\subsection{Bulge-to-Disk mass ratio}

From an observational perspective such as in the Hubble sequence \citep{1926ApJ....64..321H}, 
the bulge-to-disk mass ratio ($B/D$) is related to the pitch angle. 
Sa galaxies have a larger bulge-to-disk mass-ratio
compared to Sb and Sc galaxies \citep{1961hag..book.....S}.
To test this hypothesis, we simulate an extra model (md1mb10), with a much larger $B/D$ than the 
the other models.

The disk-to-halo mass ratio of this model is relatively large  ($B/D=1.0$),
but the disk-to-total mass ratio ($f_{\rm d}=0.35$) is not 
as large as for models which form a bar before forming spiral arms.
The S0--Sa galaxies, for example,
NGC\,1167 \citep{2008ARep...52...79Z} and M\,104
\citep{2006MNRAS.371.1269T}, have such a massive bulge and also many
narrow spiral arms. 

\begin{figure*}
\includegraphics[width=55mm]{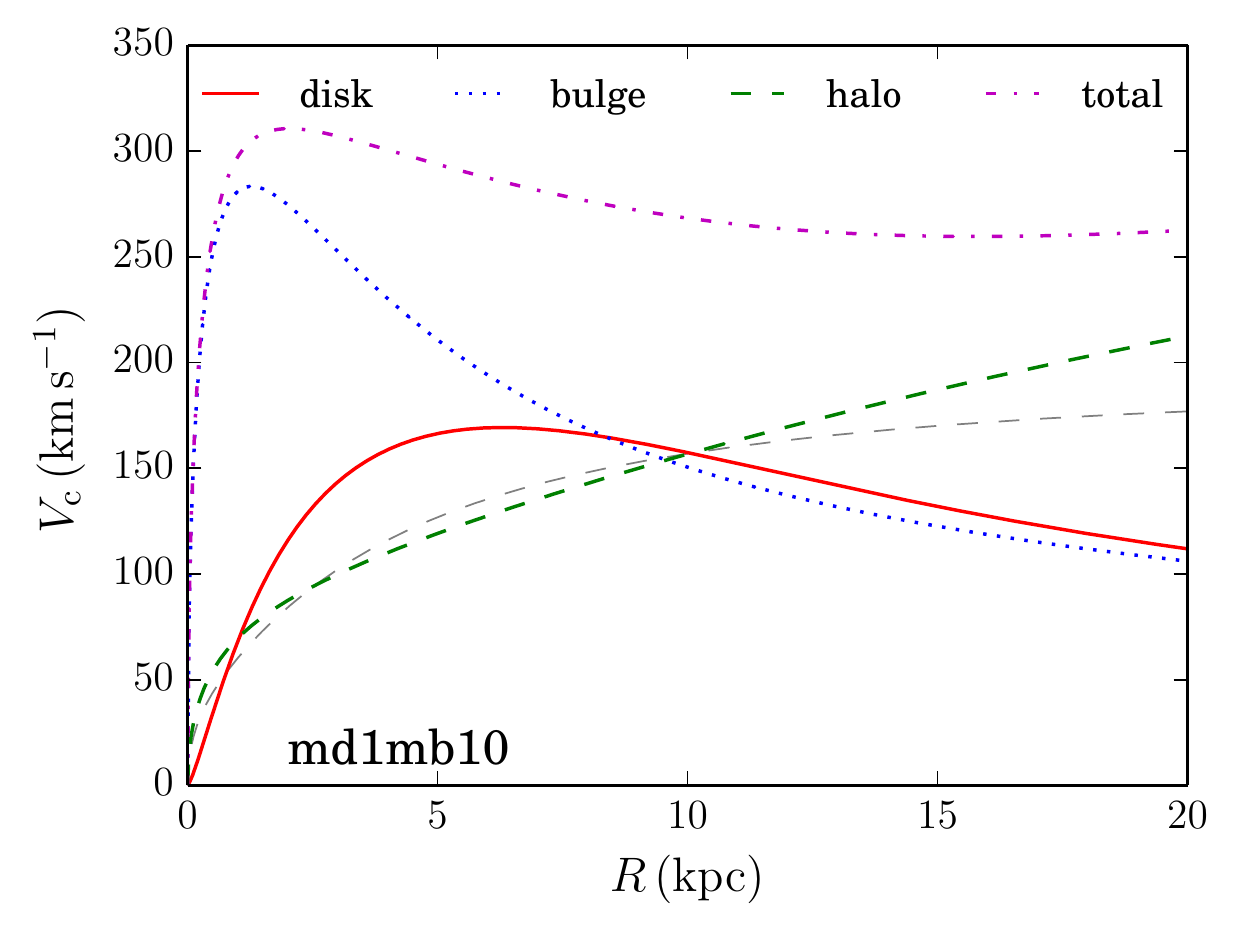}\includegraphics[width=42mm]{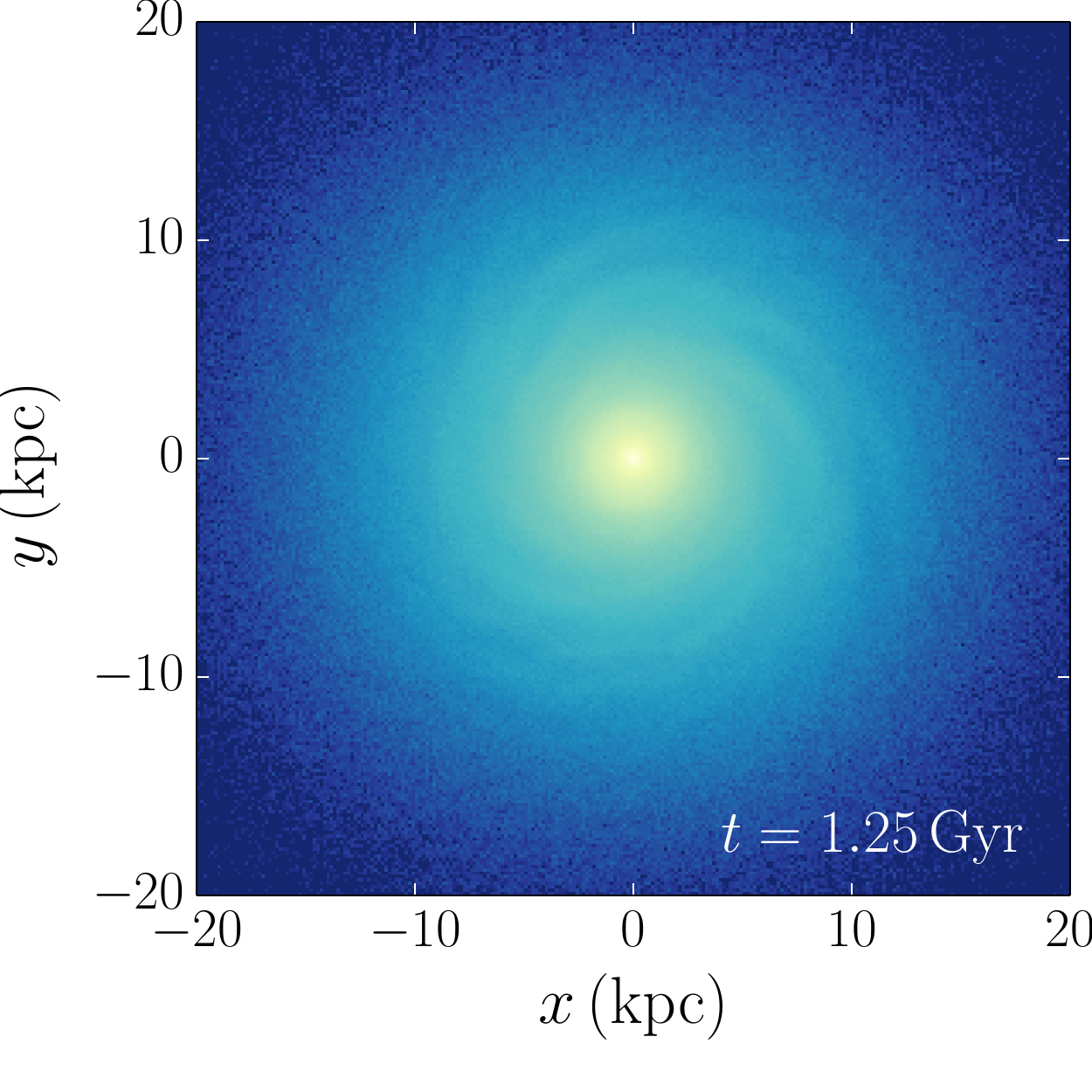}\includegraphics[width=42mm]{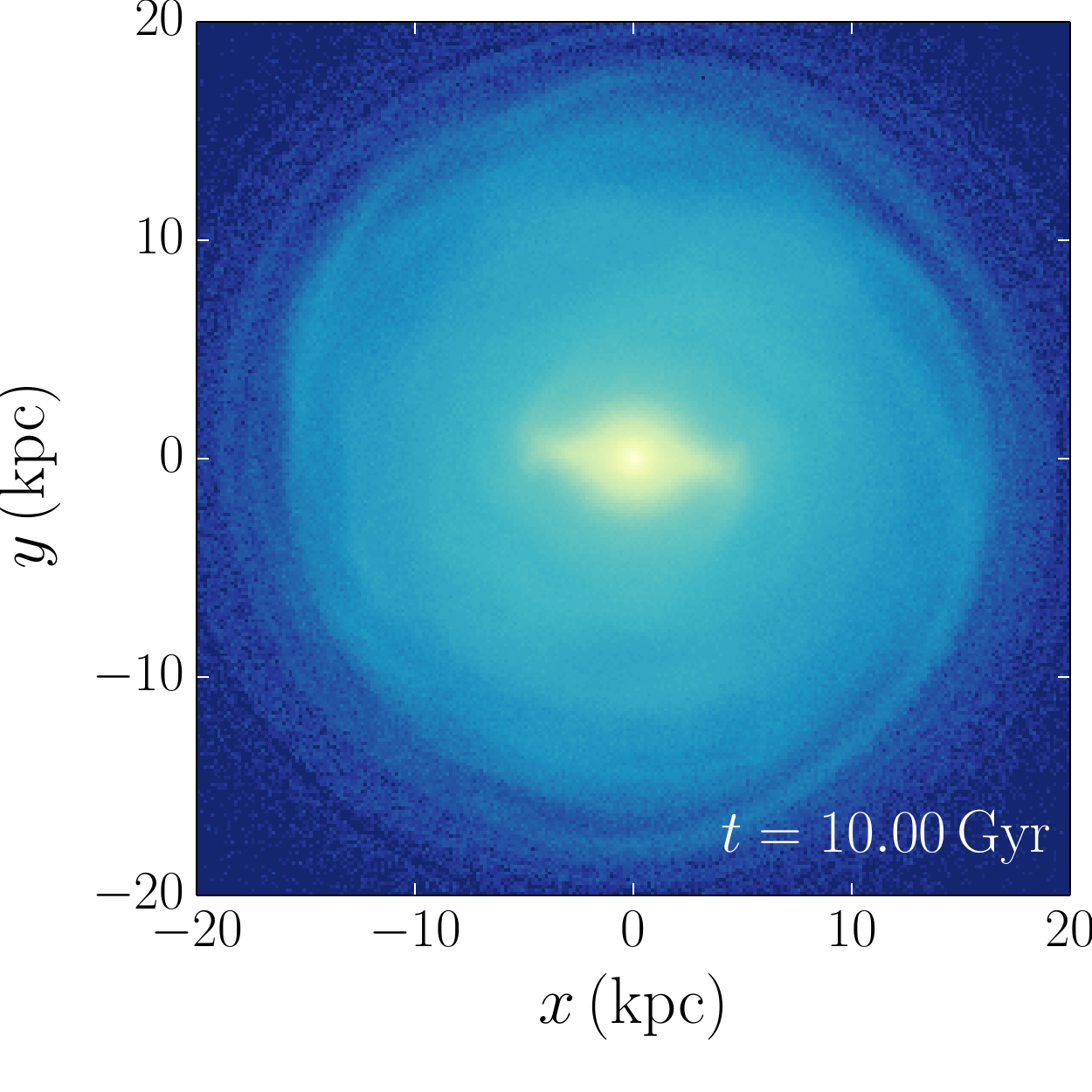}\\
\caption{Rotation curve (left) and surface density at 1.25\,Gyr (middle) and 10\,Gyr for model md1mb10. Gray dashed curve in
the left panel indicates the rotation curve of the halo for model md1mb1. \label{fig:snapshots_mb10}}
\end{figure*}

In Fig.~\ref{fig:snapshots_mb10} we present the rotation curves (left
panel) and the surface-density images (middle and right panels) for model
md1mb10.  This model formed multiple spiral arms similar to Sa galaxies 
before it developed a bar. 
The measured pitch angle was $\sim 20^{\circ}$ within 10\,kpc 
but less than 10$^{\circ}$ at $R>10$\,kpc. 

\begin{figure}
\includegraphics[width=\columnwidth]{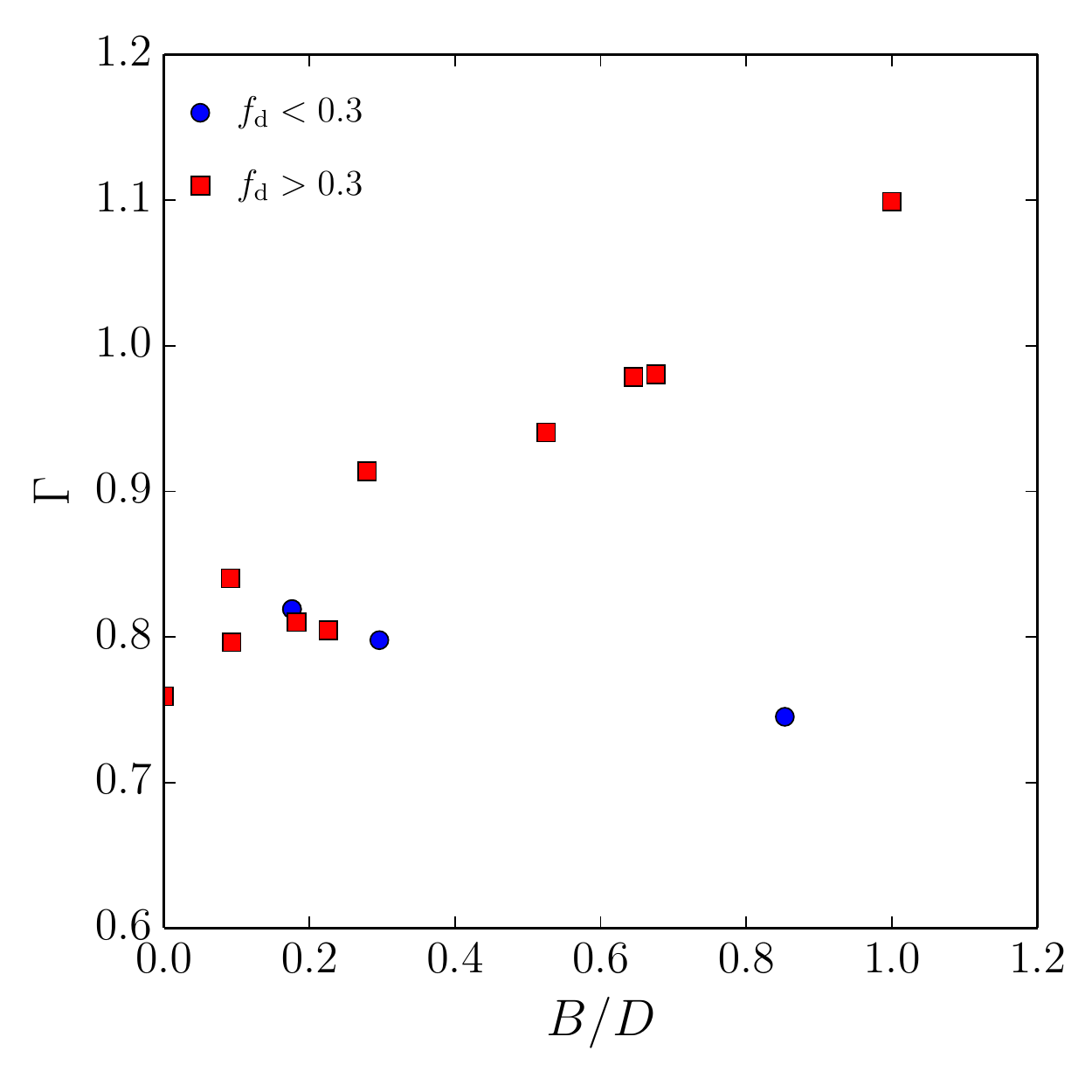}
\caption{The relation between the shear rates at $2.2R_{\rm d}$ and the bulge-to-disk mass ratio of our models. 
Circles and squares indicate models with $f_{\rm d}(2.2R_{\rm d})<0.3$ and $f_{\rm d}(2.2R_{\rm d})>0.3$,
respectively. \label{fig:shear_rate}}
\end{figure}

In the previous subsection, we demonstrated the relation between the 
pitch angle and the shear rate ($\Gamma$). Here, in
Fig.~\ref{fig:shear_rate}, we present the relation between $\Gamma$
and the bulge-to-disk mass ratio ($B/D$). 
For our models with $f_{\rm d}>0.3$,
$\Gamma$ appears to correlate with $B/D$, however the models with a small
$f_{\rm d}$ tend to have a small $\Gamma$. 
In the latter sequence, the bulge mass is similar but the disk mass 
decreases as $B/D$ increases and as a result $f_{\rm d}$ decreases.

In order to understand how the 
initial parameters affect the combination of $B/D$ and $\Gamma$,
we create an additional set of models. Here we measure the properties in
the initial realizations without actually running the simulations. 
When we keep the disk mass fraction ($f_{\rm d}$) and the bulge scale-length
($r_{\rm b}$) fixed then $\Gamma$ increases as $B/D$ increases.
Even when we compare models that have similar $B/D$ and $f_{\rm d}$,
we find that models with a larger $r_{\rm b}$ have a larger $\Gamma$.
And when we increase $f_{\rm d}$, while keeping $B/D$ fixed, then $\Gamma$ increases.
We also tested the effect of the halo scale length parameter but found
that this did not affect $B/D$ or $\Gamma$ substantially.
These sequential changes in the relation between $B/D$ and $\Gamma$
are summarized in Fig.~\ref{fig:Gamma_BD} of Appendix C, and the
detailed parameters of the individual models are summarized in
Tables~\ref{tb:models_add} and \ref{tb:models_add2}. Thus, galaxies
with a massive bulge tend to form tightly-wound spirals, but the shear
rate ($\Gamma$) is more essential to the pitch angle than $B/D$.

In addition, we look at the relation between $B/D$ and the bar formation
epoch ($t_{\rm b}$). 
We present this using the red symbols for models with $B/D>0.5$ in Fig.~\ref{fig:bfe}.
Models with a large $B/D$ tend to take a shorter time before the bar formation, 
but compared to the dependence on $f_{\rm d}$, the effect of $B/D$ on $t_{\rm b}$ is 
unclear. 

Summarizing all our simulated results,
we conclude that the disk-to-total mass fraction ($f_{\rm d}$) and the shear rate 
($\Gamma$) are important parameters that decide the disk galaxy morphology,
such as the number of spiral arms, the pitch angle, and the formation of a bar.

\section{Discussions}

\subsection{Hubble sequence and galaxy morphology}\label{Sect:Hubble}

We performed $N$-body simulations of disk galaxies
which start in an axisymmetric equilibrium state and which form spiral and 
bar structures. 
Using these simulations we investigated the relation between changes 
in the initial state of the bulge, disk, and halo (initial conditions in the simulations)
and the resulting morphology of the disk galaxies.

Especially the disk-mass fraction ($f_{\rm d}$) and shear rate ($\Gamma$)
are important parameters which influence the morphology of the simulated galaxies.
While $f_{\rm d}$ determines the number of spiral arms and
the bar formation epoch, $\Gamma$ has some influence on the bar formation
epoch and more strongly affects the spiral arms pitch angle.

Observationally, disk galaxies are classified using the
Hubble sequence~\citep{1926ApJ....64..321H}. In the Hubble sequence,
spiral galaxies are classified either as spirals (SA) or barred spirals (SB).
Each of these two branches has sub-types a to d; where the bulge-to-disk 
mass ratio ($B/D$) decreases and the pitch angle increases as we go from a to d~\citep{1961hag..book.....S}. 
In this section,
we discuss how the Hubble sequence can be understood by the disk's initial configuration
and secular evolution.

\begin{figure*}
\includegraphics[width=2.0\columnwidth]{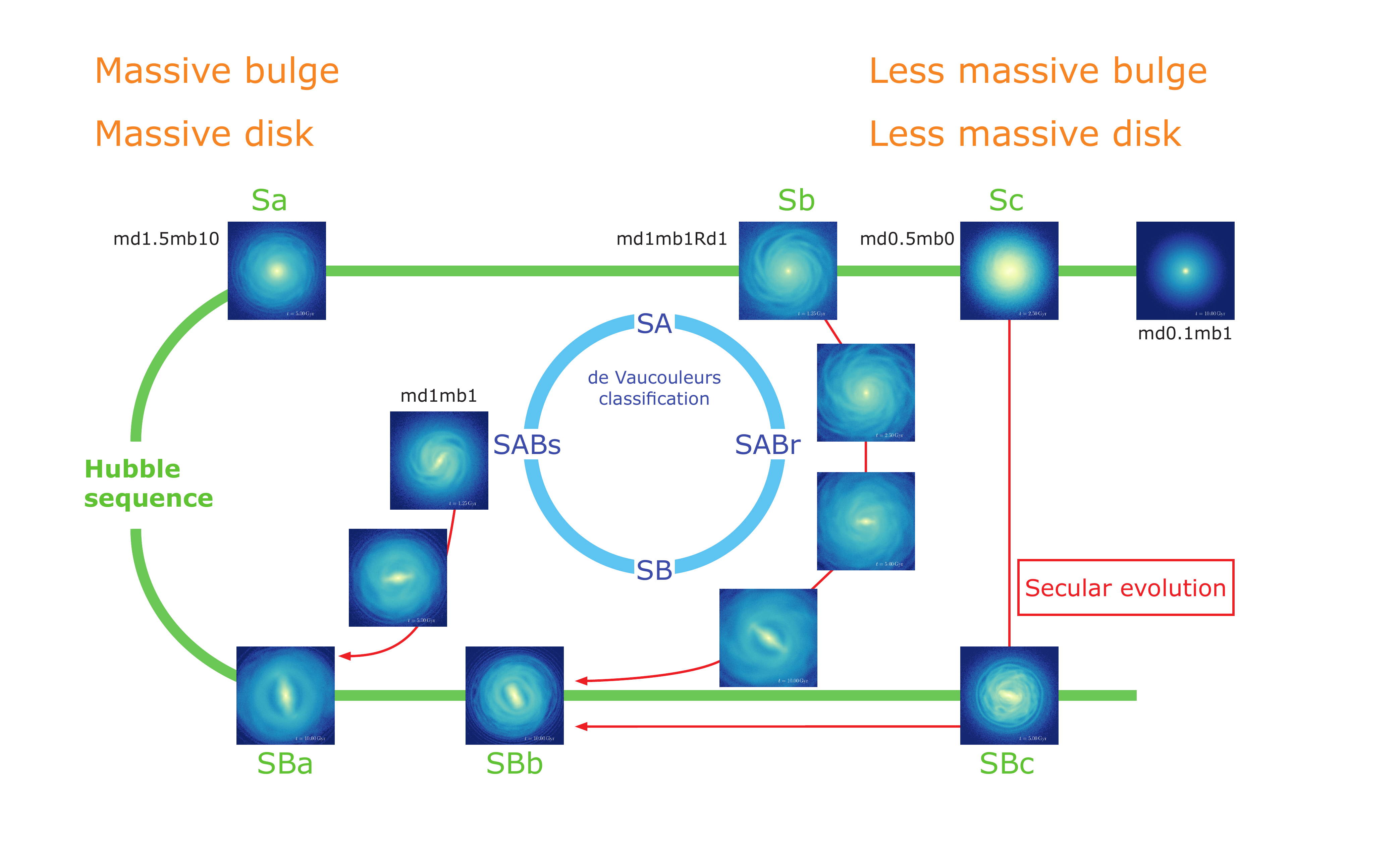}
\caption{The Hubble sequence and de Vaucouleurs classification overlaid with snapshots of our models. 
The red arrows show the models secular evolution.\label{fig:Hubble}}
\end{figure*}

In Fig.~\ref{fig:Hubble} we present a subset of 
snapshots from our simulation on the Hubble sequence.
In the Hubble sequence, the spirals are more loosely wound and the bulge becomes
fainter when moving from Sa to Sc. This is connected with the results we see in Section 3 where 
galaxies with a massive bulge have more tightly wound spiral arms due to the larger 
shear rate when the disk mass fraction is kept similar. Indeed, selecting
some models, we see that the sequence of spiral galaxies from Sa to Sc 
originates from changes in the initial distribution of the disk, bulge
and dark matter halo. 
However, $B/D$ does not linearly change the shear rate ($\Gamma$) as we
demonstrated in Section 3.4 (see also Appendix C). The shear rate also increases with
the disk mass fraction ($f_{\rm d}$). 
Therefore, $\Gamma$ is a more direct parameter to influence the pitch angle rather than $B/D$.
Of course galaxies with a massive bulge tend to have tightly winding spirals
because they tend to have a larger $\Gamma$.

Flocculent galaxies emerge
from models with a small disk to total-mass fraction
(such as in model md0.1mb1). 
Even after 10\,Gyr this model did not form a bar and 
the data in Fig.~\ref{fig:bfe} indicates that
it will take more than a Hubble time before the bar forms.

Once the bar formation criteria is satisfied the spiral galaxies leave the 
spiral sequence and move into the SB sequence. If the barred galaxy started as 
a Sc galaxy then in the early stages it resembles the barred-spiral structures 
as seen in SBc galaxies.
These galaxies continue to evolve towards SBb galaxies. For example, model md0.5mb0 has a strong bar and 
the spirals become more tightly wound in the later phases (see Figs.~\ref{fig:snapshots_5Gyr}
and \ref{fig:snapshots_10Gyr}).

The de Vaucouleurs classification~\citep{1959HDP....53..275D} appears 
when spiral galaxies evolve into barred galaxies. In the models 
where the disk is massive enough to form a bar, but less massive than
the models md0.5mb1 and md1mb1Rd1.5, a ring structure just outside 
of the bar appears after the bar has formed. 
In Fig.~\ref{fig:snapshots_10Gyr} and \ref{fig:snapshots_Rdisk} we see 
that these models still retain some 
spiral structures in the outer parts of the disks.
For models md1mb1 and md1.5mb5, which have a disk mass with
$f_{\rm d}>0.5$, a bar forms directly after the start of the 
simulations (see Fig.~\ref{fig:bfe}). They further form strong s-shaped 
and outer ring structures after the bar fully developed 
(right most of Fig.~\ref{fig:snapshots_10Gyr}). For these models we do not 
observe any clear spiral structure in the outer regions of the disk. 
This may be due to the absence of gas in our simulations \citep{2003MNRAS.344..358B}.

Although we did not take gas into account in our simulations, the fraction of gas is an important factor for the morphology of
disk galaxies.
\citet{1993ApJ...414..474S} suggested that bar formation is prevented if more than 10\% of the disk's mass 
is in the form of gas. \citet{2003MNRAS.344..358B} also 
performed a series of simulations that included gas and concluded that the bar stability
depends on the presence of gas.
We ignore the formation of disks that 
start from gas rich models. Such disk galaxies have been observed at a redshift of $\sim 1$
\citep{2017Natur.543..397G}. The effects of gas, however, must be considered in order to fully understand
the morphology of disk galaxies.
In addition, the galaxies in our study are not positioned in a cosmological framework,
which would be necessary to also take the merging history of disk galaxies into account. 
However, disk galaxies seem to be formed with a relatively quiet merging history 
\citep[e. g., ][]{2002NewA....7..155S, 2003ApJ...591..499A}.
Therefore, our study could be a testbed for studying the origin of disk galaxy morphology.

\subsection{Non-barred grand-design spirals}

As described in Section~\ref{Sect:bf}, swing amplification theory 
predicts that galaxies with massive disks (large disk-to-halo 
mass fraction) typically develop two spiral arms. This condition at the
same time satisfies the constraints for the rapid formation of a bar.
Both two-armed spirals and bars are structures of $m=2$.
In our models, galaxies with a massive disk often directly
form a bar rather than first forming a two-armed spiral disk. This implies
that $m=2$ structures in galactic disks are mostly bars. 
However, from observations we know that non-barred grand-design spiral
galaxies do actually exist.
One possible cause is that perturbations induced by a companion galaxy leads to 
the formation of such a spiral galaxy. This was tested by \citet{1972ApJ...178..623T} 
who, using simulations, showed that tidal interactions can lead to the 
formation of two spiral arms without a bar. Recently, \citet{2018MNRAS.474.5645P}
also showed this using simulations that included gas.
If accompanying galaxies are indeed the 
driver for the formation of two armed spirals then the 
number of grand-design spirals with companions must exceed the number of 
isolated grand-design galaxies. 
\citet{1979ApJ...233..539K} and \citet{1982MNRAS.201.1021E} observationally 
showed that disk galaxies with companions consist of a larger fraction of grand-design
spirals (0.6--1.0) compared to isolated galaxies (0.2--0.3).
However, not all non-barred grand-design spiral galaxies have companions. 
M\,74, for example, has no apparent companion~\citep{2011MNRAS.414..538K}.
In the following paragraphs we explore the formation of non-barred grand-design 
spirals without a companion. 

\begin{figure*}
\includegraphics[width=55mm]{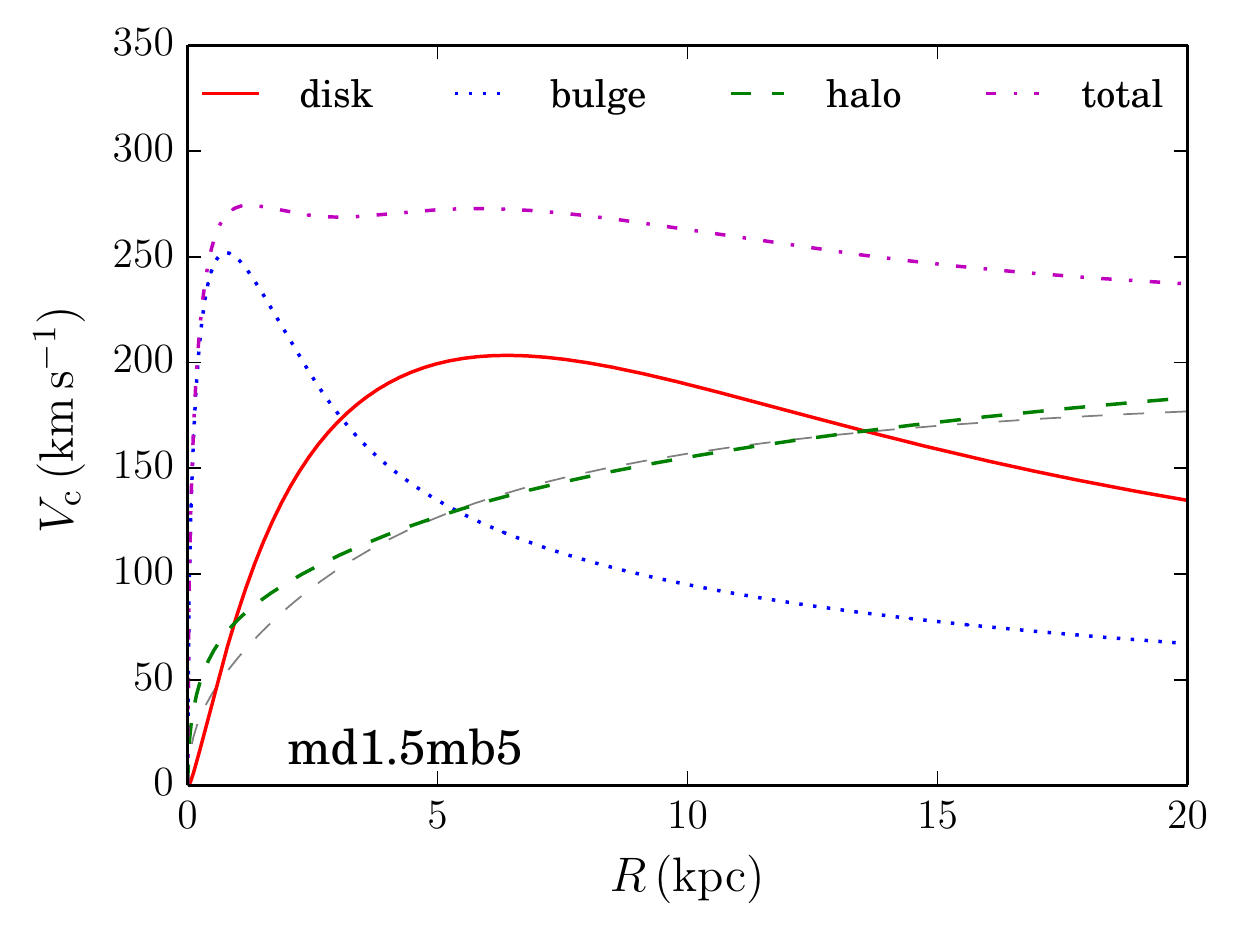}\includegraphics[width=42mm]{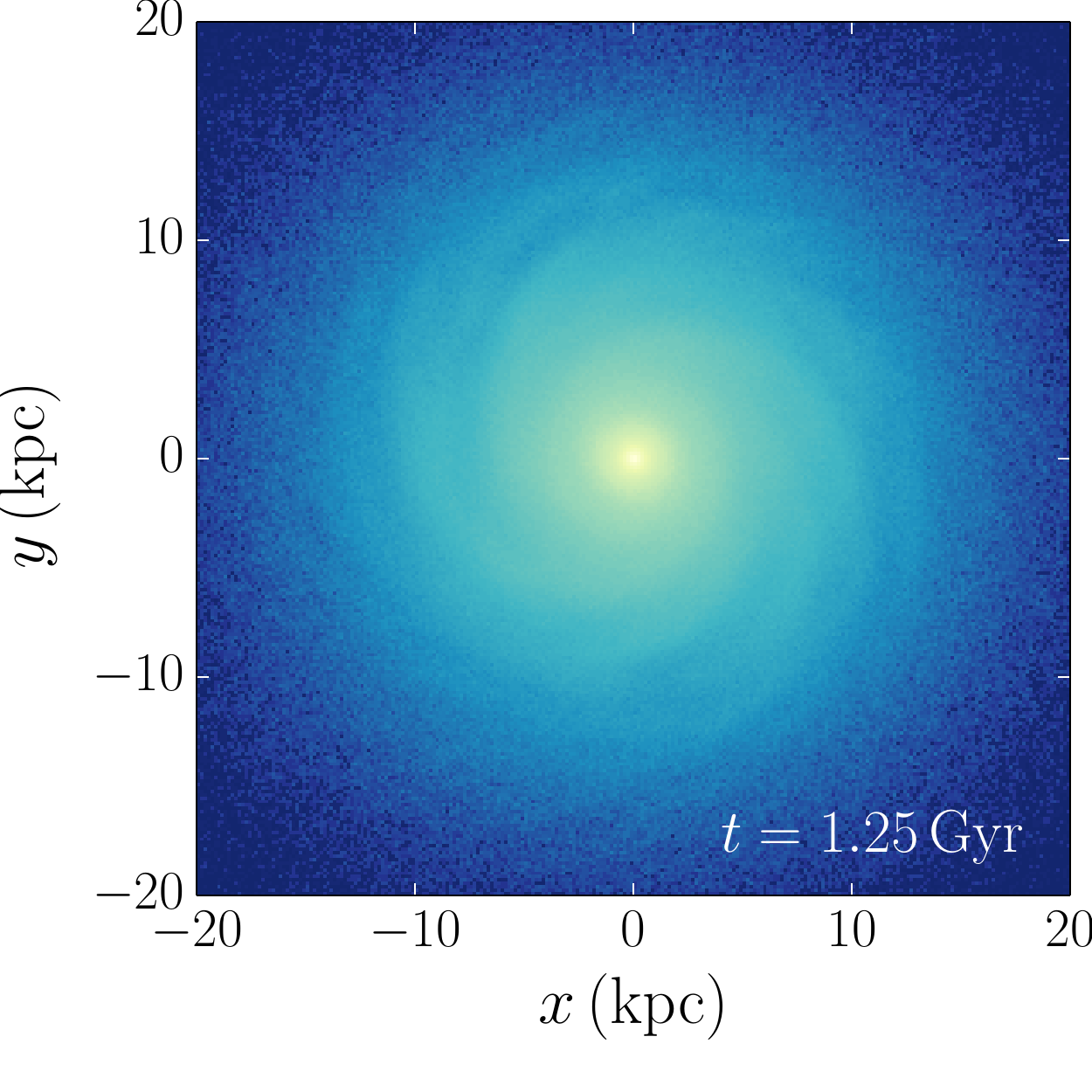}\includegraphics[width=42mm]{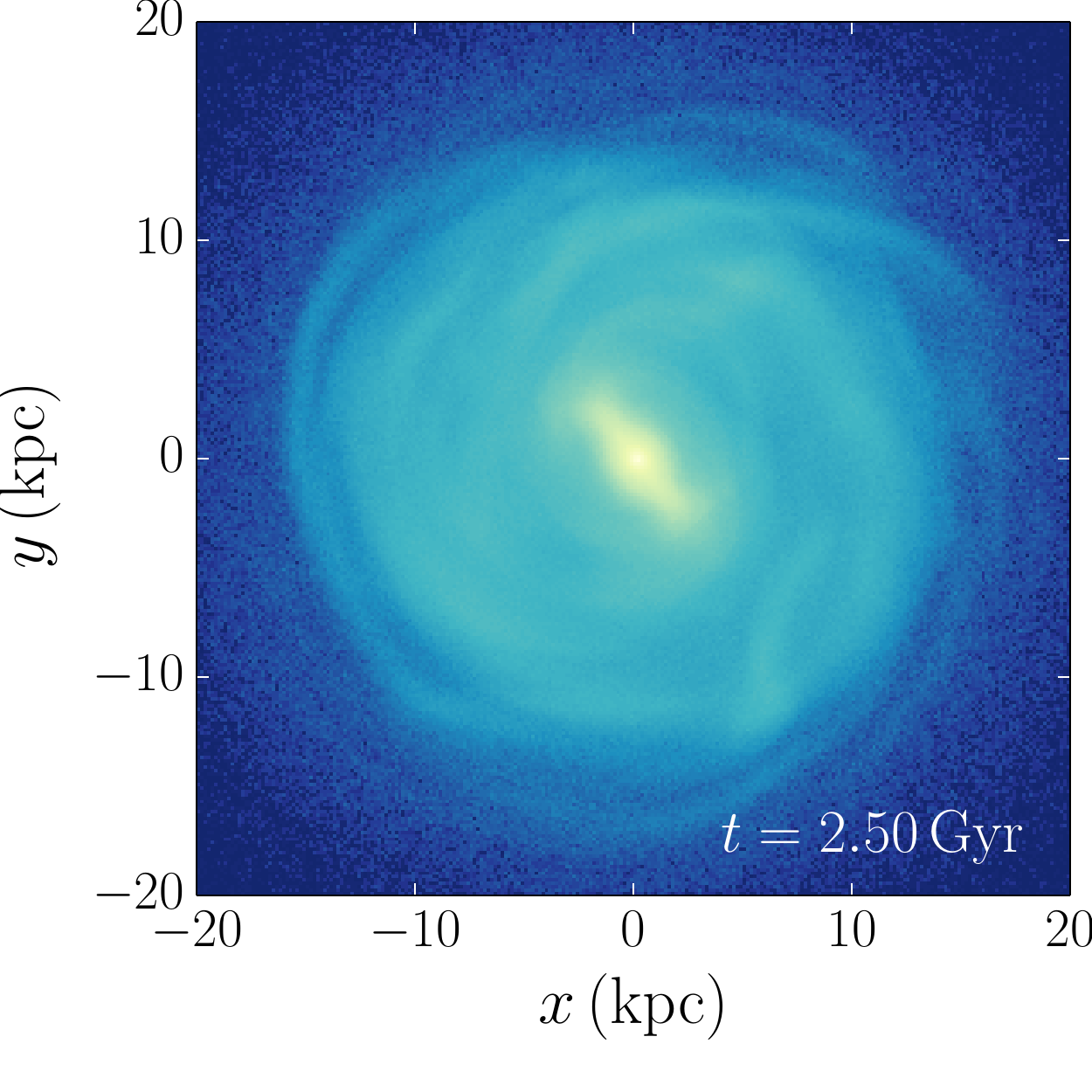}\\
\caption{Initial rotation curves (left) and density snapshots for model md1.5mb5 at $t=1.25$ Gyr (middle) and $t=2.5$ Gyr (right).
    The gray dashed curve in the left panel is the same as the one in Fig.~\ref{fig:snapshots_mb10}.
\label{fig:snapshots_md1.5}}
\end{figure*}

In the previous sections we saw that a massive disk leads to $m=2$ structures.
On the other hand, a massive bulge suppresses the formation of a bar, but massive bulges
tend to increase the number of spiral arms (e.g. model md1mb10).
To create a non-barred grand-design spiral we therefore setup a new model, md1.5mb5, which 
has the largest disk mass-fraction of all our models ($f_{\rm d}\sim 0.6$),  and a 
moderately massive bulge, $B/D\sim 0.3$. This model is expected to form a bar.
In Fig.~\ref{fig:snapshots_md1.5} we present the initial rotation curve (left panel) and the
density snapshots (right panels) for this model. 
At an age of $t=1.25$\,Gyr this model shows structure comparable to that observed in non-barred grand-design spirals, 
but only in the short time before the bar is formed. 
We conclude that non-barred grand-design spirals can form without companion, but that the structure is short-lived and disappears as soon as the bar forms. 
In our simulations we ignore the presence of gas, which may change the results.
\citet{2003MNRAS.344..358B}, however, performed a series of simulations 
that included gas dynamics, and they also concluded that a companion is necessary for non-barred grand-design spirals
\citep[see also a review by ][]{2014PASA...31...35D}.

\section{Summary}

We performed a series of galactic disk $N$-body simulations
to investigate the formation and dynamical evolution of spiral arm 
and bar structures in stellar disks which are embedded in live 
dark matter halos.
We adopted a range of initial conditions where the models have similar halo 
rotation curves, but different masses for the disk and bulge components, 
scale lengths, initial $Q$ values, and halo spin parameters.
The results indicate that the bar formation epoch increases exponentially 
as a function of the disk mass fraction with respect to the total mass at the 
reference radius (2.2 times the disk scale length), $f_{\rm d}$.
This relation is a consequence of swing amplification~\citep{1981seng.proc..111T},
which describes the amplification rate of the spiral arm when it transitions from 
leading arm to trailing arm because of the disk's differential rotation.
Swing amplification depends on the properties that characterize the disk, 
Toomre's $Q$, $X$, and $\Gamma$. The growth rate reaches its maximum
for $1<X<2$,  although the position of the peak slightly depends on $Q$ as well as on
$\Gamma$. We computed $X$ for 
$m=2$ ($X_2$), which corresponds to a bar or two-armed spiral, 
for each of our models and found that this value is related to the bar's
formation epoch.

The bar amplitude grows most efficiently when $1<X_2<2$. For models 
with $1<X_2<2$ the bar develops immediately after the start 
of the simulation. As $X_2$ increases beyond $X_2=2$, the growth rate
decreases exponentially. We find that the bar formation epoch increases
exponentially as $X_2$ increases beyond $X_2=2$, in other words
$f_{\rm d}$ decreases. The bar formation epoch exceeds a Hubble time
for $f_{\rm d}\lesssim 0.35$.

Apart from $X$, the growth rate is also influenced by $Q$ where
a larger $Q$ results in a slower growth. This indicates that the bar formation
occurs later for larger values of $Q$. 
Our simulations confirmed this and showed that for the bar ($m=2$) the growth rate
is predicted by swing amplification and becomes visible when it grows beyond a certain amplitude.

Toomre's swing amplification theory further predicts that
the number of spiral arms is related to the mass of the disk, with
massive disks having fewer spiral arms. In addition, larger $\Gamma$
predicts a smaller number of spiral arms.
We confirmed these relations in our simulations. 
The shear rate ($\Gamma$) also affects the pitch angle of spiral
  arms. We further confirmed that our result is consistent with previous
studies.

We found that the disk-to-total mass fraction ($f_{\rm d}$)
and the shear rate ($\Gamma$) are the most important parameters that determine the
morphology of disk galaxies. 
When juxtaposing our models with the Hubble sequence,
the fundamental subdivisions of (barred-)spiral galaxies with 
massive bulges and tightly wound-up spiral arms from S(B)a to S(B)c is 
also be observed as a sequence in our simulations. Where the models 
with either massive bulges or massive disks have more tightly
wound spiral arms. This is because having both a massive disk and bulge results in 
a larger $\Gamma$, i.e., more tightly wound spiral arms.

Once the
bar is formed it starts to heat the outer parts of the disk.
From this point onwards, 
the self-gravitating spiral arms disappear.
This may be in part caused by the 
lack of gas in our simulations. 
After the bar grows, we no longer discern  
spiral arms in the outer regions of the disk. This could imply
that gas cooling and star formation are required in order to 
maintain spiral structures in barred spiral galaxies for over 
a Hubble time~\citep{1981ApJ...247...77S,1984ApJ...282...61S}.

Our simulations further indicate that non-barred grand-design spirals are
transient structures which immediately evolve into barred
galaxies. Swing amplification teaches us that a massive disk is
required to form two-armed spiral galaxies. This condition, at the
same time, satisfies the short formation time of the bar structure.
Non-barred grand-design spiral galaxies therefore must evolve into barred
galaxies.  We consider that isolated non-barred grand-design spiral galaxies 
are in the process of developing a bar.

\section*{Acknowledgments}

We thank the anonymous referee for the very helpful comments.
This work was supported by JSPS KAKENHI Grant Number 26800108, 
HPCI Strategic Program Field 5 'The origin of matter and the universe,' 
and the Netherlands Research School for Astronomy (NOVA).
Simulations are performed using GPU clusters, HA-PACS at the
University of Tsukuba, Piz Daint at CSCS and Little Green Machine II
(621.016.701). Initial development has been done using the Titan
computer Oak Ridge National Laboratory.  This work was supported by a
grant from the Swiss National Supercomputing Centre (CSCS) under
project ID s548 and s716.  This research used resources of the Oak Ridge
Leadership Computing Facility at the Oak Ridge National Laboratory,
which is supported by the Office of Science of the U.S. Department of
Energy under Contract No. DE-AC05-00OR22725 and by the European
Union's Horizon 2020 research and innovation programme under grant
agreement No 671564 (COMPAT project).
\appendix

\section[]{The effects of other parameters}

We discussed the effect of the bulge and disk masses on the
development of bars and spiral arms in the main text. Here we briefly
summarize the effects of the other parameters we investigated.

\subsection{The halo spin}

The spin of the halo is known to be an important parameter that 
affects the bar's secular evolution. 
\citet{2014ApJ...783L..18L} showed that a co-rotating disk and halo 
speed up the bar formation, but decrease its final length. This 
is due to the angular momentum transfer between the disk and halo.
If the halo does not spin it absorbs the bar's angular momentum, 
which slows down the bar and increases its length. 
A co-rotating halo, however, returns angular momentum to the disk instead of 
just absorbing it. 
This stabilizes the angular momentum transfer, and the bar evolution ceases.

We setup a few models, based on model md1mb1, but now with a rotating halo. 
In order to give spin to the halo we change the sign of the angular momentum $z$ component, $L_{\rm z}$.
For models md1mb1s0.65 and md1mb1s0.8, 65 and 85\,\% of the halo particles are rotating in the same 
direction as the disk. For models without rotation, this value is 50\,\%. 

To compare our results with previous studies, we measure the spin 
parameter~\citet{1969ApJ...155..393P,1971A&A....11..377P}:
\begin{eqnarray}
\lambda = \frac{J|E|^{1/2}}{GM_{\rm h}^{5/2}},
\end{eqnarray}
where $J$ is the magnitude of the angular momentum vector, and $E$ is the total 
energy.
In our models, $\alpha_{\rm h}=0.65$ (0.8) correspond 
to $\lambda\sim0.03$ (0.06).

\begin{figure}
\begin{center}
  \includegraphics[width=40mm]{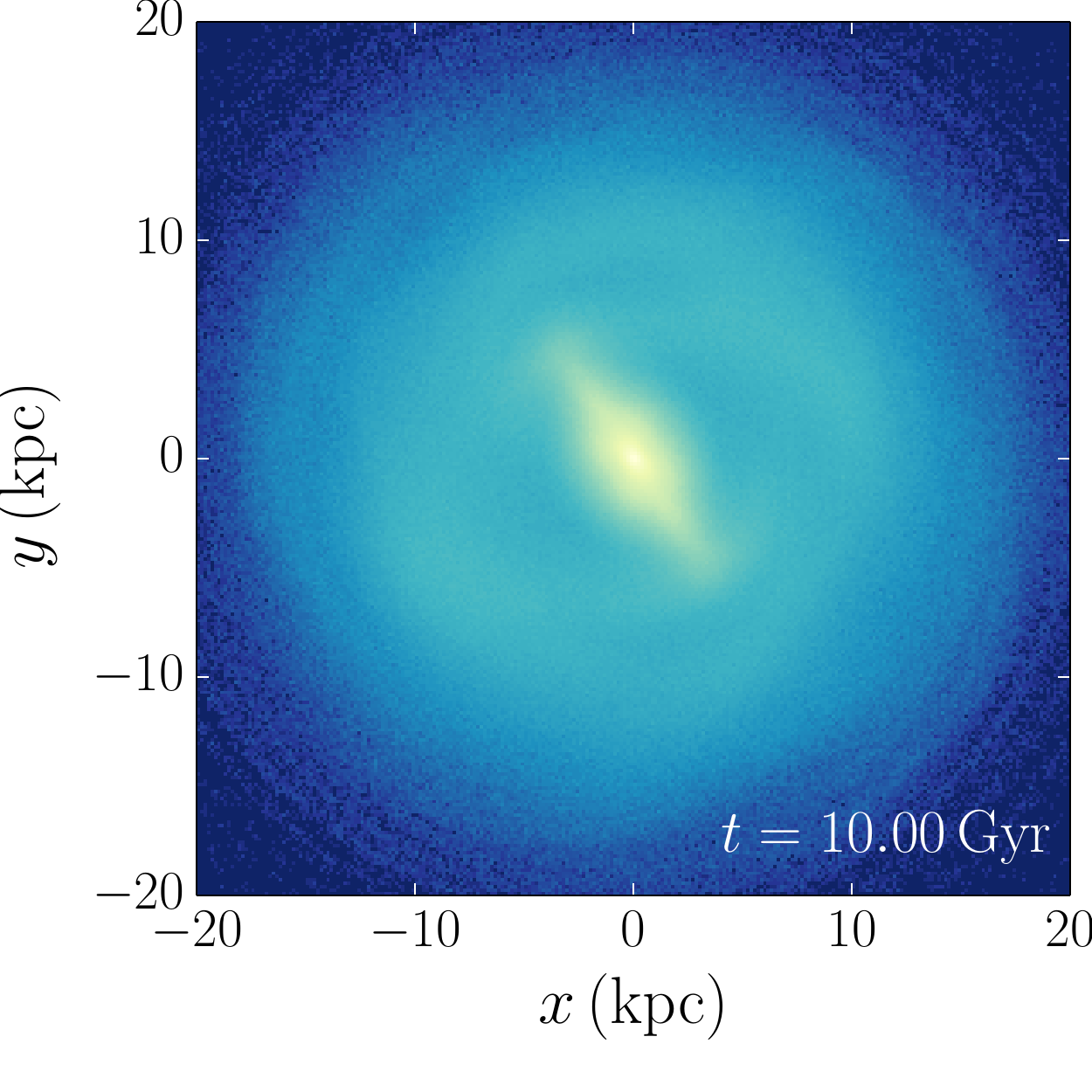}
  \includegraphics[width=40mm]{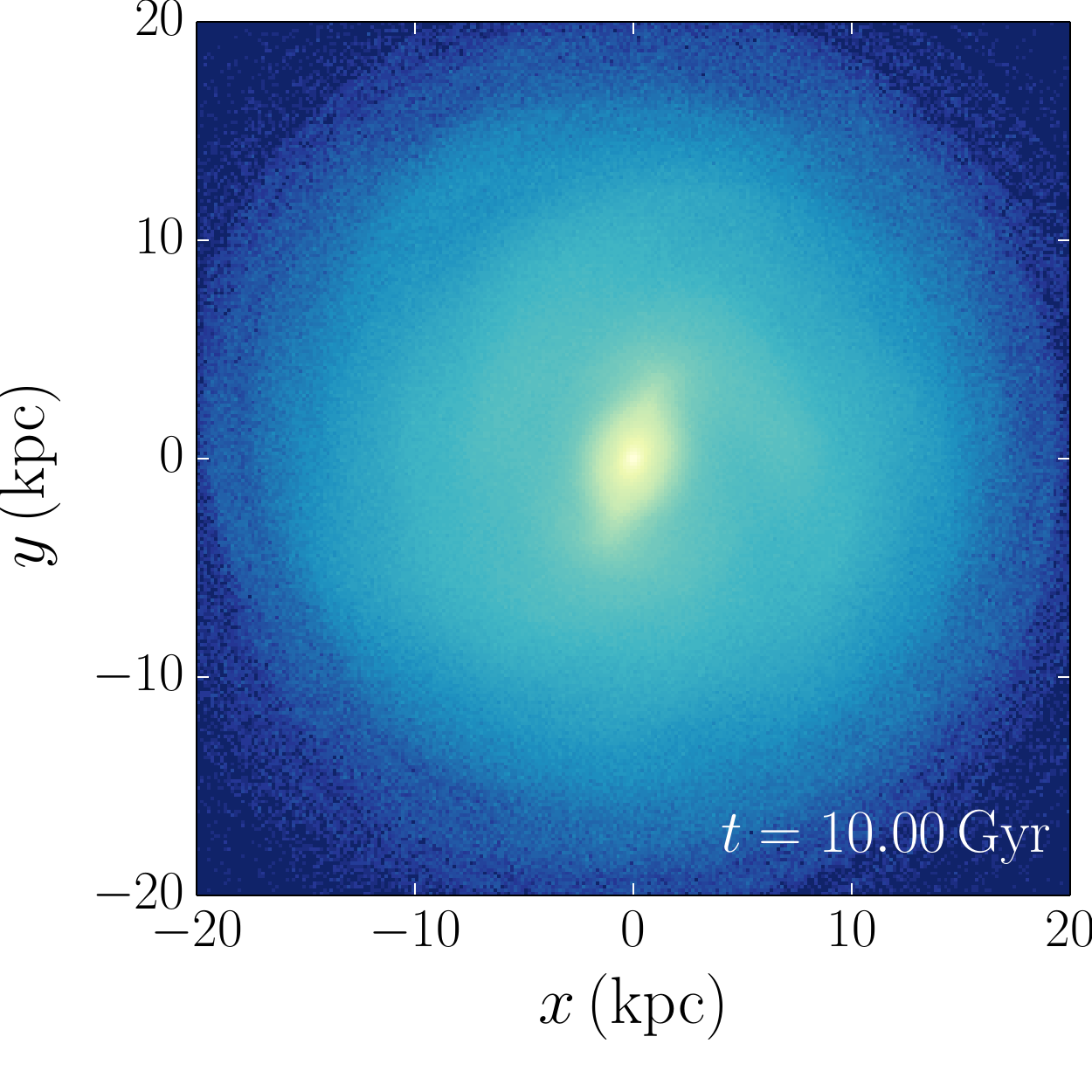}\\
    \caption{Snapshots for models md1mb1s0.65 (left) and md1mb1s0.8 (right), which are the same as model md1mb1 (Fig.~\ref{fig:snapshots_10Gyr},far most right panel) but now including halo spin.}\label{fig:snapshots_spin_b}
\end{center}
\end{figure}

In Fig.~\ref{fig:snapshots_spin_b} we present the effect that the halo spin
has on models md1mb1s0.65 and md1mb1s0.8. 
The results indicate that  
the bar is shorter for the models with a stronger halo spin.

\begin{figure*}
\includegraphics[width=\columnwidth]{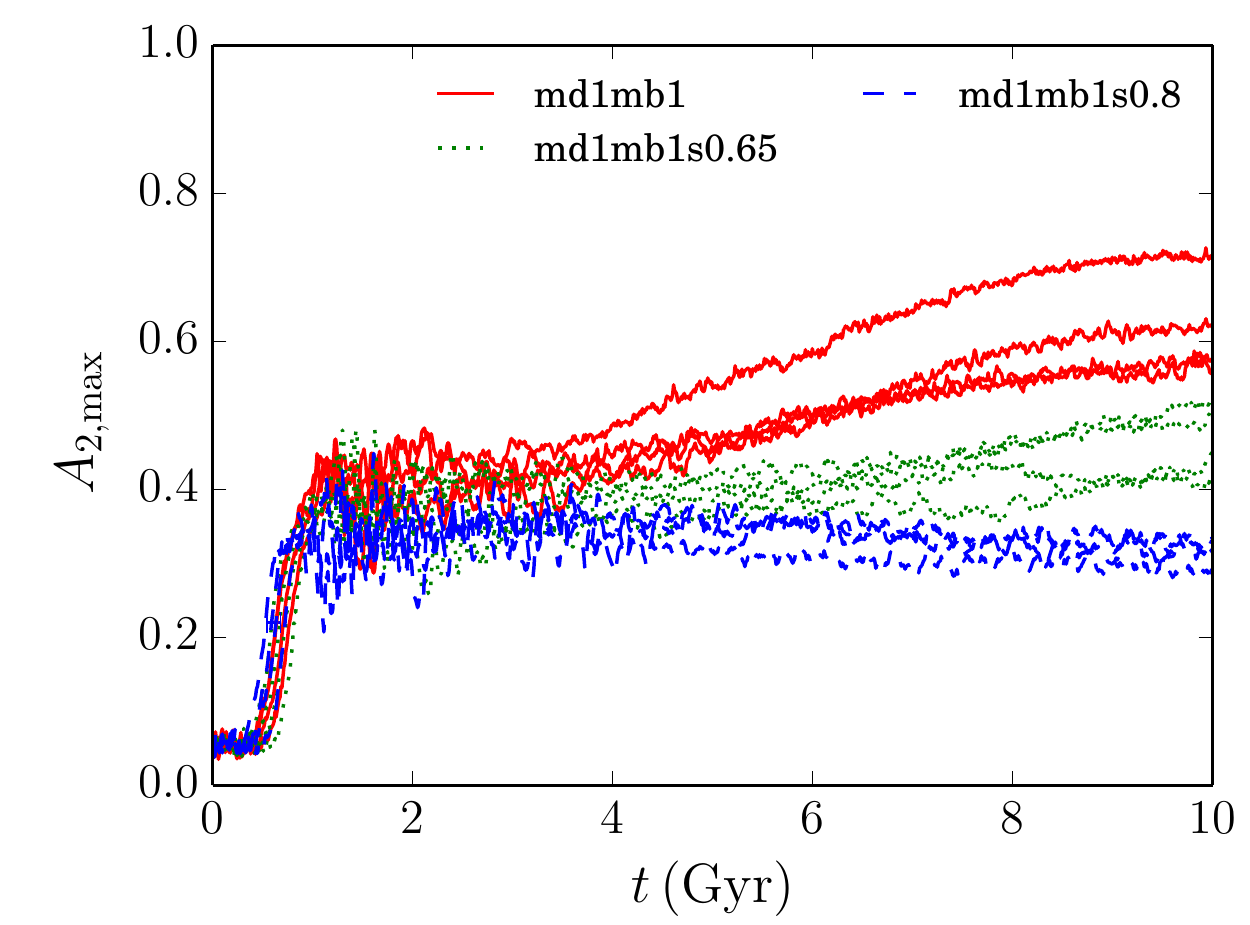}\includegraphics[width=\columnwidth]{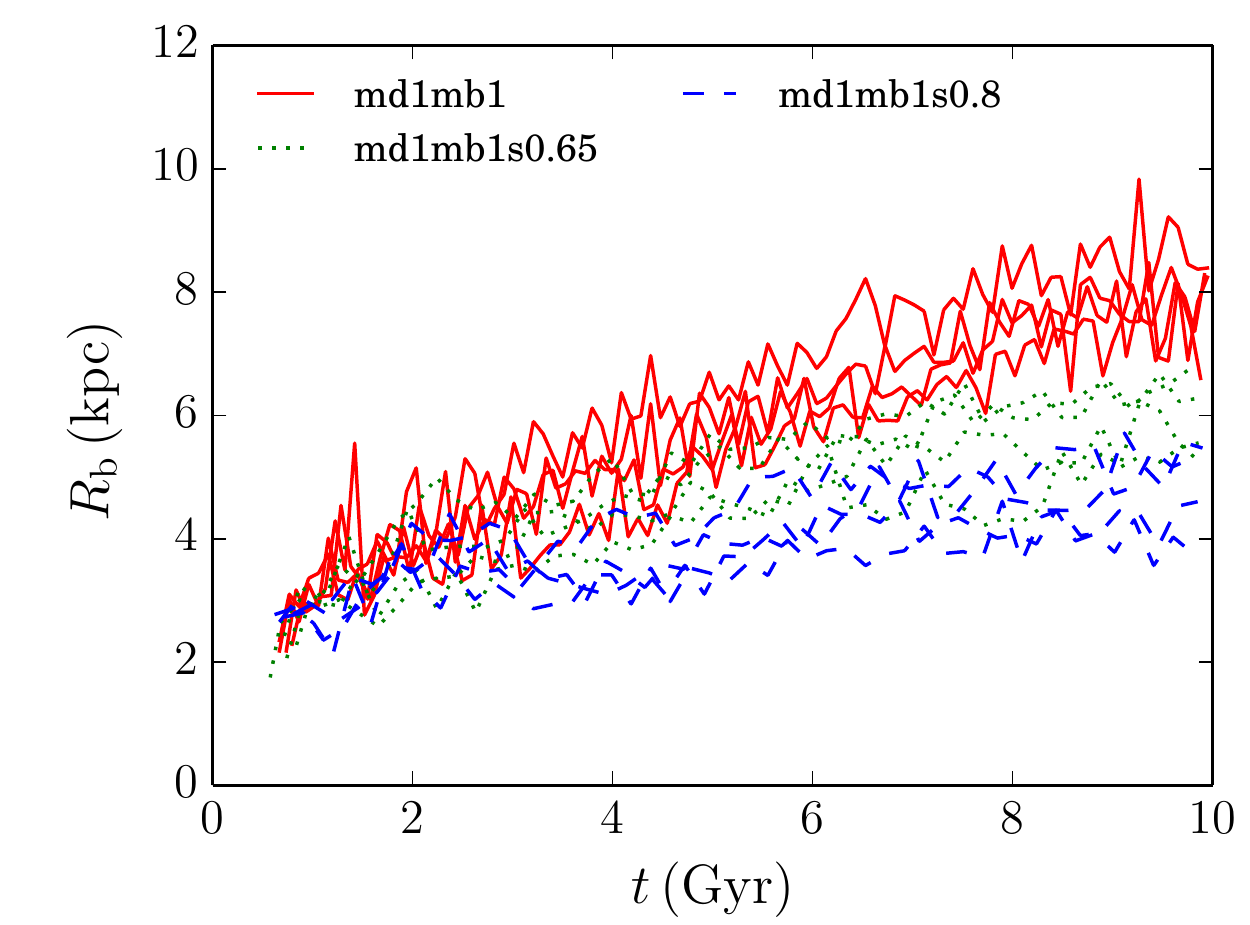}
\caption{Time evolution of the maximum amplitude for $m=2$ (left) and the bar length (averaged for every $\sim 0.1$\,Gyr) for models md1mb1, md1mb1s0.65, and md1mb1s0.8. For each model, we performed four simulations changing the random seed (varying positions and velocities of the particles) when generating the initial realizations.
\label{fig:A2_max_spin}}
\end{figure*}

In Fig.~\ref{fig:A2_max_spin} we show the length and maximum amplitude of
the resulting bars.
These results are consistent with previous results which show that
the length of the bar and its amplitude decay when the halo spin increases.
However, in contrast to~\citet{2013MNRAS.434.1287S} and \citet{2014ApJ...783L..18L} ,
we find that the epoch of bar formation in our models is similar, 
whereas a faster formation was expected based on the larger halo spin. 
In order to rule out the effect of run-to-run variations~\citep{2009MNRAS.398.1279S},
we performed four additional simulations for each of models md1mb1, md1mb1s0.65 and md1mb1s0.8.
For the bar formation epochs we calculated the average and standard deviation. 

The average bar formation-epoch is $0.674 \pm 0.053$,
$0.691\pm 0.083$, and $0.610\pm 0.069$\,Gyr for models md1mb1, md1mb1s0.65 and md1mb1s0.8, respectively.
This may be caused by the relatively early bar formation (within $\sim0.8$\,Gyr)
compared to the previous
studies; 1--2\,Gyr for \citet{2014ApJ...783L..18L} and 3--4\,Gyr for
\citet{2013MNRAS.434.1287S}.
Indeed, in~\citet{2014ApJ...783L..18L} the bar formation epoch starts slightly earlier when 
a moderate spin parameter ($\lambda=0.045$ and 0.06) is introduced. The
dependence of the bar formation-epoch on the halo spin is even clearer in
\citet{2013MNRAS.434.1287S}, where the formation time is longer
than in~\citet{2014ApJ...783L..18L}.
We therefore argue that the rapid bar formation in our models may hide the sequential delay
of the bar formation as caused by the halo spin.

\begin{figure}
\includegraphics[width=40mm]{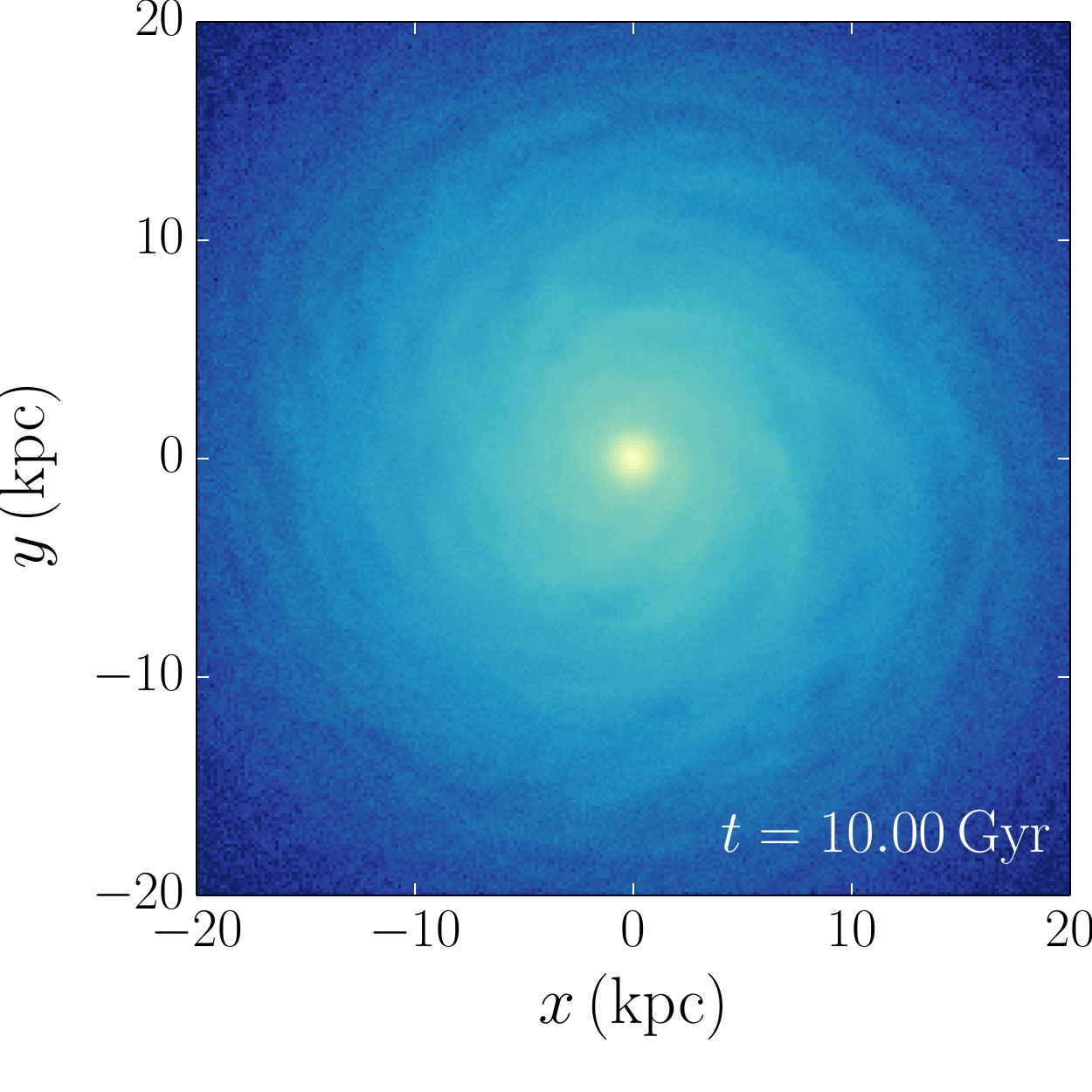}\includegraphics[width=40mm]{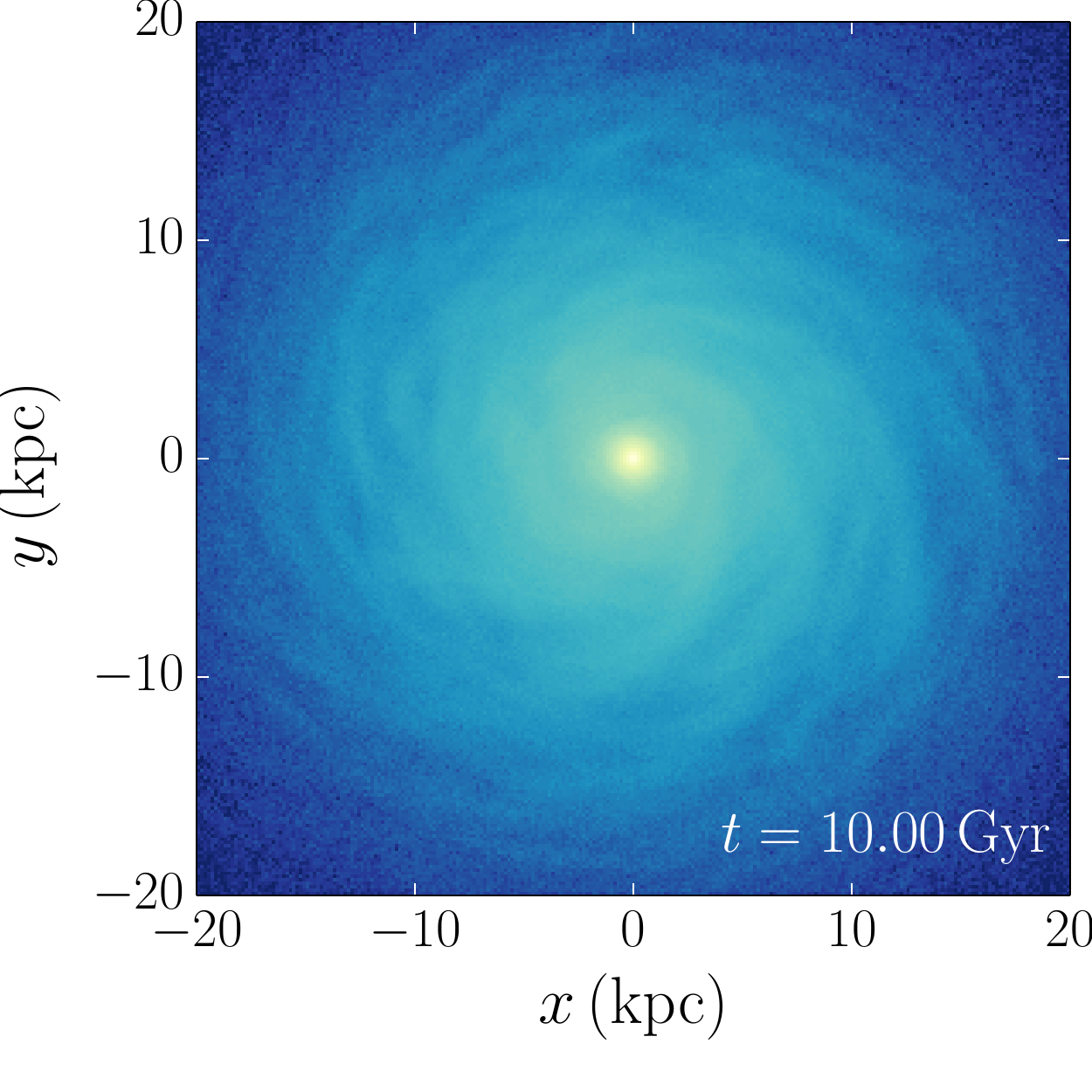}\\
\caption{Snapshots for models md0.5Rd1.5 (left) and md0.5Rd1.5s (right). \label{fig:snapshots_spin_sp}}
\end{figure}

\begin{figure*}
  \includegraphics[width=\columnwidth]{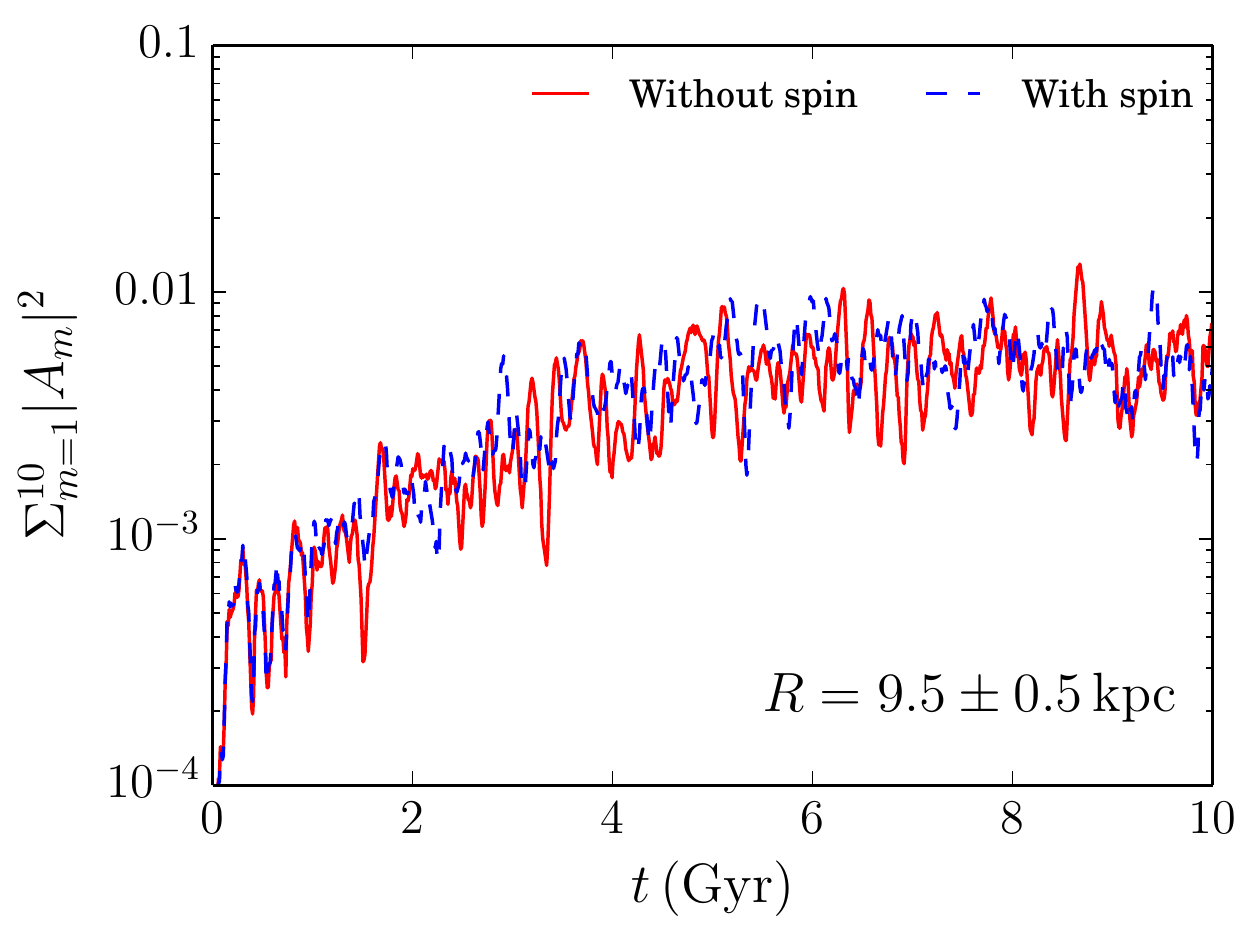}
  \includegraphics[width=\columnwidth]{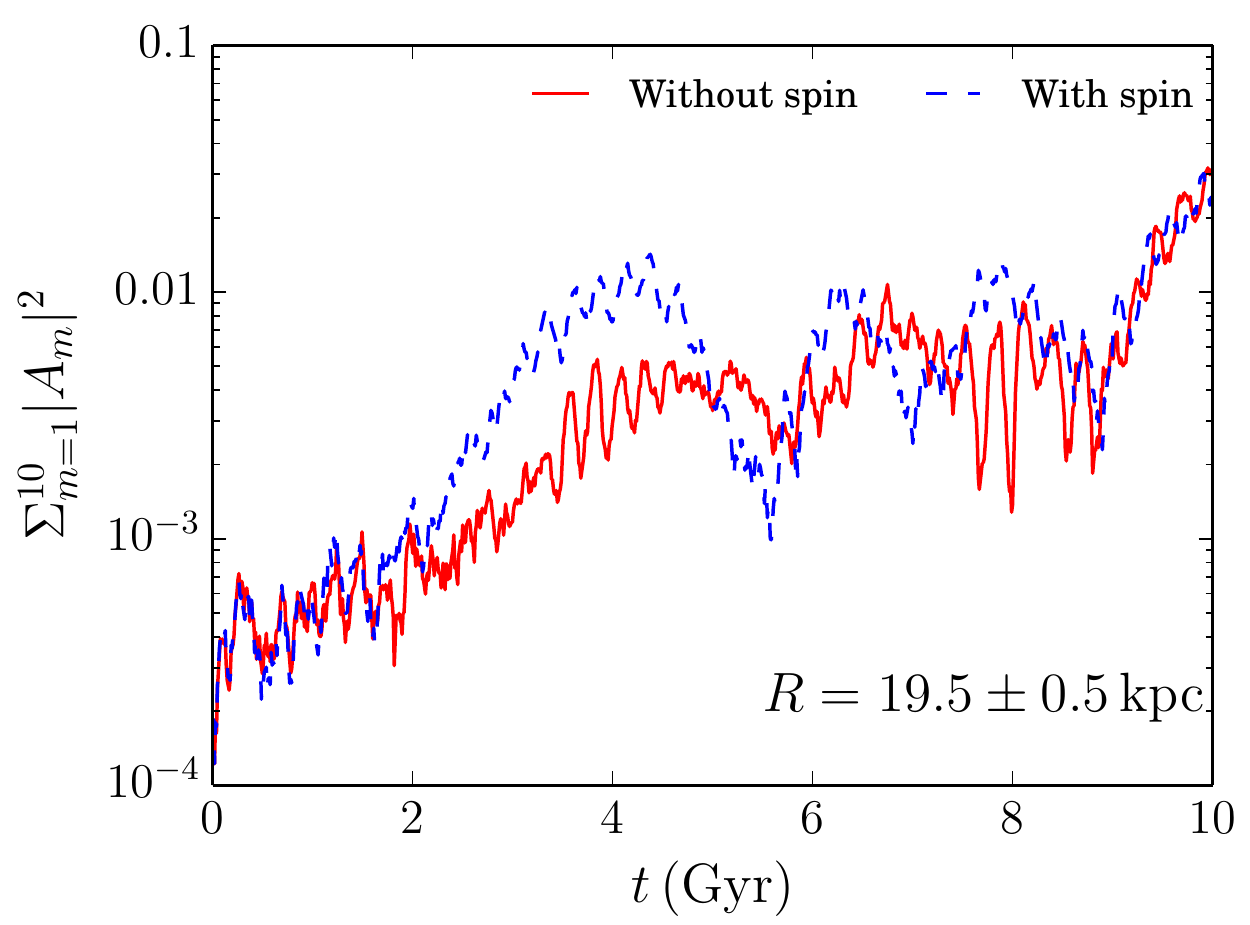}\\
\caption{Total power for models md0.5Rd1.5 and md0.5Rd1.5s at $R=9.5$ kpc (left) and 19.5 kpc (right). \label{fig:mode_spin_sp}}
\end{figure*}

In addition to the bar forming models above, we also added halo spin to a model that 
shows no bar formation within 10\, Gyr. This model, md0.5Rd1.5s, is based on md0.5Rd1.5
but now with a halo spin of 0.8. 
In Fig.~\ref{fig:snapshots_spin_sp}, we present the snapshots of the above models at $t=10$\,Gyr. 
In contrast to the barred galaxies, their spiral structures look quite similar. 
To quantitatively compare the spiral amplitudes we use the total amplitude of the spiral arms 
given by $\sum ^{10}_{m=1} |A_m|^2$, where $A_m$ is the Fourier amplitude (Eq.~\ref{eq:Fourier}).
Instead of the bar amplitude, we measured the spirals total amplitude at 
$2.2R_{\rm d}$ and at $4.5R_{\rm d}$ (for this model 9.5 and 19.5\,kpc, respectively), 
the results are shown in Fig.~\ref{fig:mode_spin_sp}. 
The evolution of the spiral amplitudes are quite similar for both models, 
just like the pitch angle  $24^{\circ}$--$29^{\circ}$ (with)
$24^{\circ}$--$26^{\circ}$ (without halo spin) and the number of 
spiral arms $m=7$--8 for $R=10$--14\,kpc (see Table~\ref{tb:pitch_angle}).

\begin{figure*}
  \includegraphics[width=\columnwidth]{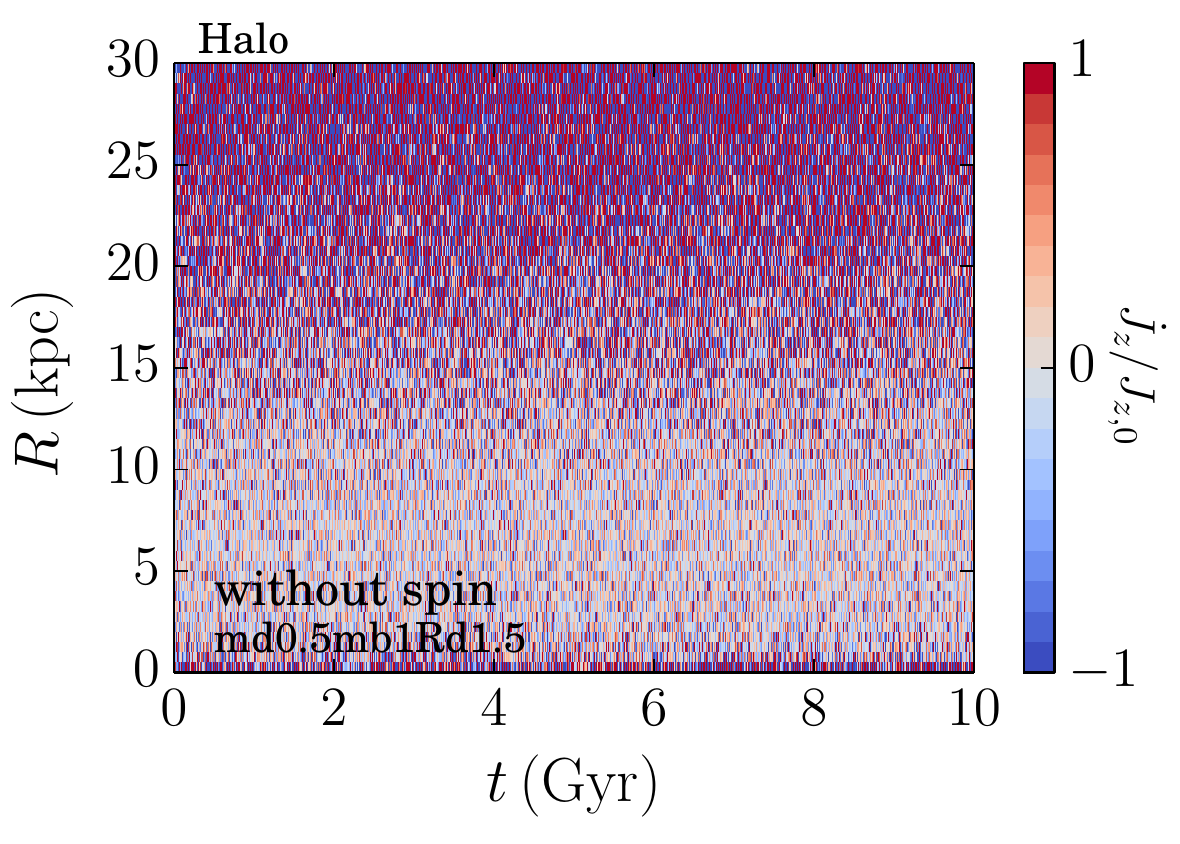}\includegraphics[width=\columnwidth]{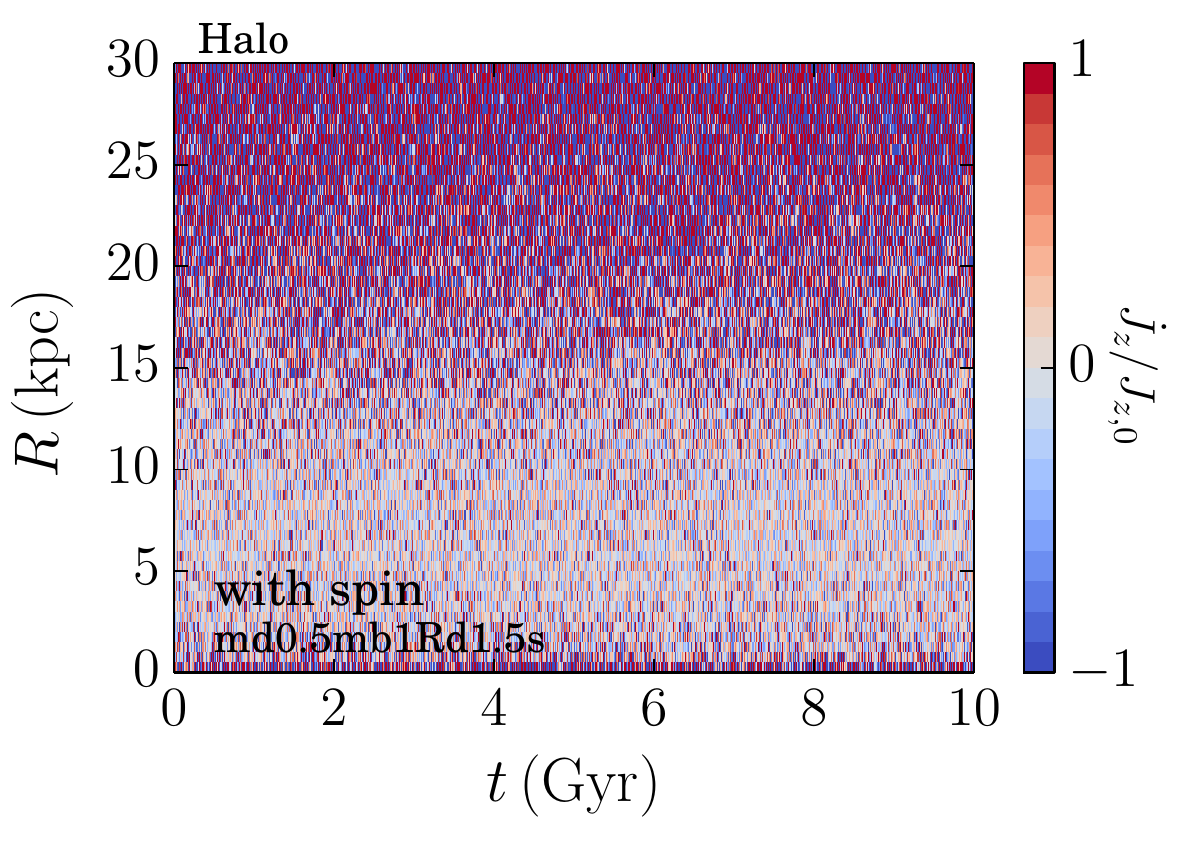}\\
  \includegraphics[width=\columnwidth]{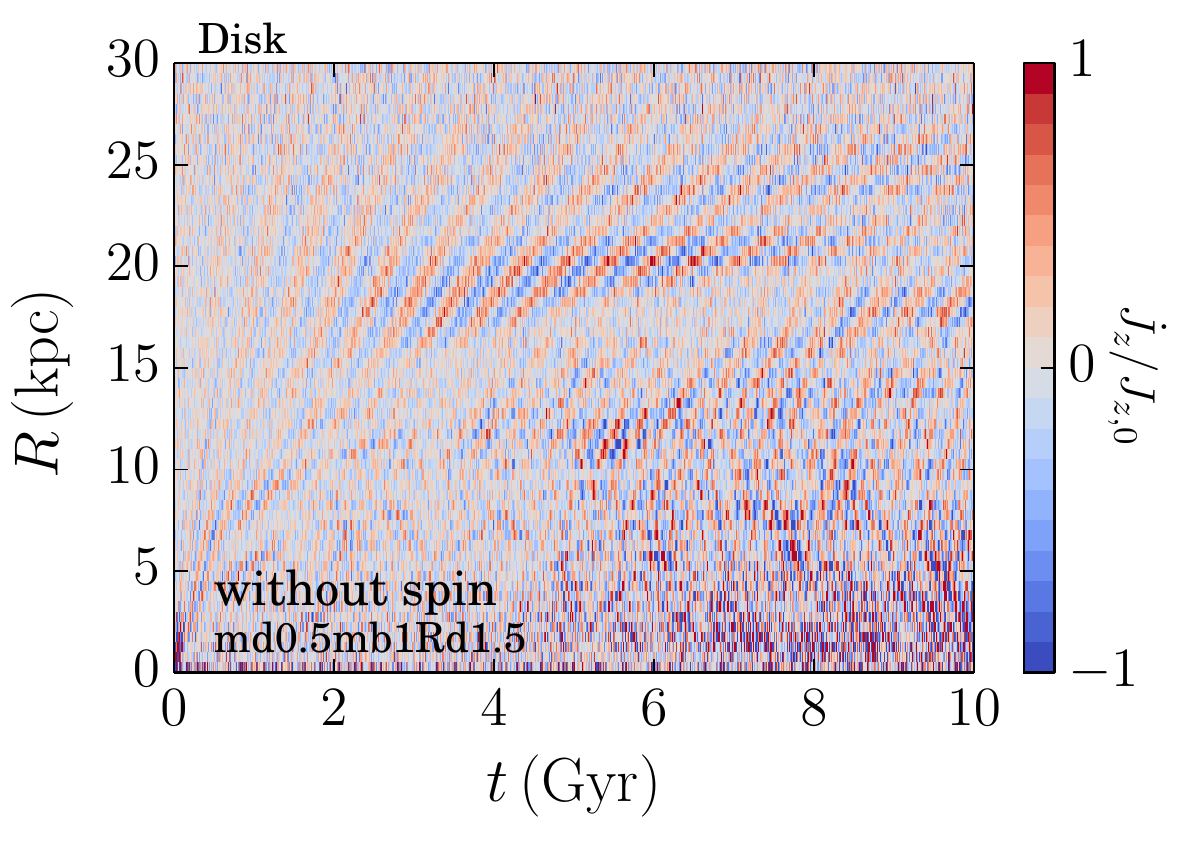}\includegraphics[width=\columnwidth]{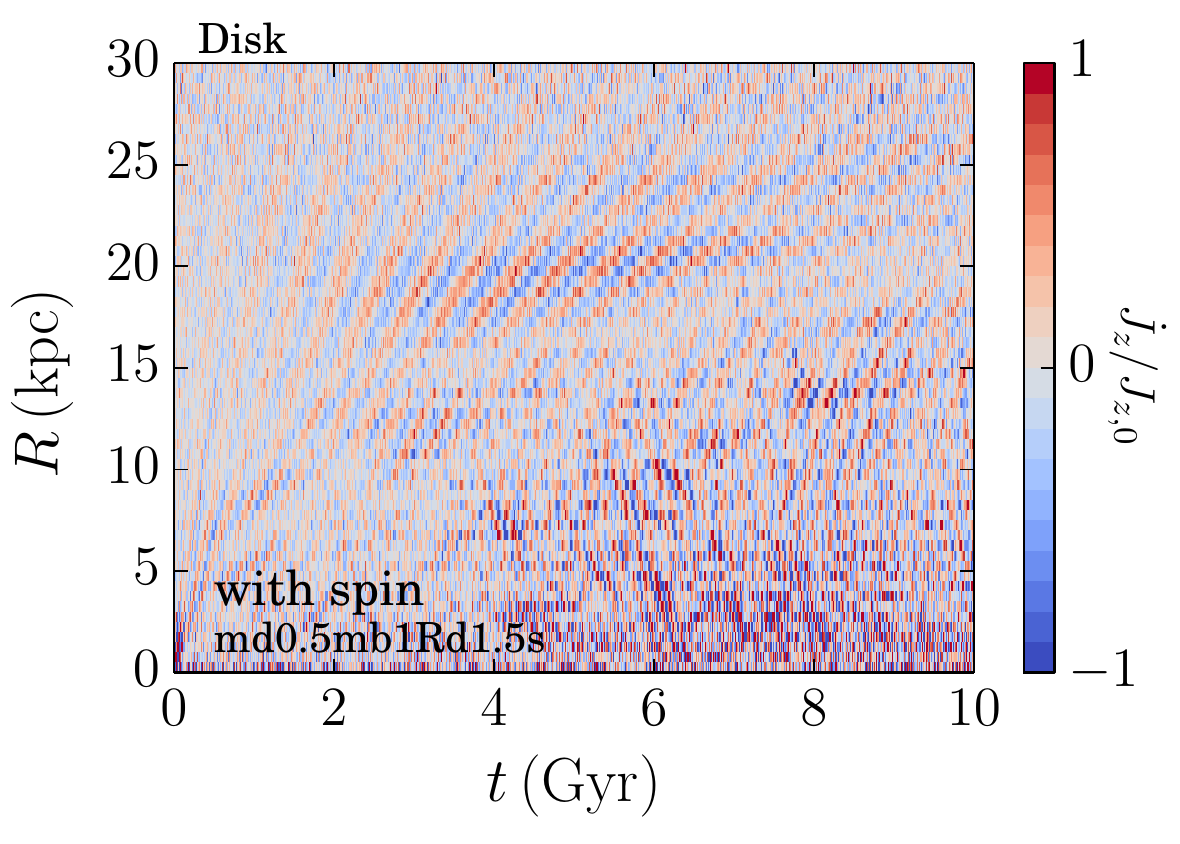}\\
\caption{Angular momentum flow of the halo (top) and the disk (bottom) as a function of cylindrical radius and time for models md0.5mb1Rd1.5 (left) and md0.5mb1Rd1.5s (right). The angular momentum flow is calculated from the angular momentum's change in the $z$-component for every $\sim 10$\,Myr. The value (color) is scaled to the initial angular momentum of the disk at each radius for both the disks and halos.}
\label{fig:AM}
\end{figure*}

In addition, in Fig.~\ref{fig:AM} we investigate the angular-momentum flow for 
the disk and halo as a function of time and cylindrical radius.
Following \citet{2014ApJ...783L..18L} and \citet{2009ApJ...707..218V},
we measure the change in angular momentum of the $z$-component at
every $\sim 10$\,Myr.
For the halos (top panels) there is no continuous angular
momentum transfer from the disk to the halo, but we only discern random variations
in the angular momentum. These fluctuations look stronger at outer
radii, but this is because the angular momentum changes are normalized by
the disk' angular momentum, which is smaller in the outer regions.

\begin{figure*}
  \includegraphics[width=\columnwidth]{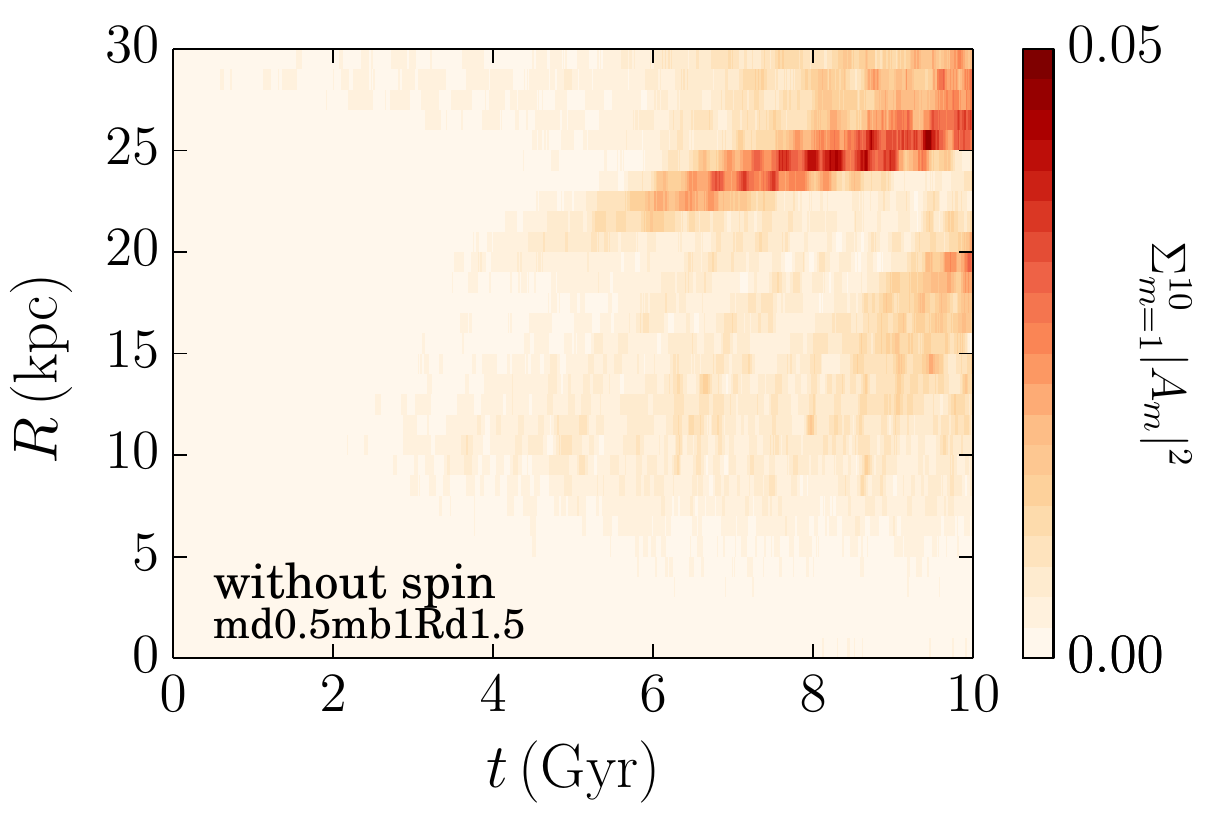}
  \includegraphics[width=\columnwidth]{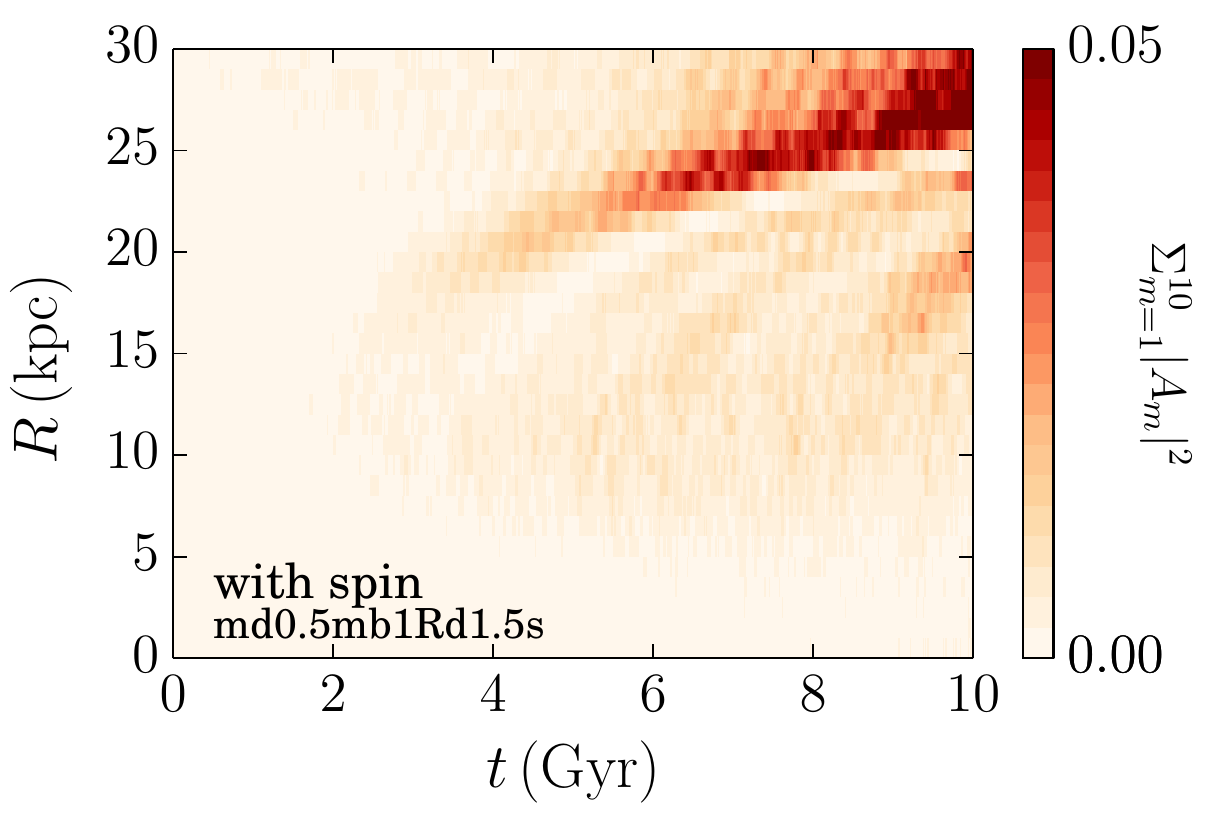}\\
\caption{Total power as a function of cylindrical radius and time for models md0.5Rd1.5 (left) and md0.5Rd1.5s (right).}\label{fig:amplitude_ev}
\end{figure*}

The angular momentum of the disks vary with time (see the red and blue
stripes in the bottom panels), but overall the disk loses only 1.9\,\%
of its initial angular momentum for models with spin and 1.7\,\% for
models without.
The amplitude of the stripes for the disks roughly corresponds to the
amplitude of the spiral pattern. In Fig.~\ref{fig:amplitude_ev}, we show the
total power as a function of cylindrical radius and time for models
md0.5Rd1.5 (left) and md0.5Rd1.5s (right).
From this we conclude that for spiral arms the angular momentum transfer between the disk and 
the halo is not efficient.
On the other hand, for barred galaxies the angular momentum flow 
from the disk to the halo is considerably smaller for models with 
a larger halo spin \citep[see Fig. 3 in][]{2014ApJ...783L..18L}.

\subsection{Initial Q value}

To verify the expectation that the initial value of Toomre's $Q$ parameter 
($Q_0$) influences the bar and spiral structure, we created a set of models in 
which we varied this parameter. 

\begin{figure*}
\includegraphics[width=\columnwidth]{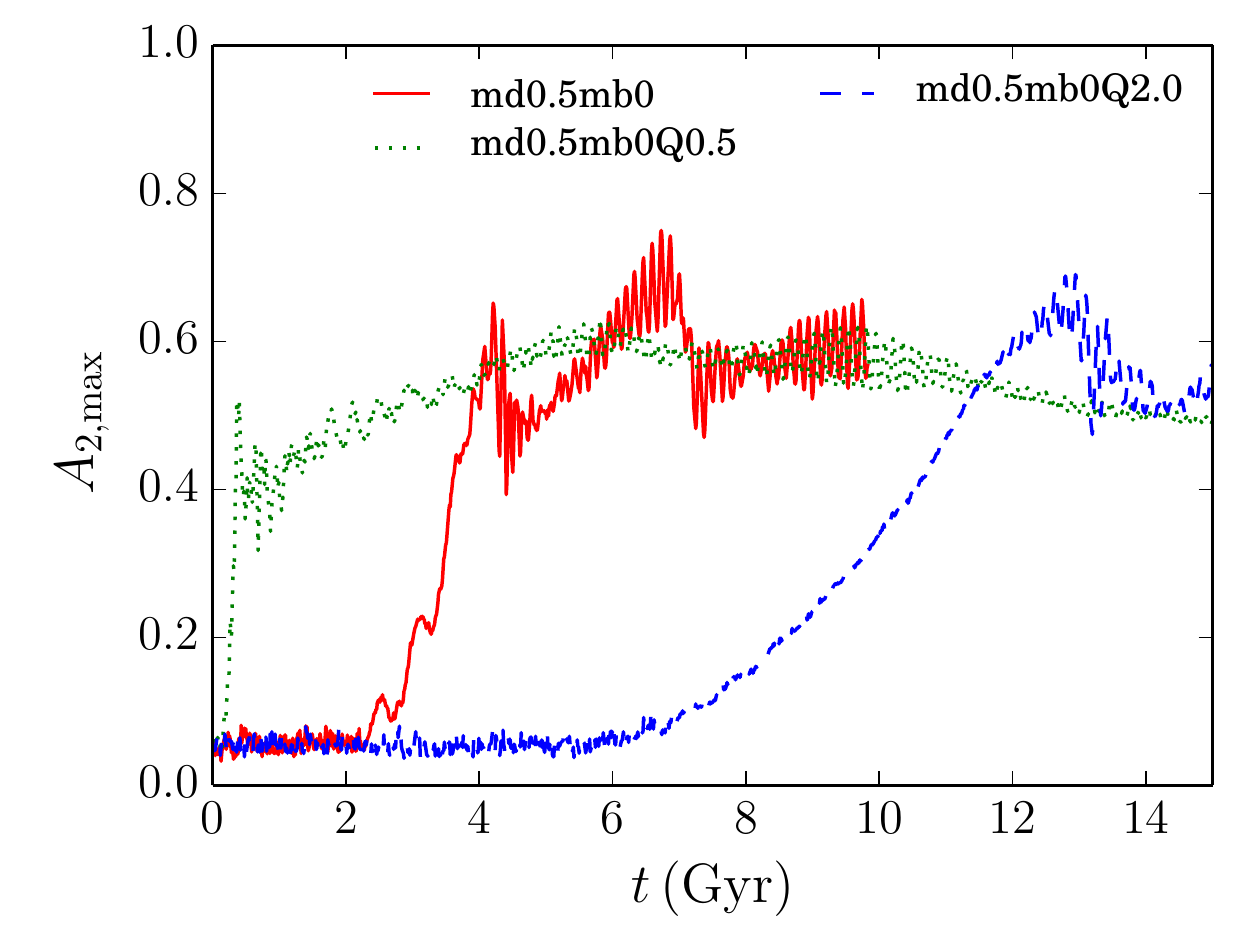}\includegraphics[width=\columnwidth]{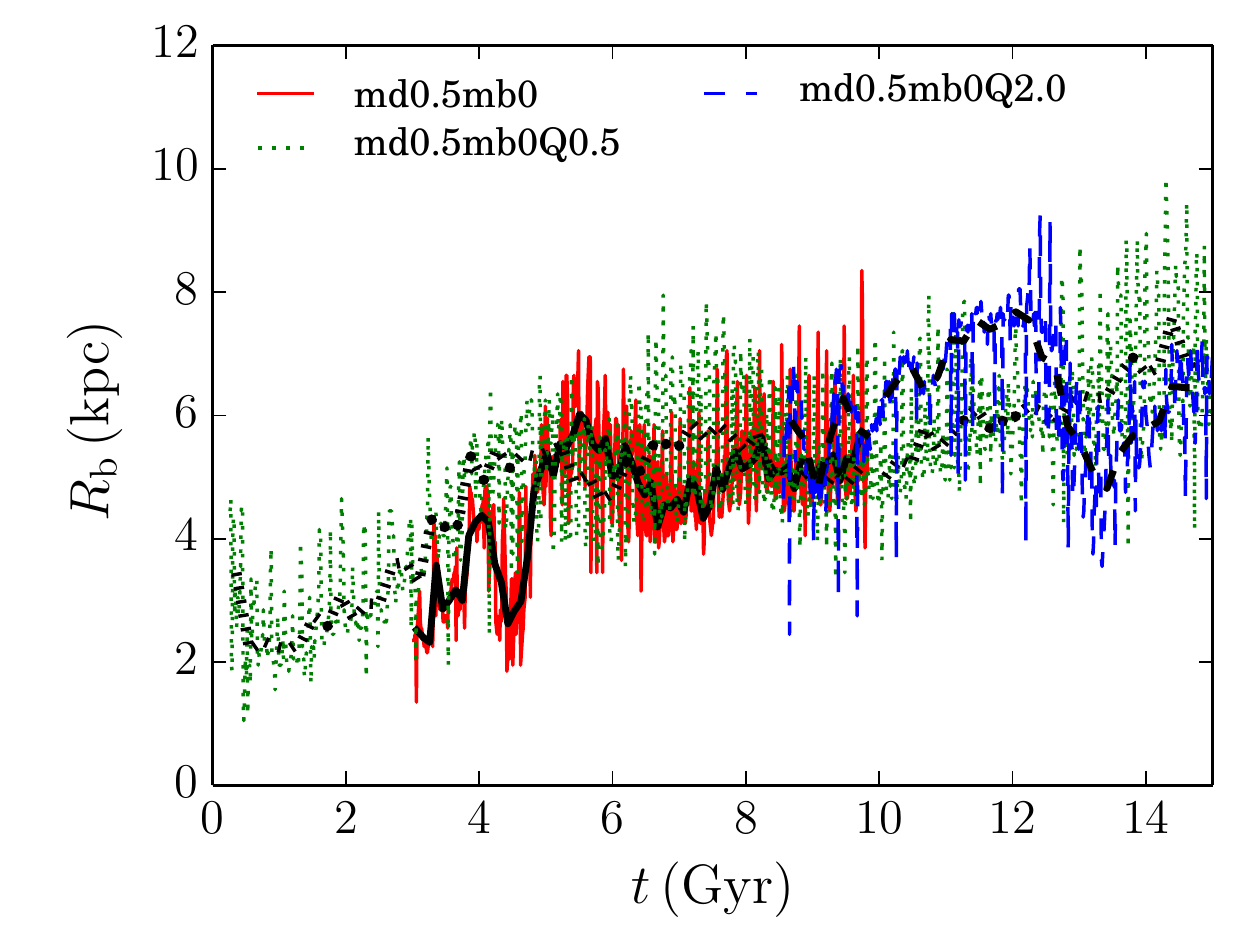}
\caption{Same as Fig.~\ref{fig:A2_max_mdisk}, but for models md0.5mb0Q0.5 and md0.5mb0Q2.0.
\label{fig:A2_max_Q}}
\end{figure*}

\begin{figure}
\begin{center}
\includegraphics[width=40mm]{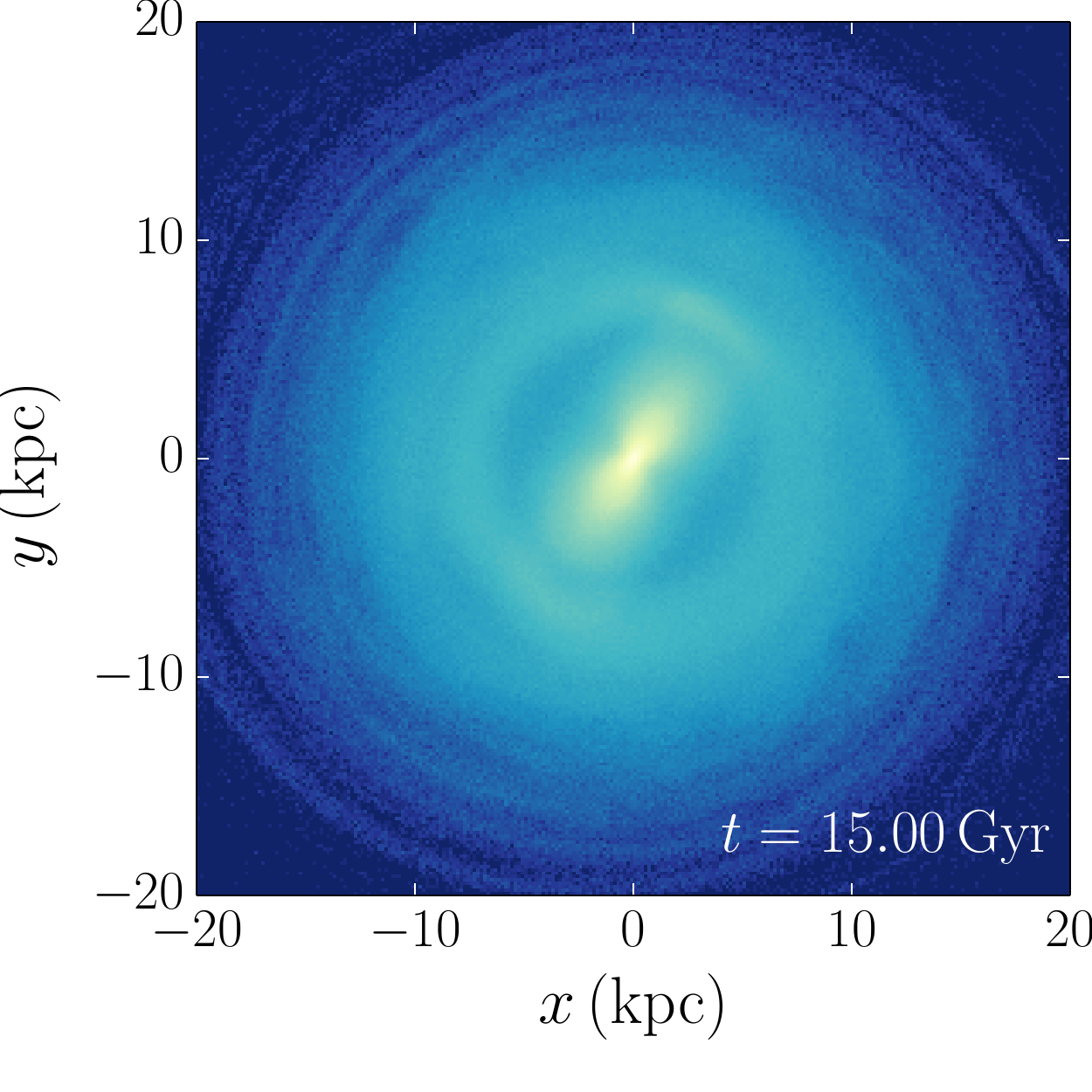}\includegraphics[width=40mm]{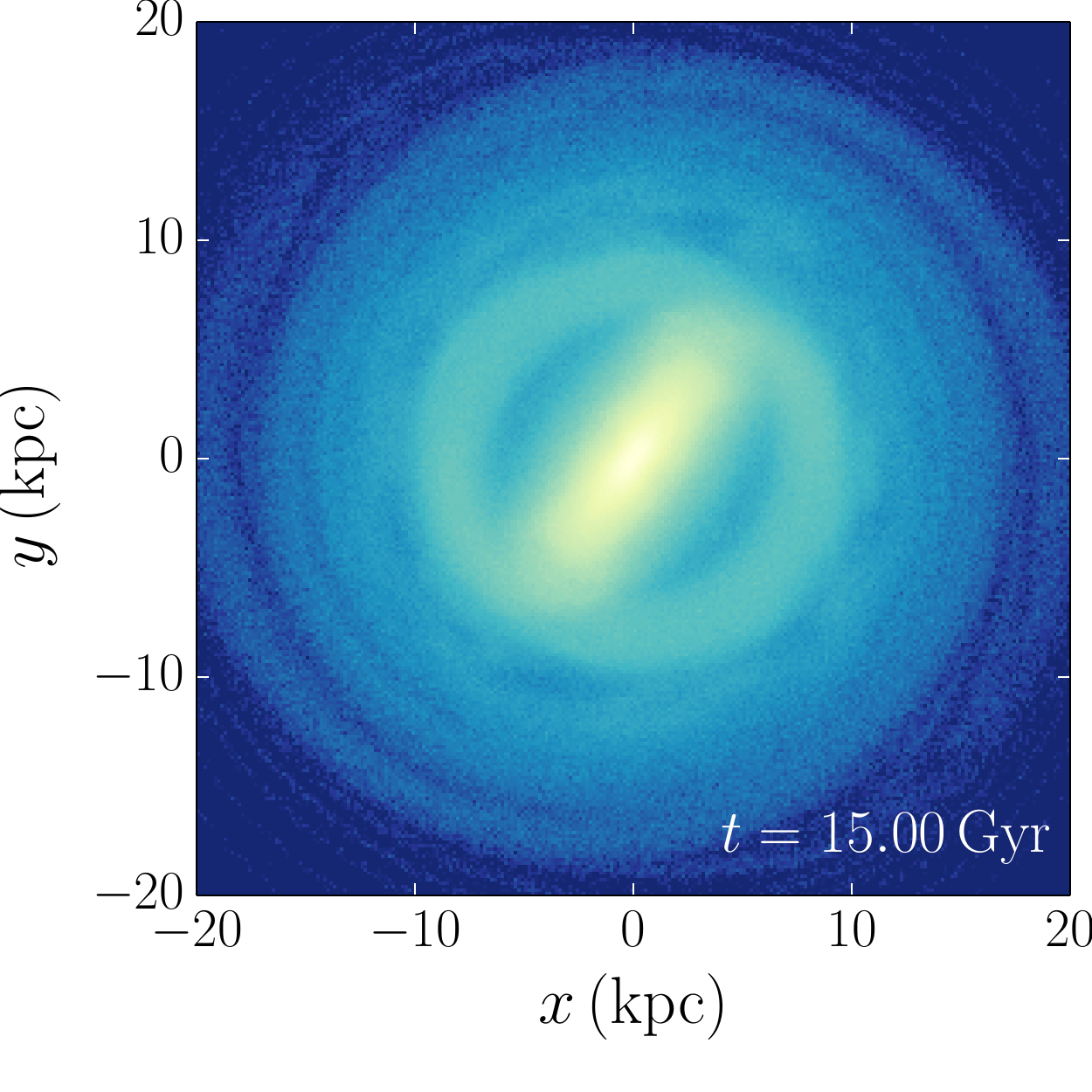}\\
    \caption{Snapshots for models md0.5mb0Q0.5 (left) and md0.5mb0Q2.0 (right).\label{fig:snapshots_Q}}
\end{center}
\end{figure}

The models are based on md0.5mb0, with one having an initially unstable disk
(md0.5mb0Q0.5) and 
the other having a large $Q_0$, in which no spiral arms develop (md0.5mb0Q2.0).
The time evolution of the bar's amplitude and length is presented in
Fig.~\ref{fig:A2_max_Q} 
and the surface densities are shown in Fig.~\ref{fig:snapshots_Q}.
For md0.5mb0Q2.0 there is no sign of spiral or bar structure until $\sim 5$\,Gyr, but 
a bar develops shortly after that (left panel of Fig.~\ref{fig:A2_max_Q}).
This matches with the 
expectation that $Q_0$ influences the bar formation epoch,
the smaller the $Q_0$ value the faster the bar forms.
The peak amplitude just after the bar formation is higher for
the larger $Q_0$, but the final amplitude is similar
(see the left panel of Fig.~\ref{fig:A2_max_Q}).
We also confirmed that the final bar length does not depend on $Q_0$
(see the right panel of Fig.~\ref{fig:A2_max_Q}). 
However, the radius that gives the maximum amplitude is different for 
the models with a large or a small value of $Q$.
The radius for $A_{\rm 2, max}$ is 2.6 and 4.9\,kpc for models with $Q_0=0.5$
and $2.0$, respectively. This result is qualitatively consistent with 
\citet{2012PASJ...64....5H} where an initially colder disk forms
a weaker and more compact bar due to the smaller velocity dispersion of the disk
(although they stopped their simulation just after the first amplitude peak).

This further proves (as discussed in Section~3.3) that the growth 
rate of swing amplification governs the bar formation timescale.
The growth rate decreases 
as $Q$ increases \citep{1981seng.proc..111T} which is confirmed by our simulations. 
With $Q_0=2.0$, the disk is initially stable and hence the spiral structure has 
to be induced by the bar. These ring-like spiral arms are sometimes seen in SB0--SBa 
galaxies such as NGC\,5101 \citep{2011ApJS..197...21H}.

\subsection{Disk scale length}

\begin{figure}
\includegraphics[width=50mm]{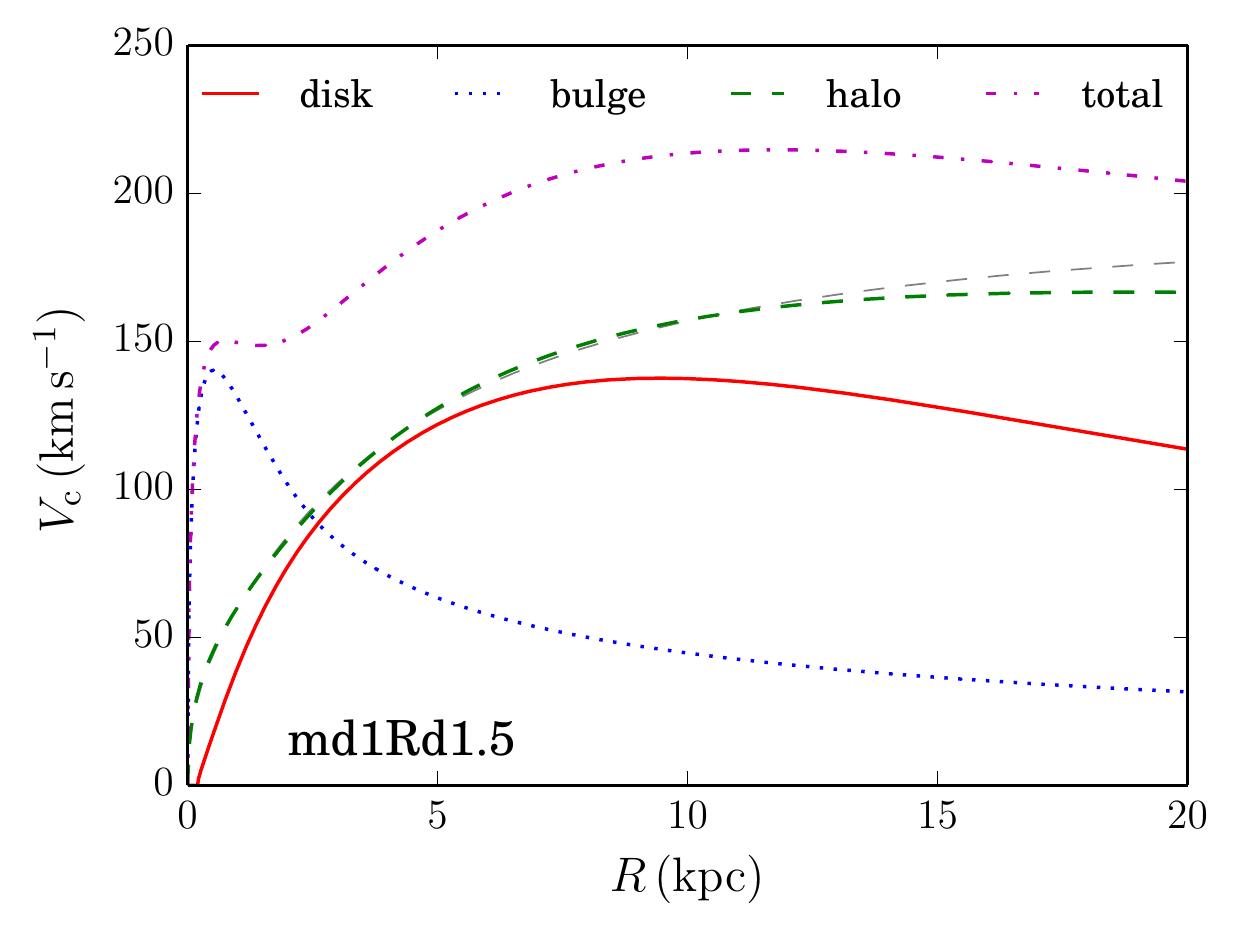}\includegraphics[width=38mm]{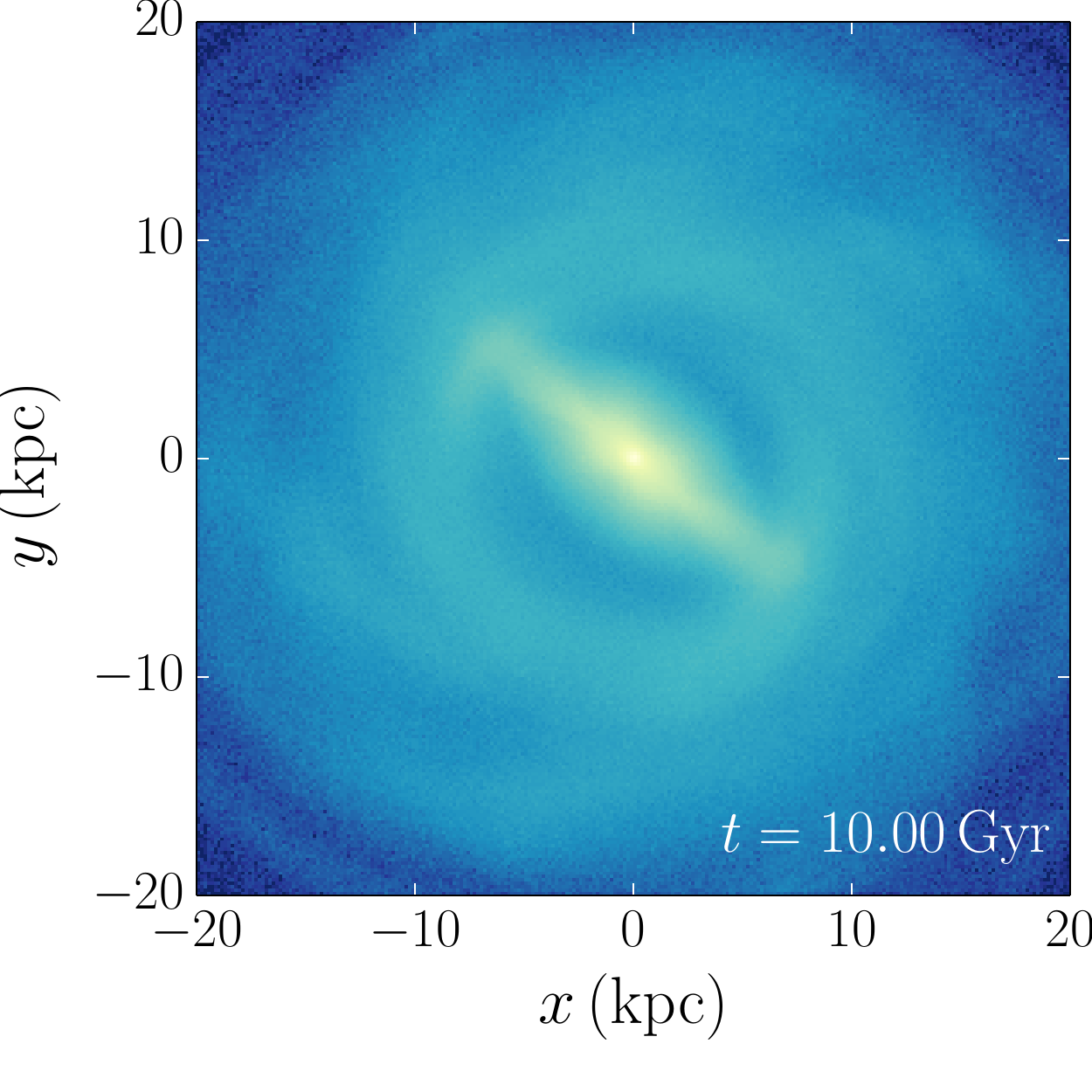}\\
\includegraphics[width=50mm]{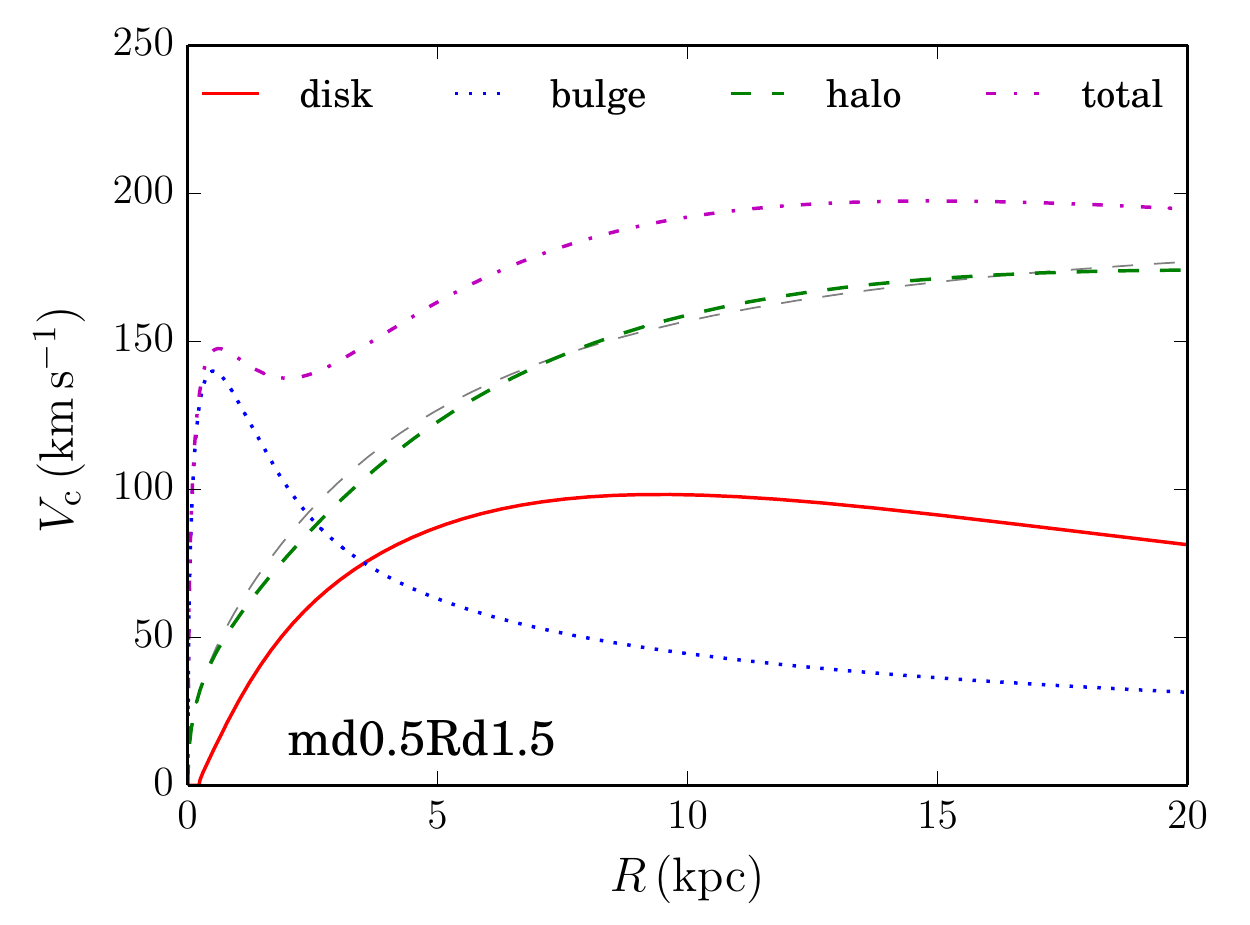}\includegraphics[width=38mm]{figures/md0.5_Rd1.5_110M_1024c.pdf}\\
    \caption{Rotation curves (left) and snapshots at 10 Gyr (right) for models md1mb1Rd1.5 (top) and md0.5mb1Rd1.5 (bottom). 
    The gray dashed curve is the same as the one in Fig.~\ref{fig:snapshots_mb10}. \label{fig:snapshots_Rdisk}}
\end{figure}

\begin{figure*}
\includegraphics[width=\columnwidth]{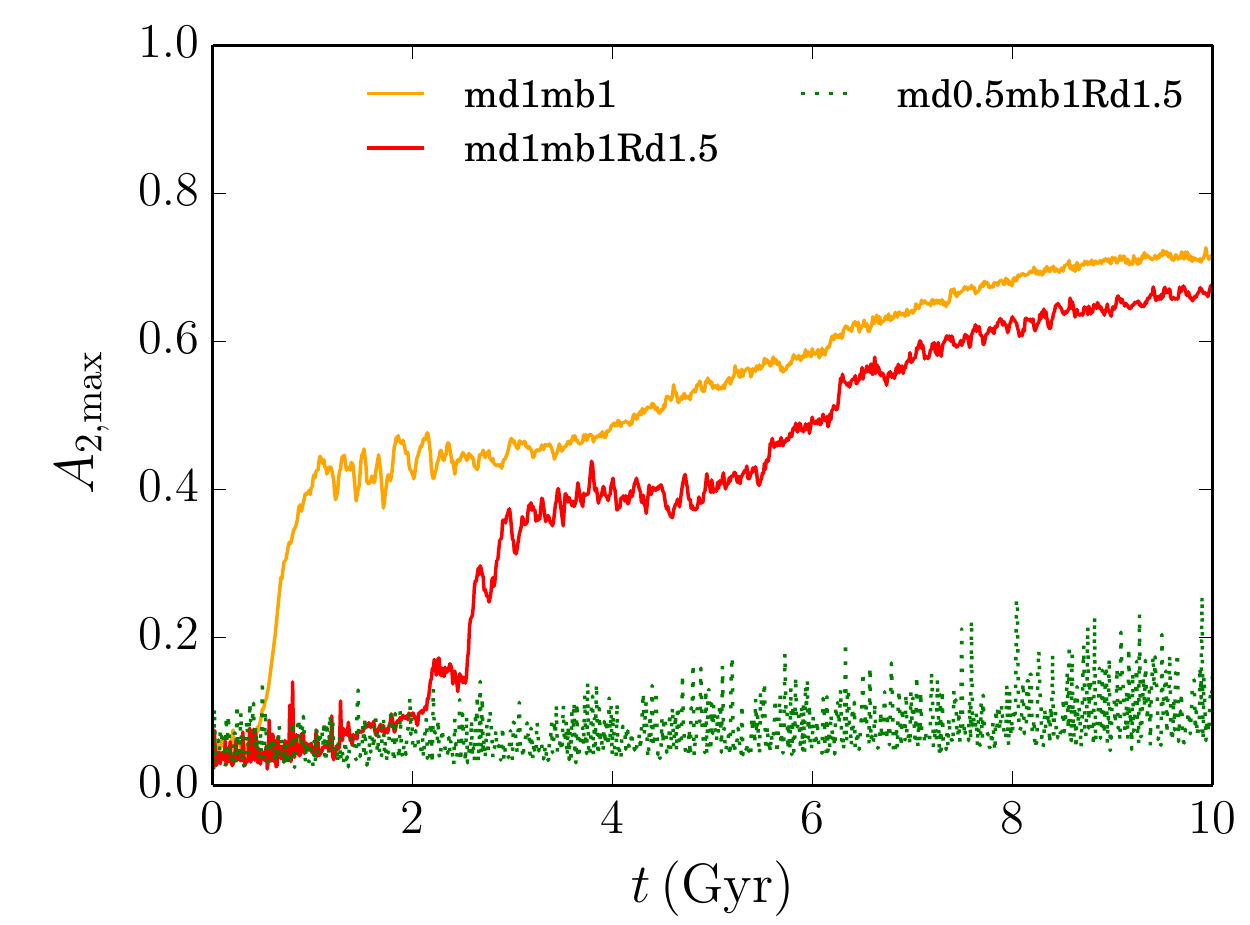}\includegraphics[width=\columnwidth]{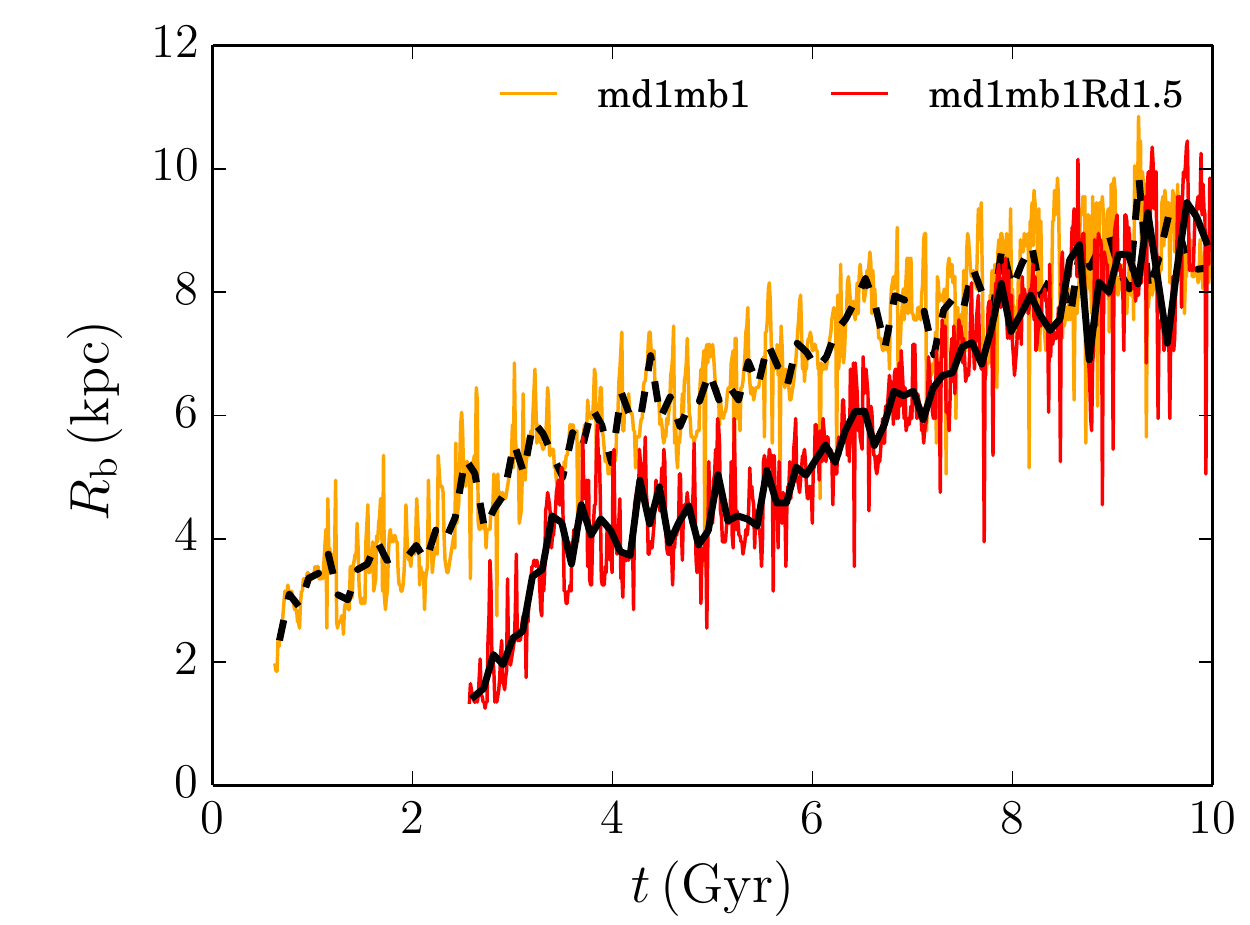}
\caption{Same as Fig.~\ref{fig:A2_max_mdisk}, but now for models md1mb1Rd1.5 and md0.5mb1Rd1.5 with md1mb1 shown as reference.
\label{fig:A2_max_Rd}}
\end{figure*}

We further examine models md1mb1Rd1.5 and md0.5mb1Rd1.5, which
have a larger disk length scale. For these models the total disk mass is the same 
as that of models md1mb1 and md0.5mb1, but the disk scale length 
is larger. The changed disk scale length results in different rotation 
curves (see Fig.~\ref{fig:snapshots_Rdisk}). Given  Eq.~\ref{eq:mX} we expect 
that this leads to fewer spiral arms. The top views of these models are presented in
Fig.~\ref{fig:snapshots_Rdisk} (right panels) and the evolution of 
the bar's amplitude and length in Fig.~\ref{fig:A2_max_Rd}. 
The bar formation epoch of model md1mb1Rd1.5 (2\,Gyr) is
later than that of model mdmb1 (1\,Gyr). Model md0.5mb1Rd1.5 did not form 
a bar within 10\,Gyr, although model md0.5mb1 formed a bar at $\sim6$\,Gyr.
The difference
between these models is that the disk mass fraction ($f_{\rm d}$) for model
md1mb1R1.5 and md0.5mb1R1.5 is smaller than those for model md1mb1 
and md0.5mb1 (see Table~\ref{tb:bar_crit}).
Although the bar formation starts later for model md1mb1Rd1.5, the bar grows
faster, and 
the final bar length at 10\,Gyr is comparable for these models.
The bar's secular evolution, however, may continue further. 
In order to understand what decides the final bar length further simulations
are required.

\section[]{Evolution of spiral arms in a live halo}
\label{Sect:AppB}

For the formation and evolution of bars, the effect of a
live halo has been investigated in previous work, and
it has been shown that the angular-momentum transfer from bars to
live halos helps the growth of bars
\citep{2000ApJ...543..704D,2002ApJ...569L..83A,2003MNRAS.341.1179A}.
Previous work, however, focused on the evolution of bars but
not on spiral arms. 
In most of the previous work rigid halo simulations were used
to study the dynamical evolution of spiral arms. As all
our simulations are performed using a live halo, we made
a comparison to the results in \citet{2011ApJ...730..109F}
to test the effect of a live halo on the spiral structure.

\begin{figure*}
\includegraphics[width=\columnwidth]{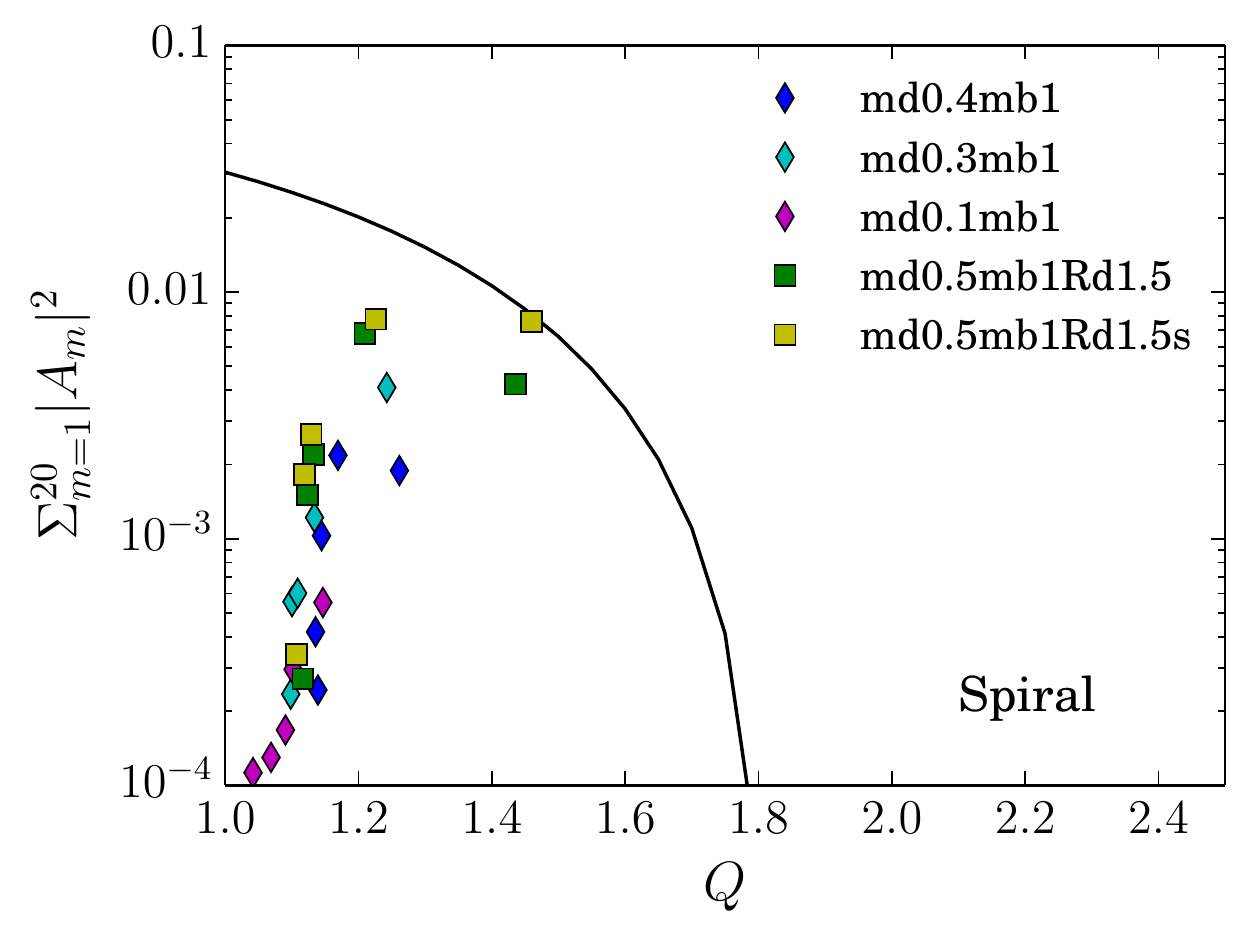}\includegraphics[width=\columnwidth]{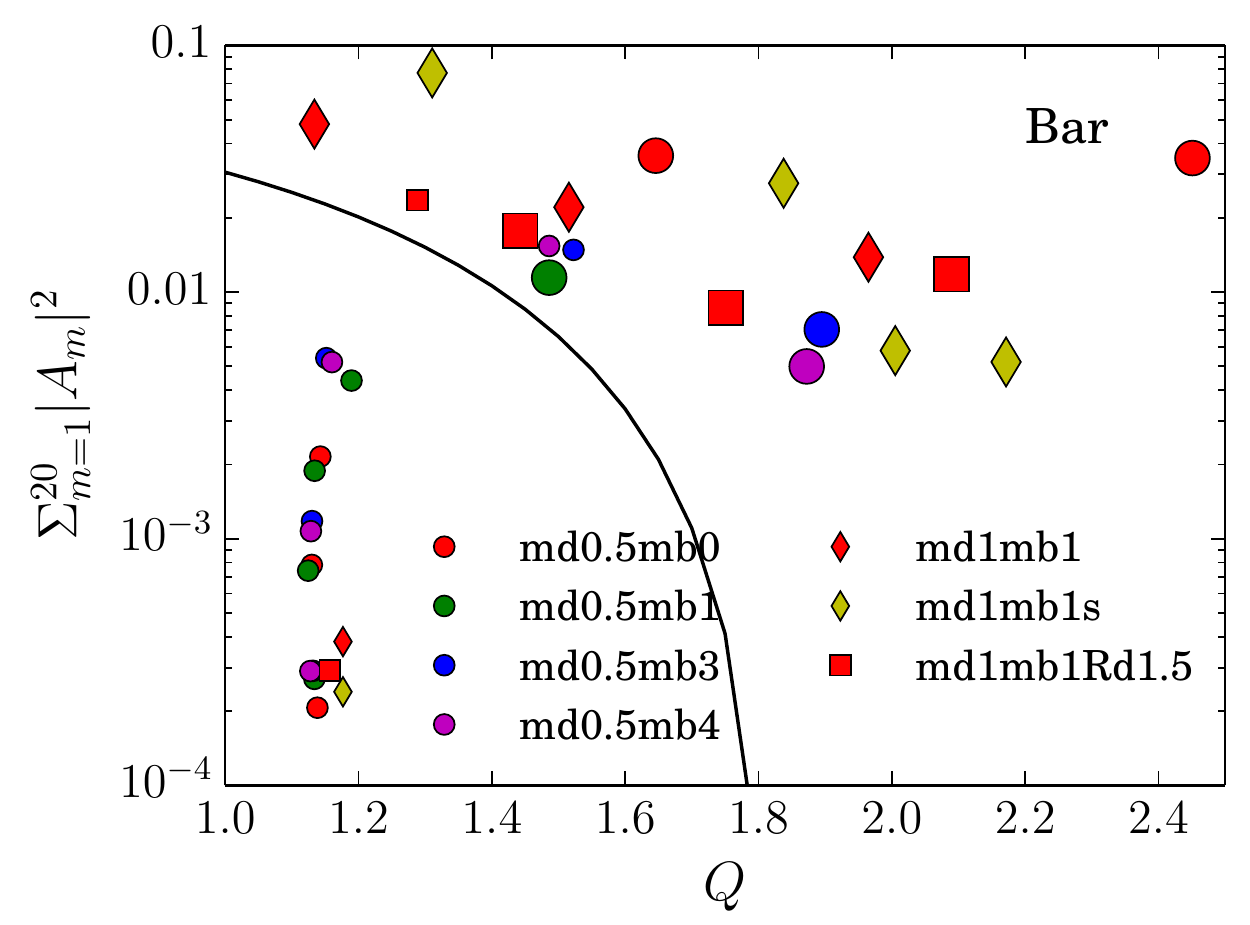}
\caption{Relation between $Q$ and the total power of the Fourier amplitudes at the reference radii ($R=6.5\pm0.5$\,kpc, except for models md1mb1Rd1.5, 
md0.5mb1Rd1.5 and md0.5mb1Rd1.5s where $R=9.5\pm0.5$\,kpc) at $t=0, 0.125, 2.5, 5$, and 10\,Gyr. 
Large symbols indicate the data points from after the bar formation.
The total power is averaged over 20 snapshots ($\sim 200$\,Myr).
Black curves indicate
Eq.~\ref{eq:amp} for $|0.1A_{\rm m}|^2$. We adopt this value in order to 
  compare with Fig.~12 in \citet{2011ApJ...730..109F}.
  \label{fig:Q_amp}}
  
\end{figure*}

In Fig.~\ref{fig:Q_amp} the relation between $Q$ and
the total amplitude ($\sum _{m=1}^{20}|A_m|^2$) at  $t=0, 0.125, 2.5, 5$, 
and 10\,Gyr is presented for our models at $R=2.2R_{\rm d}$.
The left panel shows the models which did not form a bar until 10\,Gyr. 
In \citet{2011ApJ...730..109F} we found that the spiral amplitudes
grow up to a maximum given by the following equation:
\begin{eqnarray}
A_{m} = 3.5C - 1.0 - 0.75Q^2,
\label{eq:amp}
\end{eqnarray}
where $C$ depends on the shape of the spiral arms. 
Following \citet{2011ApJ...730..109F}, we use $C=3/5$ because we assume that the local 
density enhancement due to spiral formation is described by the collapse 
of a homogeneous sphere.
If we consider that $|A_m|^2$ is similar to
the total amplitude, then the amplitudes of self-gravitating 
spiral arms can be analytically obtained as a function of $Q$. 
\citet{2011ApJ...730..109F} further found that after the spiral arm has 
reached its maximum amplitude, $Q$ increases because of the heating of 
the spiral arms and, following equation~\ref{eq:amp}, the amplitude starts to 
decay (black curve in Fig.~\ref{fig:Q_amp}).
For models without a bar we obtain the same results as for our 
models with live halos, although most of the data points are from before 
the amplitude decrease (see the left panel of Fig.~\ref{fig:Q_amp}).
Results for bar forming models are shown in the right panel of Fig.~\ref{fig:Q_amp}.
Before bar formation (small symbols), the relation between the
amplitude and $Q$ is similar to that in spiral galaxies, i.e., the amplitude
is smaller than the maximum predicted amplitude (black curve).
There are three points which exceed the limit, which was also seen in
\citet{2011ApJ...730..109F}.
After the bar forms, the developed amplitudes exceed that of the 
maximum given by $Q$, this is indicated by the larger symbols in the figures. 
The symbols are clearly above the theoretical line and $Q$ keeps increasing due
to the bar induced heating.

Snapshots of the bar models indicate that in the outer regions of the disk the spiral structure 
seems to disappear after the bar evolves (see Figs.~\ref{fig:snapshots_10Gyr} and
  \ref{fig:snapshots_spin_sp}), but the Fourier amplitudes are still large
compared to the models without bars, as shown in Fig.~\ref{fig:Q_amp}.
In the right panel we see that
the amplitude tends to decrease as $Q$ increases i.e., as the secular evolution
proceeds, except for models md1mb1,
md1mb1Rd1.5, and md0.5mb0, where the length of the bar reaches the reference radius
($2.2R_{\rm d}$)
by 10\,Gyr. In these models, we take the amplitude of the bar as the total amplitude
because the amplitude at $m=2$ is much larger than the others. 
  For these models, the relation between $Q$ and the spiral amplitude is not applicable
  at $2.2R_{\rm d}$ because the bar extends to this distance.
  Therefore, we also see that, even for barred spiral models, the spiral amplitude decreases as 
  $Q$ increases. The measured spiral amplitude, however,
  is much larger compared to the amplitude for spiral only models.

\begin{figure*}
\includegraphics[width=\columnwidth]{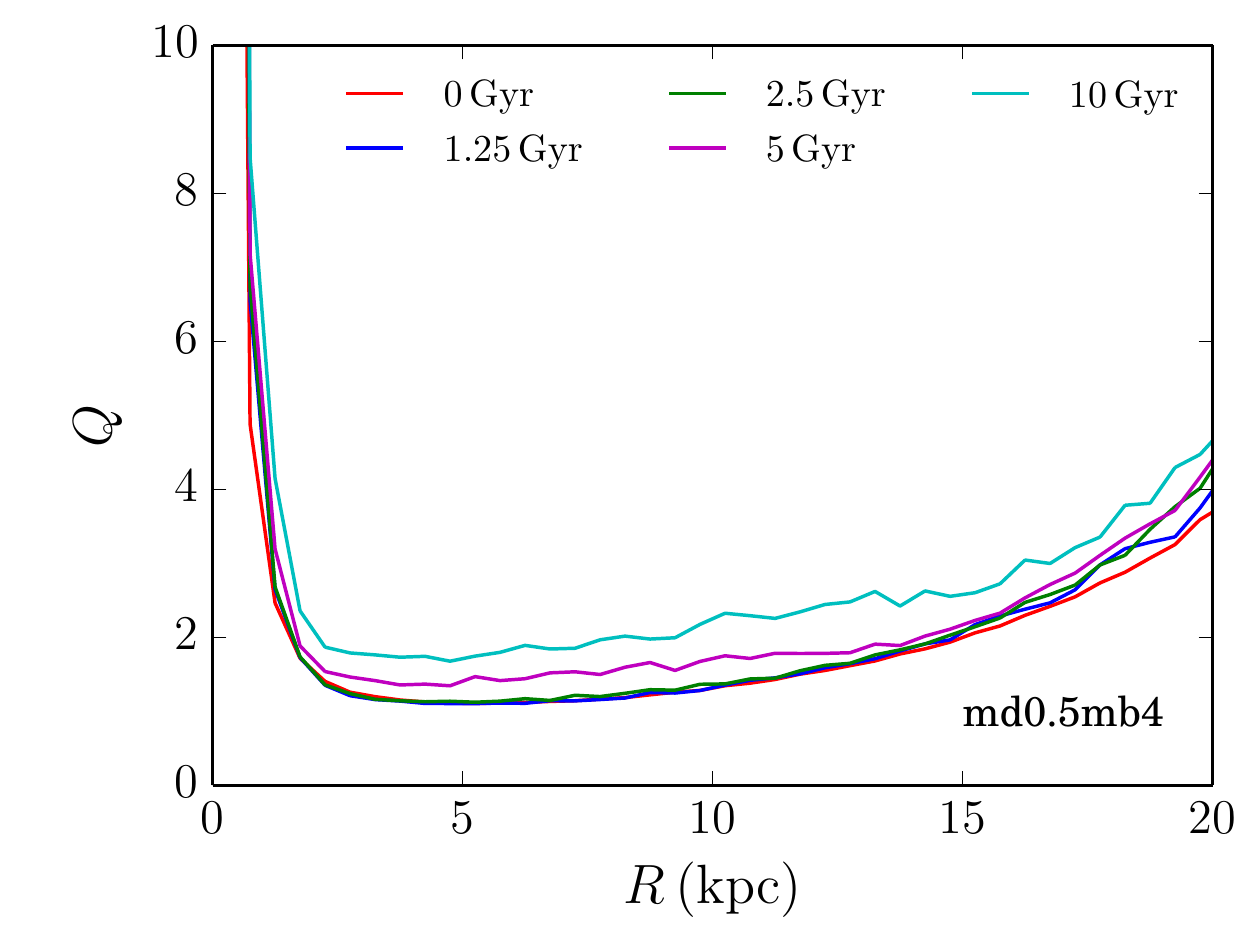}\includegraphics[width=\columnwidth]{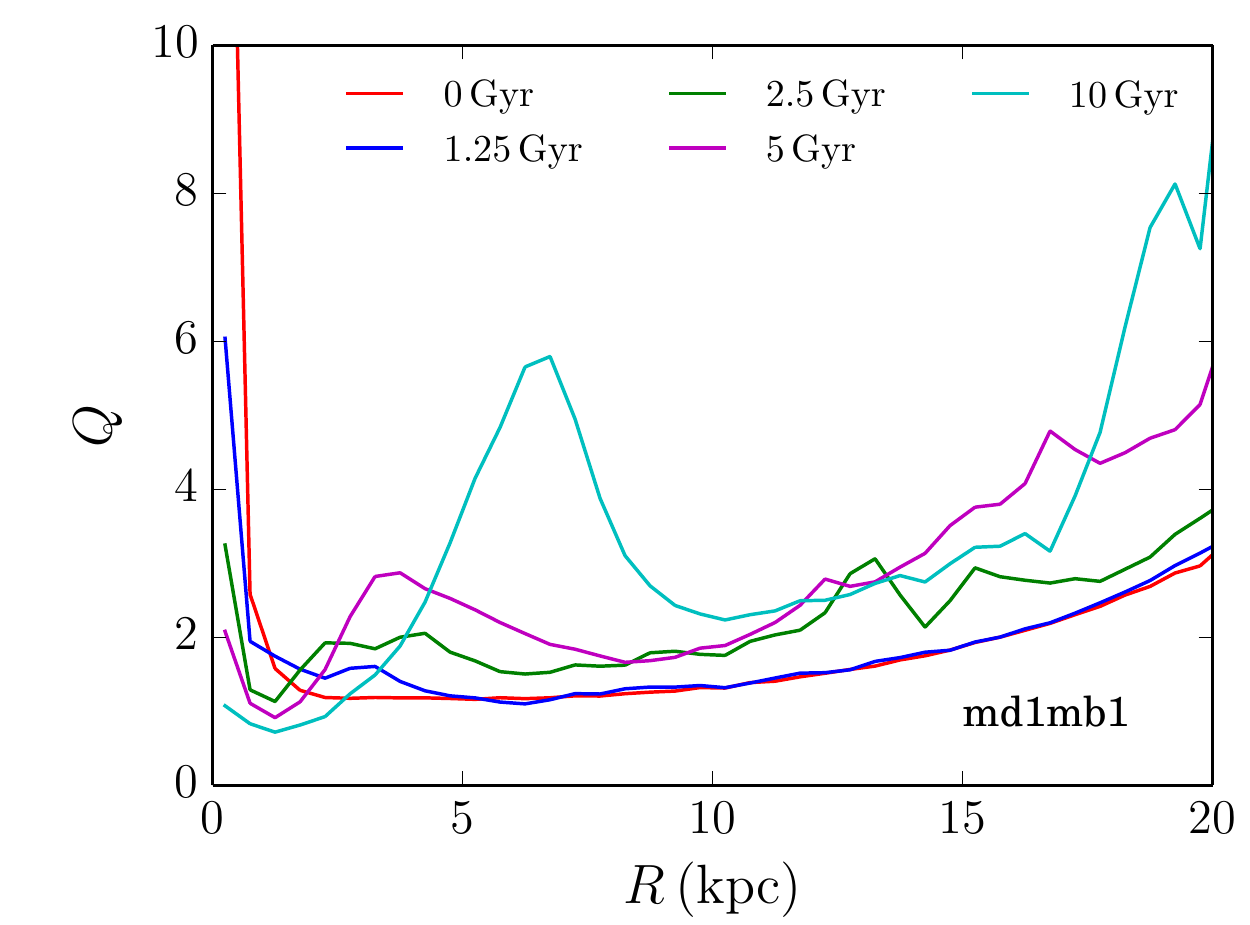}
\caption{Time evolution of Toomre's $Q$ for models md0.5mb4 (left) and md1mb1 (right). The peak in the cyan curve at $R\sim7$kpc corresponds to the end of the bar.\label{fig:Q_ev}}
\end{figure*}

The distribution of $Q$ as a function of the galactic radius for models md1mb1 (strong bar) and md0.5mb4 (without bar, although 
a bar forms after 10\,Gyr) is presented in Fig.~\ref{fig:Q_ev}. 
The spiral arms heat up the disk moderately (left panel), but the bars
heat up the disk dramatically once they are formed (right panel).

\begin{figure}
\includegraphics[width=\columnwidth]{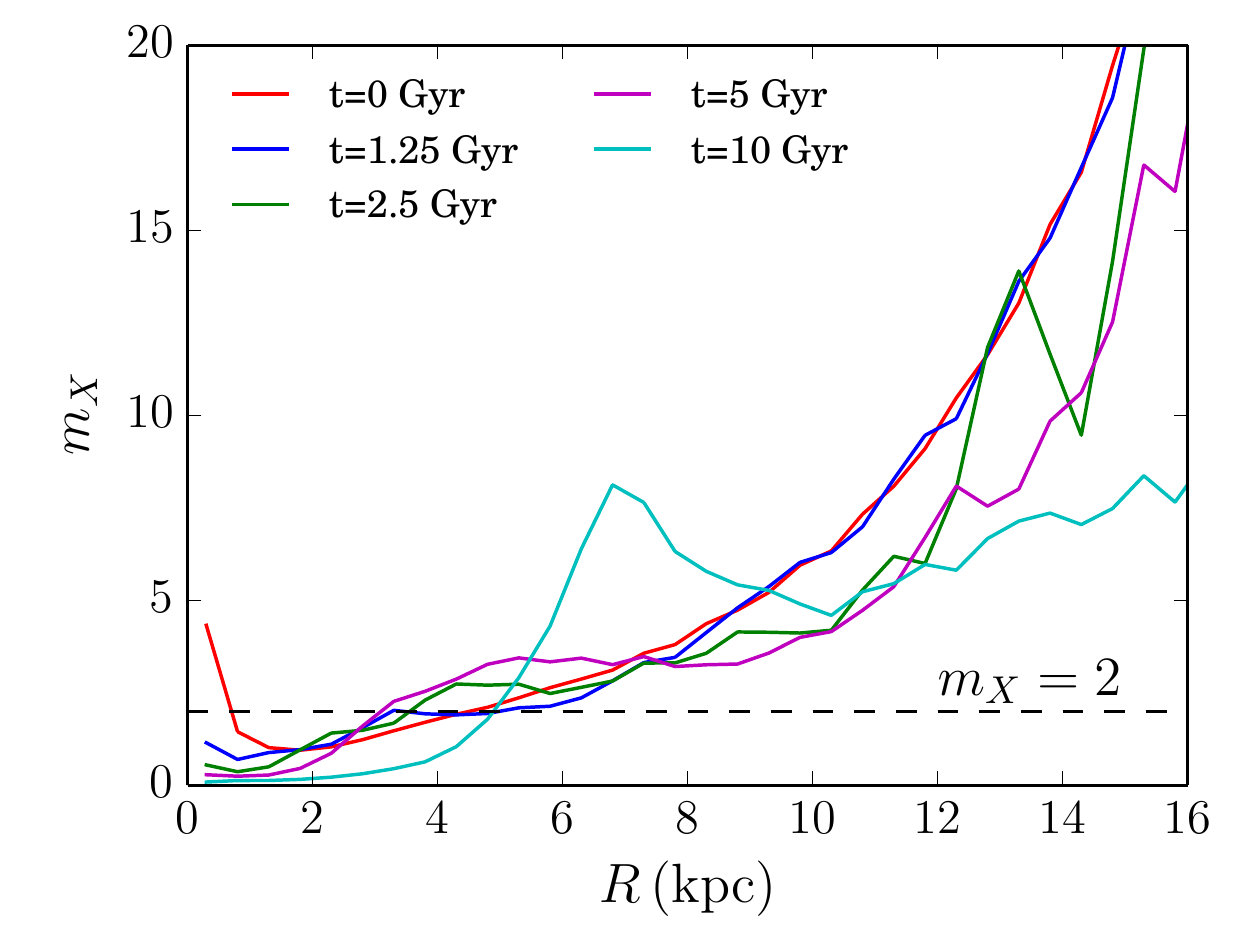}
\caption{The number of spiral arms using Eq.~\ref{eq:mX} at $t=0$, 0.125, 0.25, 5, and 10\,Gyr 
for model md1mb1. The peak in the cyan curve at $R\sim7$\,kpc corresponds to the bar end.
\label{fig:mX_md1mb1}}
\end{figure}

Another effect of the bar is that it seems to prevent the formation of self-gravitating 
spiral arms, which corresponds to the number of spiral arms expected from swing amplification.
In Fig.~\ref{fig:mX_md1mb1}, the expected number of spiral arms ($m_X$)
for model md1mb1 at $t=1.25$, 2.5, 5, and 10\,Gyr is presented.
In the outer regions of the disk, $m_{X}$ decreases as the bar develops,
but still $m_X>2$. 
However, 
when measured using Fourier decomposition (Eq. (\ref{eq:Fourier})) 
we always find $m=2$ after bar formation (see also md1mb1 in Fig.~\ref{fig:snapshots_10Gyr}).
One possible reason for this is that the high $Q$ value prevents the formation of
self-gravitating spiral arms after the bar has formed, even though 
it is known that the bar induces spirals. 
In real galaxies, we often see multiple spiral arms in the bar's outer region.
This discrepancy may be because our simulations ignore the effect of gas.

To conclude, using a live halo instead of an analytic halo potential is important when studying the 
formation and evolution of the bar and its properties in disk galaxies. Since the live halo influences the angular 
momentum of the bar, its speed and length will be rather different than when
an analytic halo is adopted. The 
effect on spiral structure without a bar, however, is less pronounced.

\section[]{Relation between bulge-to-disk mass ratio and shear rate}

In Section 3.4 we showed that the bulge-to-disk mass ratio ($B/D$) is not 
always a good indicator for the shear rate ($\Gamma$), because $\Gamma$ 
also depends on other parameters such as the disk-mass fraction ($f_{\rm d}$). 
Here, we construct additional initial conditions by sequentially changing some 
parameters in order to investigate their importance. We do not simulate 
these models, but measure $B/D$ and $\Gamma$ in the generated models at $t=0$.
All parameters of these models are summarized in Tables \ref{tb:models_add} and 
 \ref{tb:models_add2}.

\begin{figure}
\includegraphics[width=\columnwidth]{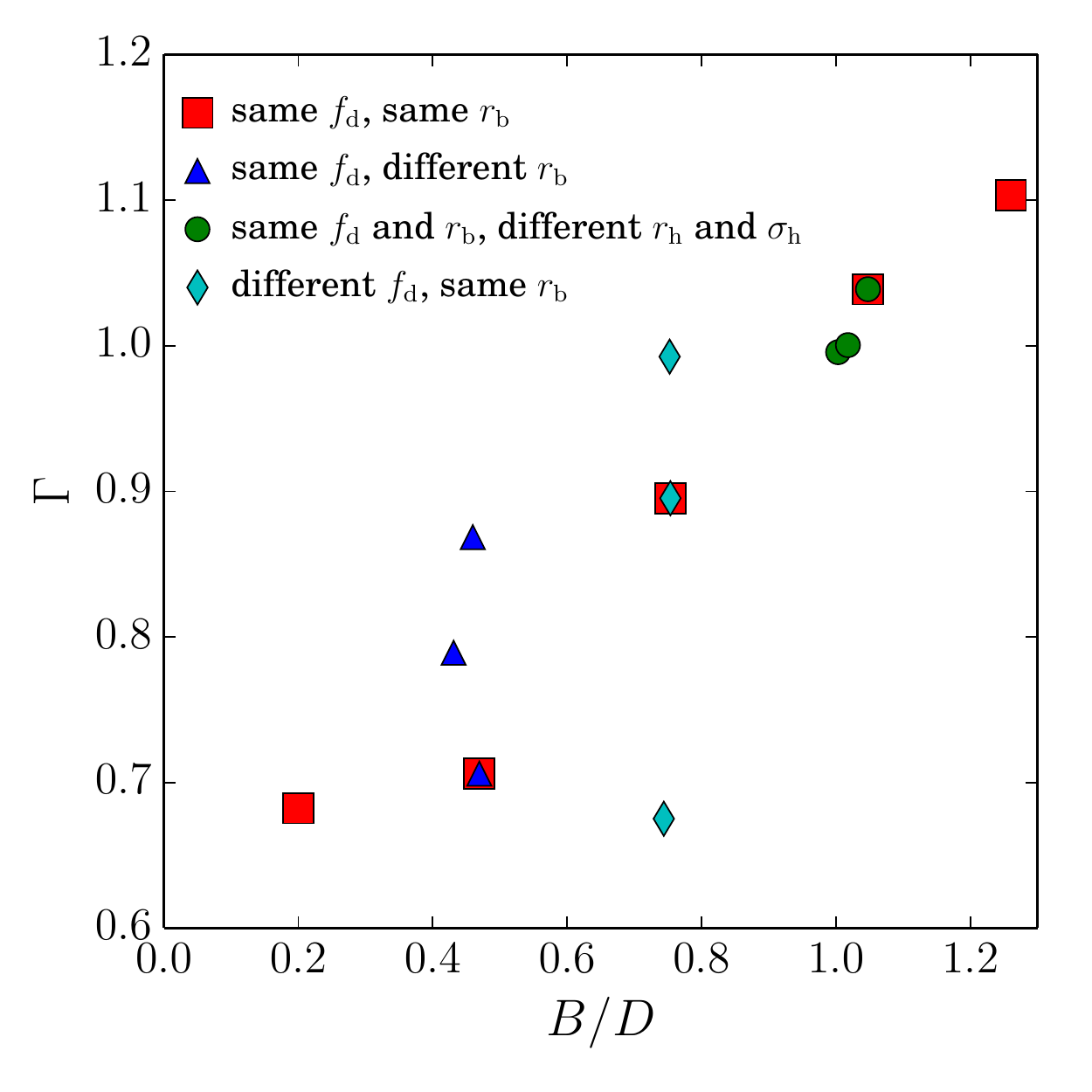}
\caption{Relation between bulge-to-disk mass ratio ($B/D$) and shear rate ($\Gamma$). \label{fig:Gamma_BD}}
\end{figure}

In Fig.~\ref{fig:Gamma_BD}, we present the relation between $B/D$ and $\Gamma$
calculated from the additional initial conditions.
If we keep both $f_{\rm d}$ and the bulge scale length ($r_{\rm b}$) constant,
$\Gamma$ monotonically increases as $B/D$ increases (square symbols).
But if we increase $r_{\rm b}$ while keeping $f_{\rm d}$ constant, then $\Gamma$ 
also increases (triangle symbols). 
If we increase $f_{\rm d}$ and keep $B/D$ and $r_{\rm b}$ constant,
$\Gamma$ increases (diamond symbols). 
The halo scale length ($r_{\rm h}$) and scale velocity 
($\sigma_{\rm h}$), on the other hand, barely affect the relation between $B/D$ and $\Gamma$
(circle symbols).

\begin{table*}
\begin{center}
\caption{Parameters for additional initial conditions\label{tb:models_add}}
\begin{tabular}{lccccccccccc}
\hline
           &  \multicolumn{3}{l}{Halo} &  \multicolumn{4}{l}{Disk} &  \multicolumn{3}{l}{Bulge} \\
Parameters &  $a_{\rm h}$ & $\sigma_{\rm h}$ & $1-\epsilon_{\rm h}$ & $M_{\rm d}$ & $R_{\rm d}$ & $z_{\rm d}$ & $\sigma_{R0}$  & $a_{\rm b}$ & $\sigma_{\rm b}$ & $1-\epsilon_{\rm b}$ \\ 
Model   &  (kpc) & ($\kms$) &  &  $(10^{10}M_{\odot})$ & (kpc) & (kpc) & ($\kms$)  & (kpc) & $(\kms)$\\
\hline \hline
Add1 & 8.2 & 350 & 0.9  & 2.45 & 2.8 & 0.36 & 105 & 0.64 & 300 & 1.0  \\ 
Add2 & 11.5 & 443 & 0.9  & 2.45 & 2.8 & 0.36 & 105 & 0.65 & 400 & 1.0  \\ 
Add3 & 8.2 & 370 & 0.9  & 2.45 & 2.8 & 0.36 & 105 & 0.64 & 500 & 1.0  \\ 
Add4 & 10 & 340 & 0.9  & 2.45 & 2.8 & 0.36 & 105 & 0.64 & 550 & 1.0  \\ 
Add5 & 8.2 & 295 & 0.9  & 2.45 & 2.8 & 0.36 & 105 & 0.65 & 600 & 1.0  \\ 
Add6 & 8.2 & 284 & 0.9  & 2.45 & 2.8 & 0.36 & 105 & 1.3 & 370 & 1.0  \\ 
Add7 & 8.2 & 330 & 0.9  & 2.45 & 2.8 & 0.36 & 105 & 0.8 & 380 & 1.0   \\  
Add8 & 8.2 & 330 & 1.0  & 2.45 & 2.8 & 0.36 & 105 & 0.64 & 550 & 1.0  \\ 
Add9 & 12 & 400 & 1.0  & 2.45 & 2.8 & 0.36 & 105 & 0.64 & 540 & 1.0  \\ 
Add10 & 8.2 & 370 & 0.9  & 1.47 & 2.8 & 0.36 & 105 & 0.64 & 390 & 1.0  \\ 
Add11 & 12 & 330 & 0.9  & 2.45 & 2.8 & 0.36 & 105 & 0.64 & 486 & 1.0  \\ 
\hline
\end{tabular}
\end{center}
\end{table*}

\begin{table*}
\begin{center}
\caption{Obtained values for additional initial conditions\label{tb:models_add2}}
\begin{tabular}{lccccccccc}
\hline
Model    & $M_{\rm d}$ & $M_{\rm b}$ & $M_{\rm h}$ & $M_{\rm b}/M_{\rm d}$& $R_{\rm d, t}$ & $r_{\rm b, t}$ & $r_{\rm h, t}$ & $f_{\rm d}$  &  $\Gamma$\\ 
   & ($10^{10}M_{\odot}$) & ($10^{10}M_{\odot}$) & ($10^{10}M_{\odot}$) & (kpc) & (kpc) & (kpc) &  &   &  \\ 
\hline  \hline
Add1 & 2.57 & 0.514 & 56.0 & 0.20 & 31.6 & 3.57 & 284 &  0.346 & 0.682 \\ 
Add2 & 2.58 & 1.21 & 137 & 0.47 & 31.6 & 5.32 & 330 & 0.343 & 0.706 \\ 
Add3 & 2.69 & 2.03 & 94.6 & 0.75 & 31.6 & 6.65 & 234 &  0.321 & 0.895 \\ 
Add4 & 2.59 & 2.74 & 124 & 1.05 & 31.6 & 8.67 & 288 &   0.340 & 1.04 \\ 
Add5 & 2.61 & 3.29 & 93.2 & 1.26 & 31.6 & 9.48 & 270 &  0.307 & 1.10 \\ 
Add6 & 2.59 & 1.19 & 45.9 & 0.46 & 31.6 & 6.28& 265 &  0.332 & 0.869 \\ 
Add7 & 2.58 & 1.11 & 61.3 & 0.43 & 31.6 & 5.31 & 251 & 0.348  & 0.789 \\ 
Add8 & 2.61 & 2.62 & 108 & 1.0 & 31.6 & 7.98 & 494 &  0.322 &  0.996\\ 
Add9 & 2.59 & 2.64 & 195 & 1.02 & 31.6 & 8.46 & 687 &  0.341 & 1.00 \\  
Add10 & 1.58 & 1.18 & 74.8 & 0.74 & 31.6 & 5.65 & 234 &  0.251  & 0.675 \\ 
Add11 & 2.75 & 2.07 & 130 & 0.75 & 31.6 & 8.00 & 324 &  0.401 & 0.992 \\ 
\hline
\end{tabular}
\end{center}
\medskip
\end{table*}

\bibliographystyle{mnras}
\bibliography{reference}

\label{lastpage}
\end{document}